\documentclass[a4paper,11pt]{article}
\usepackage{jcappub} 
\usepackage{placeins}
\usepackage{amsmath}
\usepackage{xfrac}
\usepackage{orcidlink}
\usepackage{bm}
\usepackage{multirow}

\title{\boldmath Exploring the Dark Sector: Interacting Radiation in Light of Modern Cosmological Probes}

\author[1,2]{Matteo Mescia\orcidlink{0009-0002-8166-9621},}
\author[1,2,3]{Héctor Gil-Marín\orcidlink{0000-0003-0265-6217}}
\affiliation[1]{Institut de Ci\`encies del Cosmos (ICCUB), Universitat de Barcelona (UB), c. Mart\'i i Franqu\`es, 1, 08028 Barcelona, Spain.}

\affiliation[2]{Departament de F\'{\i}sica Qu\`{a}ntica i Astrof\'{\i}sica, Universitat de Barcelona, Mart\'{\i} i Franqu\`{e}s 1, E08028 Barcelona, Spain}

\affiliation[3]{Institut d'Estudis Espacials de Catalunya (IEEC), c/ Esteve Terradas 1, Edifici RDIT, Campus PMT-UPC, 08860 Castelldefels, Spain}

\emailAdd{matteo.mescia@icc.ub.edu, hectorgil@icc.ub.edu}

\abstract{We constrain a phenomenological dark radiation (DR) framework consisting of free-streaming and fluid-like components, providing a model-independent extension of the standard radiation sector. Using \textit{Planck} CMB data, DESI DR2 BAO measurements, and Pantheon+ and DES Y5 (Dovekie) supernova samples, we derive constraints on additional relativistic degrees of freedom and assess their impact on cosmological tensions.

We obtain $N_{\rm fld}<0.66$ (95\% C.L.) from CMB data alone, while the combination with BAO yields $N_{\rm fs}=2.93\pm0.23$ and $N_{\rm fld}=0.36^{+0.16}_{-0.21}$ (68\% C.L.), consistent with Standard Model expectations for free-streaming radiation.
The DR framework significantly alleviates the Hubble tension, shifting the inferred value of $H_0$ upward and increasing its uncertainty. This is achieved through an enhanced early-time expansion rate, which reduces the sound horizon scale. The tension with SH0ES is reduced from highly significant in $\Lambda$CDM to statistically non-significant for CMB+BAO data according to the $\mathcal{T}$-statistic.
Bayesian model comparison shows no decisive preference for DR over $\Lambda$CDM when SH0ES is excluded, with results in the regime of weakly disfavoured. However, including SH0ES data leads to decisive Bayesian evidence in favour of the DR scenario.

Overall, DR provides a compelling framework for resolving the Hubble tension. When CMB, BAO, Pantheon$+$ and SH0ES data are considered, we find an increased effective radiation content, $N_{\rm tot}=3.63^{+0.13}_{-0.15}$, with a fraction of free-streaming radiation, $f_{\rm fs}=0.392\pm 0.026$, a reduced sound horizon scale, $r_d = 141.8^{+1.3}_{-1.2}\,\mathrm{Mpc}$, and a higher primordial helium fraction, $Y_{\rm He}=0.2530 \pm 0.0017$, which lies at the level of approximately $\sim 2.0$–$2.9\sigma$ above direct determinations from metal-poor H\,{\sc ii} regions, while remaining broadly consistent with other abundance measurements within current uncertainties.}

\begin{document}
\maketitle
\flushbottom

\section{Introduction}
\label{sec:intro}

The term Dark Radiation (DR) typically refers to the relativistic energy density in the early universe arising from weakly interacting particles beyond the Standard Model (BSM), which adds an extra radiation density budget --new light relics-- to the usual photons and the standard model neutrinos. These light relics, which can range from sterile neutrinos to axions and dark photons \cite{Turner_1987_axions,Baumann_2016_axions, Abazajian_2012_Sterile}, are often produced via thermal processes in the high-temperature environment of the early universe\footnote{Note that the standard model neutrinos could be also considered Dark Radiation (as mathematically they behave identically as DR), but in this paper, as it is customary in the literature, we reserve this term only for exotic species beyond the Standard Model.}. The fundamental constraints on the total amount of radiation in the universe can be established through the Big Bang Nucleosynthesis (BBN), via the observational limits on the amount of primordial helium, which restrict the effective number of relativistic degrees of freedom. This accounts for the standard neutrinos and any additional light relics \cite{Steigman_Schramm_Gunn77}. More recently, observational hints from the Atacama Cosmology Telescope \cite{ACT} and early WMAP data \cite{WMAP} led to a resurgence of interest in these degrees of freedom, formalising the term ``Dark Radiation" to describe these BSM signatures \cite{Archidiaconoetal2011,Calabreseetal2011}. While subsequent \textit{Planck} data \cite{Planck_2018} aligned more closely with the Standard Model, the discovery space for light relics remains a primary driver for Stage IV surveys. Within this context, cosmological probes exploiting the early-time cosmic microwave background (CMB) epoch, such as the future LiteBird mission \cite{LiteBird}, or the Simons Observatory \cite{SO}, are extremely powerful probes of light particles and BSM physics \cite{Bashinsky_Seljak04,Baumannetal2016}.

\vspace{0.2cm}

In analogy to photons, some DR particles, such as dark photons, could act as dark boson mediators for hidden interactions within the dark sector (a dark electromagnetism), causing exotic phenomena such as dark acoustic oscillations (DAOs, \cite{CyrRacine_2014_DAO, Garny_2025_DAO}), analogous to the baryonic acoustic oscillations (BAOs) that arise from the interaction between photons and baryons. Complementary to these CMB missions, the late-time universe large-scale structure (LSS) probes, such as the Dark Energy Spectroscopic Instrument (DESI) \cite{desi}, Euclid \cite{euclid}, the Vera Rubin Observatory \cite{Vera_rubin} or the Nancy Roman Telescope \cite{wfirst}, offer the possibility to employ the BAO, redshift space distortions (RSD), the growth of structure and weak lensing signatures, to break degeneracies with CMB data and put stringent constrains on the nature of these potential DR particles. For simplicity in this paper, we solely focus on the implication of the recent BAO data from DESI Data Release 2 (DESI DR2 \cite{DESI_DR2a,DESI_DR2}) in combination with CMB information. In some cases, we also include Type Ia supernovae (SNe) data from Pantheon Plus (Pantheon+ \cite{Scolnic22,Brout22}) and Dark Energy Survey (DES, \cite{DESSNe}), and in some cases we opt to callibrate SNe via a direct distance Ladder (SH0ES, \cite{Riess_2022_SH0ES}).

\vspace{0.2cm}

Light relics are broadly classified as either fluid-like or free-streaming radiation. The former refers to particles with a short mean free path due to a strong self-interaction, while the latter refers to BSM particles that are effectively decoupled, possessing a long mean free path, with negligible self-interaction. To maintain the CMB power spectrum signal within observational constraints while varying the total radiation density of the universe, both types of light relics must be simultaneously considered accounting for their distinct impacts on the acoustic peak phases and the damping tail \cite{Baumannetal2016}. The presence of this additional radiation density directly shifts the inferred values of other cosmological parameters, such as $\Omega_m$ or $H_0$, away from their standard $\Lambda$-cold-dark-matter model best-fitting ranges. Consequently, DR is a compelling candidate for alleviating cosmological open tensions in $\Lambda$CDM, such as the Hubble tension \cite{Planck_2018, Verde_2024_H0, Riess_2022_SH0ES} and the recently reported mild discrepancy in $\Omega_m$ values between the CMB and DESI DR2 BAO measurements \cite{DESI_DR2}. 

\vspace{0.2cm}

In this paper, we aim to perform a phenomenological study on the cosmological constraints of the effective relativistic degrees of freedom by exploiting the combination of CMB data with LSS BAO data and SNe data.  While different types of radiation contribute equally to the total effective number of relativistic degrees of freedom, they differ in their underlying symmetries and their specific impacts on the matter power spectrum. Previous works on DR, such as DAO models, analyse a framework that includes only fluid-like radiation, created by the latent heat of a supercooled phase transition in the dark sector after BBN, as BBN constitutes one of the strongest constraints on DR \cite{Garny_2026_DAO}. Other ways to bypass BBN constraints are stepped self-interacting dark radiation models (stepped SIDR models) \cite{Schoeneberg_2023_SteppedDR, Aloni_2021_Step}. They consist of two strongly self-interacting species: a massless and a massive species. A boost of the total effective number of relativistic species is accomplished when, at some transition redshift, the massive species becomes non-relativistic. 

\vspace{0.2cm}

Novel results of the paper compared to previous literature come from the following reasons. First, in the paper we consider both types of radiation and use the most recent BAO and SNe datasets. Thus, producing novel and updated results. We also show that the two SNe datasets used in the paper yield results that are consistent with each other and do not improve the constraining power compared to the CMB and BAO data combination. Second, we analyse BBN predictions since we assume dark radiation is present in the BBN epoch. With this, we have been able to identify that alleviating the Hubble tension with a dark radiation framework can generate a new observable tension in the Helium fraction $Y_{\rm He}$, depending on the BBN data considered. Finally, model selection tools have been applied, showing that the DR framework is decisively preferred over $\Lambda$CDM when using the SH0ES direct measurement of $H_0$. We also compare our DR framework to Evolving Dark Energy models. Motivated by these considerations, this paper is organised as follows. In Section \ref{models}, we specify the datasets and the computational setup used for the DR framework under consideration. In Section \ref{sec:constraints_DR}, we begin by illustrating and discussing the physical effects of dark radiation on the CMB power spectrum and the observational requirements for a consistent fit. In Section \ref{sec:constraints_DRBAOCMB} present the constraints on free-streaming and fluid-like radiation derived from our numerical analysis. In Section~\ref{sec: DR_tensions}, we evaluate the potential for DR to alleviate the aforementioned tensions in the standard cosmological model. Additionally, we discuss how these components impact the expansion history. In Section~\ref{sec:EDE_vs_DR}, we compare the DR framework with evolving dark energy models by using different model selection tools. Finally, in Section \ref{discussion and conclus}, we conclude by discussing the obtained results and outlining prospects for future work.

%Novelty of the paper: Nfs+Nfld combination, model selection with w0wa. Updated results with DESI DR2 and Dovekie recalibration. Study of the importance of the Supernovae sample choice. Quantification of the Hubble tension with the T-stat? Relation with the Nils BBN covmat?

\FloatBarrier 
\section{Models and Datasets}
\label{models}

Throughout this paper, we mainly employ an early-time CMB-based dataset based on \textit{Planck}'s PR4; and a late-time BAO-based dataset on DESI DR2. In certain cases, we implement SNe datasets, and when possible, we consider the direct ladder calibration based on Cepheids. Below, we specify how these datasets are defined.
\begin{itemize}
    \item {\bf CMB}: We employ the \textit{Planck}'s PR4 dataset \cite{Tristram_2024_PlanckPR4, Carron_2022_PlanckPR4_lensing}. We incorporate the temperature (\textit{TT}), polarisation (\textit{EE}), and cross-correlation (\textit{TE}) power spectra, as well as the CMB lensing signal. Specifically we use the \texttt{Camspec} likelihood for $\ell\ge30$ for both $E$- and $T$-modes \cite{Rosenberg_2022_CamSpecPR4}, \texttt{LoLiPoP} for the $\ell<30$ $E$-modes data \cite{Tristram_2024_PlanckPR4}, and \texttt{Commander} for the $\ell < 30$ $T$-modes data \cite{Planck_2018_V}, the latter of which is based on the 2018 data release. For CMB lensing we use the \texttt{PlanckPR4Lensing} likelihood from \cite{Carron_2022_PlanckPR4_lensing}.

\item {\bf BAO}: We include the BAO data from DESI DR2 \cite{DESI_DR2a,DESI_DR2}. We utilise the seven $z$-bins spanning the range $0.1<z<4.2$, employing both isotropic and anisotropic signals when available. These correspond to the bright galaxy sample (BGS at $z_{\rm eff}=0.295$), the luminous red galaxy samples (LRG1 at $z_{\rm eff}=0.510$; LRG2 at $z_{\rm eff}=0.706$; LRG3 at $z_{\rm }=0.922$), the emission line galaxy sample (ELG1 at $z_{\rm eff}=0.955$; ELG2 at $z_{\rm eff}=1.321$), the quasar sample (QSO at $z_{\rm eff}=1.484$) and the Lyman-$\alpha$ sample (Ly-$\alpha$ $z_{\rm eff}=2.330$). As done in the official DESI analysis, the LRG3 and ELG1 are combined both in a single sample, LRG3+ELG1 with $z_{\rm eff}=0.934$.

\item {\bf Type Ia supernovae (SNe)}: We make use of two SNe datasets. First, we use the Pantheon$+$ dataset \cite{Scolnic22,Brout22}, consisting of 1701 spectroscopically classified SNe ranging in redshift from $z = 0.001$ to $2.26$. Pantheon$+$ makes use of the SALT2 light-curve fitting model \cite{Guy_2007_SALT2}. Second, we make use of the DES-Year5 Supernovae dataset (DESY5,\cite{DESSNe}) with the Dovekie reanalysis (DESY5-Dovekie \cite{Popovic_2025_DES_Evolving, Popovic_2025_Dovekie_Calib}), consisting of 1635 spectroscopically classified SNe ranging in redshift from $z = 0.1$ to $1.13$, and 194 SNe at $z < 0.1$. DESY5-Dovekie makes use of the SALT3 light-curve fitting model \cite{Kenworthy_2021_SALT3}.

\item {\textbf{Direct Distance Ladder (SH0ES)}: We employ the direct measurement of $H_0$ from SH0ES collaboration \cite{Riess_2022_SH0ES}. This corresponds to a value of $H_0= 73.04 \pm 1.04$ km s$^{-1}$ Mpc$^{-1}$. This result is obtained by using the calibration of Cepheid variables in the host galaxies of 42 Type Ia supernovae.}

\end{itemize}

\vspace{0.2cm}

We model the effect of DR using the \texttt{CLASS} Boltzmann solver (v3.2.5) \cite{Blas_2011_CLASS2,Lesgourgues_2011_CLASS1,Lesgourgues_2011_CLASS4}, specifically employing the \texttt{N\_idr} module for fluid-like DR and the \texttt{N\_ur} for free-streaming radiation. The standard neutrino sector is modelled via the \texttt{N\_ncdm} module, with the number of massive species fixed to one. Therefore, the DR framework considered in this work has two extra variables compared to the standard $\Lambda$CDM model, associated to \texttt{N\_ur} and \texttt{N\_idr}, as we will describe in more detail in the next section. 

\vspace{0.2cm}

The cosmological parameter inference is performed using \texttt{Cobaya} \cite{Torrado_2019_Cobaya_ASCL, Torrado_2021_Cobaya}, by employing the  Metropolis-Hastings Monte Carlo Markov Chain (MCMC) sampling, with  5 walkers, imposing the usual Gelman-Rubin convergence criterion, $R-1<0.01$ \cite{Gelman_1992_Convergence}. 

\vspace{0.2cm}

Throughout our analysis we assume a flat universe and fix the sum of neutrino masses to $\sum m_\nu=0.06$ eV. 

\vspace{0.2cm}

Furthermore, the running of the primordial power spectrum index, $\alpha_s$, is fixed to zero.  We enforce BBN consistency for the primordial helium mass fraction $Y_{\rm He}$, which is self-consistently calculated using the \texttt{PArthENoPE} code (v3.0.0) \cite{Pisanti_2008_Parthenope}. 

\vspace{0.2cm}

We adopt the usual notation for the present-day fractional energy densities in baryons, cold dark matter, matter, dark energy, radiation and photons as $\Omega_b$, $\Omega_c$, $\Omega_m$, $\Omega_\Lambda$, $\Omega_r$ and $\Omega_\gamma$, respectively, and we define their corresponding physical densities as $\omega_i=\Omega_ih^2$, where $h$ is the usual `little $h$' or reduced Hubble constant parameter, $h\equiv H_0/(100$ km s$^{-1}$ Mpc$^{-1}$). 

\vspace{0.2cm}

Unless stated otherwise, we apply uniform and wide priors on the physical densities, $\omega_i$, the spectral index $n_s$, the primordial amplitude $A_s$, and the total number of relativistic species for both free-streaming and fluid-like, with the latter having a physical bound $>0$. In Appendix~\ref{sec:priors} we explicitly display all priors used on both cosmological and nuisance parameters.

\FloatBarrier

\FloatBarrier  
\section{Dark Radiation effects on the CMB}
\label{sec:constraints_DR}

We first review the key CMB observables and the imprints of DR on them. We then use these signatures to phenomenologically characterise possible extensions of Standard Model light relics, focusing on those that remain consistent with current high-precision CMB measurements.

\vspace{0.2cm}

We consider that the radiation density of the universe is divided into three distinct components:  photons, standard model neutrinos (SM neutrinos), and BSM light relics. In this framework SM neutrinos are parametrised as free-streaming radiation, while BSM light relics are treated as fluid-like radiation. Although photons behave as fluid-like radiation prior to recombination due to interactions with electrons via Thomson scattering, they are treated as a separate and fixed component in the considered parametrisation. Following this classification, the total radiation density is given by

\begin{equation}
\rho_{r} =  \rho_{\gamma} \left[1 + \frac{7}{8} \left(\frac{4}{11}\right)^{4/3} \left(N_{\mathrm{fs}} + N_{\mathrm{fld}}\right)\right].
\label{eq:DR}
\end{equation}
We define the total effective number of relativistic species as (often also referred to as $N_{\rm eff}$),
\begin{equation}
    N_{\mathrm{tot}} \equiv N_{\mathrm{fs}} + N_{\mathrm{fld}}.
\end{equation}

Where $N_{\mathrm{fs}}$ accounts for free-streaming species, and $N_{\mathrm{fld}}$ represents fluid-like species \cite{DR_CMB}.
In the Standard Model, the relation $\left.N_{\rm tot}\right|_{\rm SM}=\left.N_{\rm fs}\right|_{\rm SM}=3.044$ holds, with $N_{\rm fld}=0$. The $N_{\rm tot}$ value is slightly higher than 3 due to plasma corrections of quantum electrodynamics, flavour oscillations and incomplete neutrino decoupling during electron-positron annihilation \cite{Baumann_2018_TASI}. Throughout this work we quantify the presence of DR by measuring deviations using $\{N_{\rm fs},\,N_{\rm fld}\}$ as extra parameters to the standard $\Lambda$CDM case. These correspond to the previously introduced modules $\texttt{N\_ur}+1$ and \texttt{N\_idr}, respectively. 

\vspace{0.2cm}

We also define the excess in relativistic species with respect to the Standard Model as $\Delta N_{\rm tot}$=$N_{\rm tot}-\left.N_{\rm tot}\right|_{\rm SM}$. Note that the factor $7/8$ in Eq.~\ref{eq:DR} follows the standard convention of parametrizing the radiation density in units of equivalent fermionic degrees of freedom. Consequently, for a bosonic DR species (such as a dark photon) with the same temperature as the neutrino background, the contribution to $N_{\rm fld}$ would be scaled by a factor $8/7$. Also, note that by keeping the $(4/11)^{4/3}$ factor, we implicitly assume that DR decouples from the thermal bath before or at the same time as the neutrinos.

\vspace{0.2cm}

Alternatively, the extra parameters $\{N_{\rm fs},\,N_{\rm fld}\}$ can be reparametrised in terms of $\{N_{\rm tot},\,f_{\rm fs}\}$, where $f_{\rm fs}$ denotes the fraction of free-streaming dark radiation, defined by
\begin{equation}
f_{\mathrm{fs}} \equiv \frac{\rho_{\mathrm{fs}}}{\rho_{r}} =
\frac{7/8 \cdot (4/11)^{4/3} N_{\mathrm{fs}}}
{1 + 7/8 \cdot (4/11)^{4/3} N_{\mathrm{tot}}}.
\end{equation}
In the Standard Model, this fraction is predicted to be $\left.f_{\rm fs}\right|_{\rm SM}=0.4087$. Throughout this work, we interchangeably present results for the DR framework in terms of either $\{N_{\rm fs},\,N_{\rm fld}\}$ or $\{N_{\rm tot},\,f_{\rm fs}\}$, depending on which parametrisation more clearly highlights the feature under discussion.
In this work, we always run our inference in the $\{N_{\rm fs},\,N_{\rm fld}\}$ parameter space, and apply uniform and wide priors, with a physical lower bound on
 $N_{\mathrm{fld}}\geq0$. In addition to the DR parameters, we also vary the rest of usual cosmological parameter, $\{\omega_{\rm c},\, \omega_b,\, H_0,\,A_s,\,n_s,\,\tau\}$, with uniform and wide priors around them (see Appendix~\ref{sec:priors} for details), where $\tau$ is the optical depth at reionisation.

\vspace{0.2cm}
Following this parametrisation of the radiation density, the angular scale of the acoustic peaks in the CMB, $\theta_s$, becomes a primary constraint on the introduction of DR. It is a nearly model-independent parameter, measured with an exquisite precision of 0.03\% by \textit{Planck} \cite{Planck_2018}, defined as
\begin{equation}
    \theta_s\equiv r_s(z_\star)/D_M(z_\star),
\end{equation}
where $z_\star$ is the redshift at photon decoupling, $r_s(z_\star)$ is the comoving sound horizon at that epoch, and $D_M(z_\star)$ is the transverse comoving distance to the surface of last scattering. The sound horizon evaluated at redshift $z$ is
\begin{equation}
    r_s(z)=\int_{z}^{\infty}\frac{c_s(z')}{H(z')}dz',
\end{equation}
where $H(z)$ is the Hubble parameter and $c_s(z)$ is the sound speed in the photon--baryon plasma. In the following, unless otherwise specified, we use the shorthand notation
\begin{equation}
    r_s \equiv r_s(z_\star).
\end{equation}

Closely related is the baryon drag epoch at redshift $z_d$, when baryons effectively decouple from the Compton drag of photons. The corresponding sound horizon,
\begin{equation}
    r_d \equiv r_s(z_d),
\end{equation}
defines the standard ruler measured by late-time LSS BAO observations, whereas CMB anisotropies are primarily sensitive to $r_s$.

\vspace{0.2cm}

When introducing dark radiation, the total radiation density increases, which in turn increases $H(z)$, especially at early times (specifically before matter-radiation equality, $z_{\rm eq}\simeq 3411$), where radiation dominates. This leads to a decrease in $r_s(z_\star)$. Consequently, to keep the observed value of $\theta_s$ within \textit{Planck}'s range, $100\theta_s=1.0411\pm0.0003$, a corresponding decrease in $D_M(z_\star)$, and thus a modification of the late-time expansion history, is required. Consequently, the introduction of DR provides a well-motivated physical mechanism to alleviate the Hubble tension between early-time-based estimates and direct measurements from the SH0ES collaboration \cite{Riess_2022_SH0ES}, as discussed in Section~\ref{sec: DR_tensions}.

\vspace{0.2cm}

In addition to $\theta_s$, other parameters must remain nearly constant when introducing DR to satisfy the stringent observational constraints from CMB data. First, the photon density, $\omega_\gamma$, is precisely fixed by measurements of the CMB blackbody distribution \cite{Fixsen_2009_CMB}. Second,  the CMB power spectrum restricts the ratio of the heights of odd and even acoustic peaks. This ratio is governed by the baryon-to-photon ratio, $R(a)=3 \rho_b/(4\rho_\gamma)= 3\omega_b a/(4\omega_\gamma)$; since $\omega_\gamma$ is fixed, the baryon density $\omega_b$ is tightly constrained by the CMB peak structure and independently by BBN, and therefore cannot vary appreciably without violating observational bounds.
Finally, the CMB data does not allow for a substantial shift in the epoch of matter-radiation equality, defined by the scale factor $a_{\rm eq}=\omega_r/(\omega_{b}+\omega_c)$ \cite{Knox_2020_Hubble}. 

\vspace{0.2cm}

\begin{figure}[htbp]
\centering
\includegraphics[width=.48\textwidth]{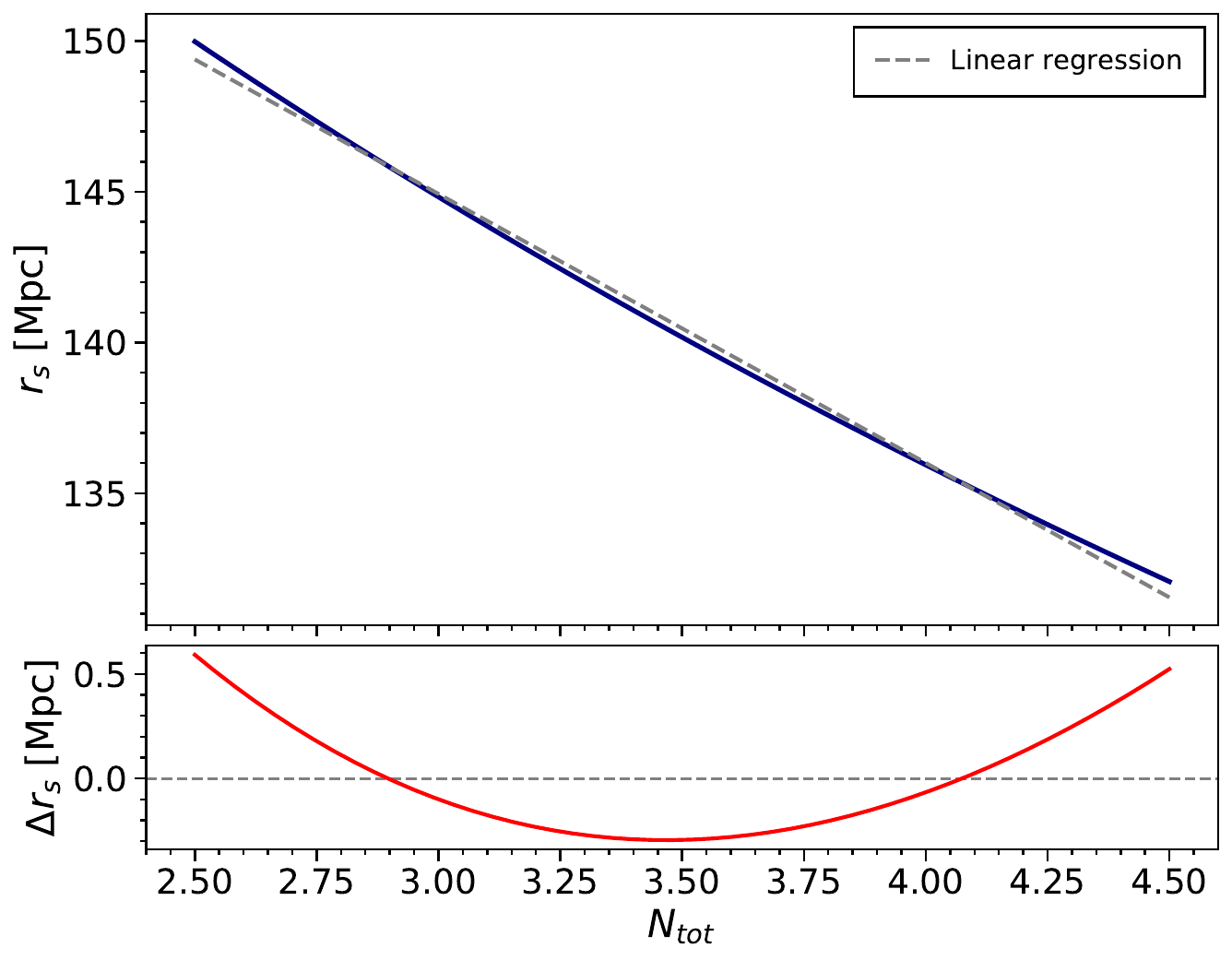}
\hfill
\includegraphics[width=.5\textwidth]{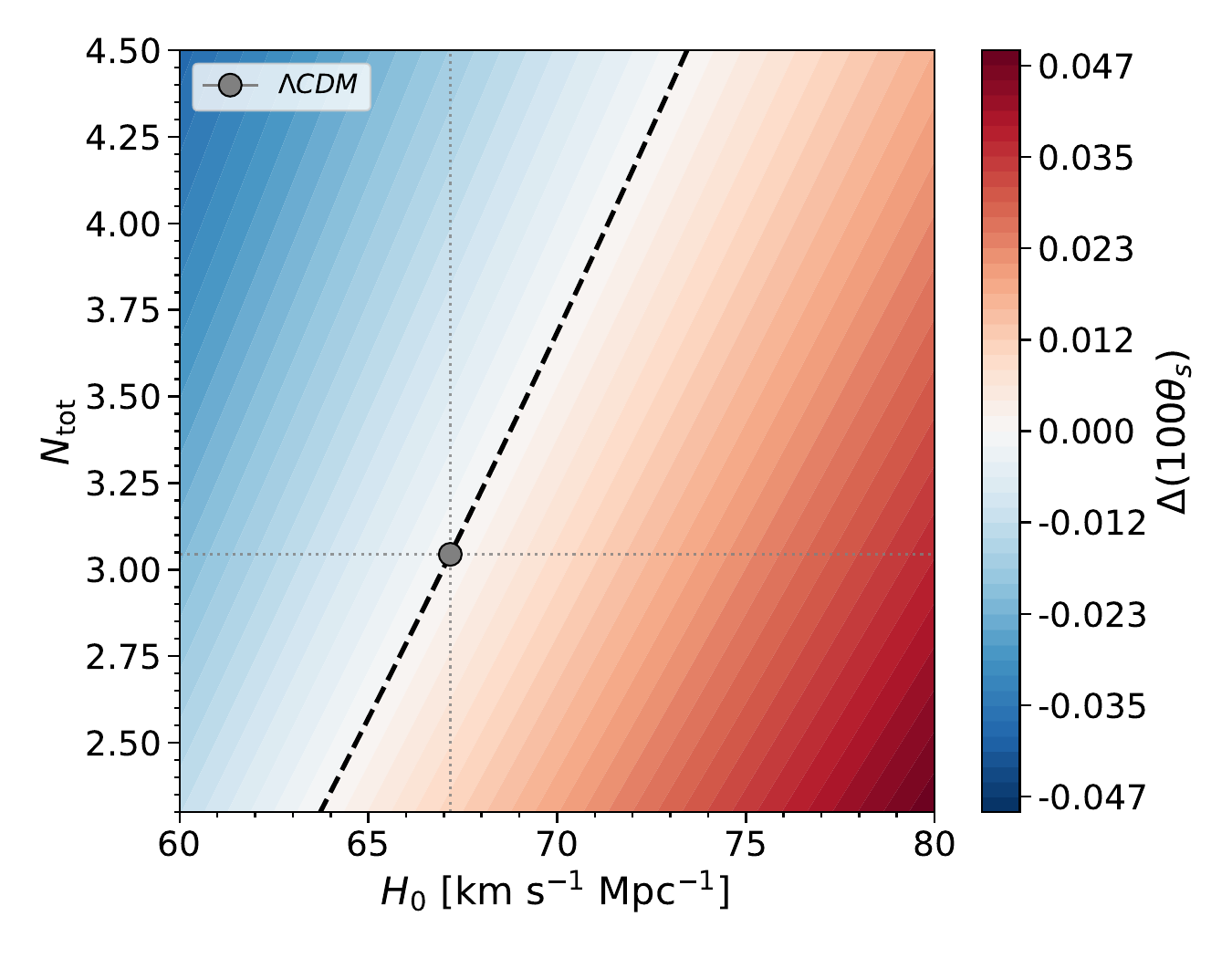}
\caption{\textit{Left panel}: The sound horizon scale, $r_s$, as a function of the total effective number of relativistic species, $N_{\rm tot}$, taking into account observational constraints. The lower sub-panel shows the non-linearity of the behaviour, relative to a linear regression displayed as a black-dashed line. \textit{Right panel}: Heatmap plot displaying the effect of changing $N_{\rm tot}$ and $H_0$ on $\theta_s$. The vertical and horizontal grey dotted lines show the fiducial cosmology values for $N_{\rm tot}$ and $H_0$ (best-fitting values from \textit{Planck}). The degeneracy direction of $\Delta(100\theta_s)\equiv\left.(100\theta_s)\right|_{\rm DR}-\left.(100\theta_s)\right|_{\Lambda{\rm CDM}}=0$ is displayed by the black dashed line.\label{fig:theta_star}}
\end{figure}

Consequently, the rest of the cosmological parameters must be adjusted: $\omega_c$ is typically modified to maintain a constant $a_{\rm eq}$ while $\omega_\Lambda$ or $H_0$ are adjusted to preserve $\theta_s$. This makes DR a compelling candidate for alleviating the Hubble tension through early-time modifications \cite{DR_CMB, DR_Notari}. As shown in the left panel of Figure \ref{fig:theta_star}, the effect of increasing $N_{\rm tot}$, makes $r_s$ smaller,  
\begin{equation}
    r_s \propto \frac{1}{\sqrt{1 + \frac{7}{8} \left(\frac{4}{11}\right)^{4/3}N_{\rm tot}}},
\end{equation}
while keeping constant $\theta_s$ can make $H_0$ larger, and thus alleviates the tension with direct measurements.

\vspace{0.2cm}

Conversely, $D_M(z_\star)$ is primarily a late-time integral and is not directly affected by the early-time radiation density. Therefore, to keep $\theta_s$ fixed, $H_0$ must be adjusted to set the required value of $D_M(z_\star)$ once $N_{\rm tot}$ is defined. This geometric degeneracy is illustrated in the right panel of Figure~\ref{fig:theta_star}.

\vspace{0.2cm}

Beyond the angular scale, the introduction of DR induces phase shifts in the high-$\ell$ acoustic peaks, arising from the different propagation sound speeds of radiation with respect to the primordial plasma \cite{Ghosh_2025_DarkMatter}. Notably, compared to a $\Lambda$CDM model without neutrinos, free-streaming DR shifts the CMB spectrum toward lower multipoles, while fluid-like DR shifts it toward higher multipoles. Including both species simultaneously is therefore essential to avoid significant alterations to the full shape of the CMB spectrum, ensuring consistency with \textit{Planck} observations.

\vspace{0.2cm}

Finally, when considering the DR framework, we must account for its role in suppressing the CMB power spectra at high-$\ell$ through Silk damping. The presence of extra radiation in the early universe increases the expansion rate, $H(z)$, which effectively makes the universe younger at a given redshift. This higher expansion rate reduces the time available for both the sound horizon, $r_s$, and the diffusion length, $r_D$, —the distance a photon travels via a random walk as it scatters off electrons— to grow. However, because $r_s$ scales linearly with time, $\propto t$, while $r_D$ scales as $\propto\sqrt{t}$, the sound horizon decreases more rapidly than the diffusion length. Consequently, the ratio $r_D/r_s$ increases with the addition of DR, resulting in an enhanced damping of the acoustic oscillations in the high-$\ell$ regime. Since this damping effect is sensitive to the total energy density, $N_{\rm tot}$, and the damping scale is independently well-constrained for a given scalar spectral index $n_s$, a strong degeneracy emerges between these parameters that must be satisfied by the observational data. Further information on the effects of DR on the damping tail can be found in Appendix~\ref{damping}.

\vspace{0.2cm}

In Appendix~\ref{sec:Cl_on_CMB}, we perform a sanity check to quantify all these effects at the same time on the $TT$, $TE$ and $EE$ power spectra $C_\ell$ when $N_{\rm tot}$ is varied at fixed $f_{\rm fs}$. Figure~\ref{fig:Cl_CMB} shows an excellent agreement with the results found in \cite{DR_CMB}, validating the employed DR implementation in \texttt{CLASS}.

\FloatBarrier  

\section{Constraints on DR from CMB and BAO}
\label{sec:constraints_DRBAOCMB}
In this section, we present the constraints on primordial cosmological parameters and the fraction of dark radiation by combining CMB with BAO data.

\subsection{Primordial power spectrum index and amplitude}
\label{sec:primordial i damping}

\begin{figure}[htbp]
\centering
\includegraphics[width=.32\textwidth]{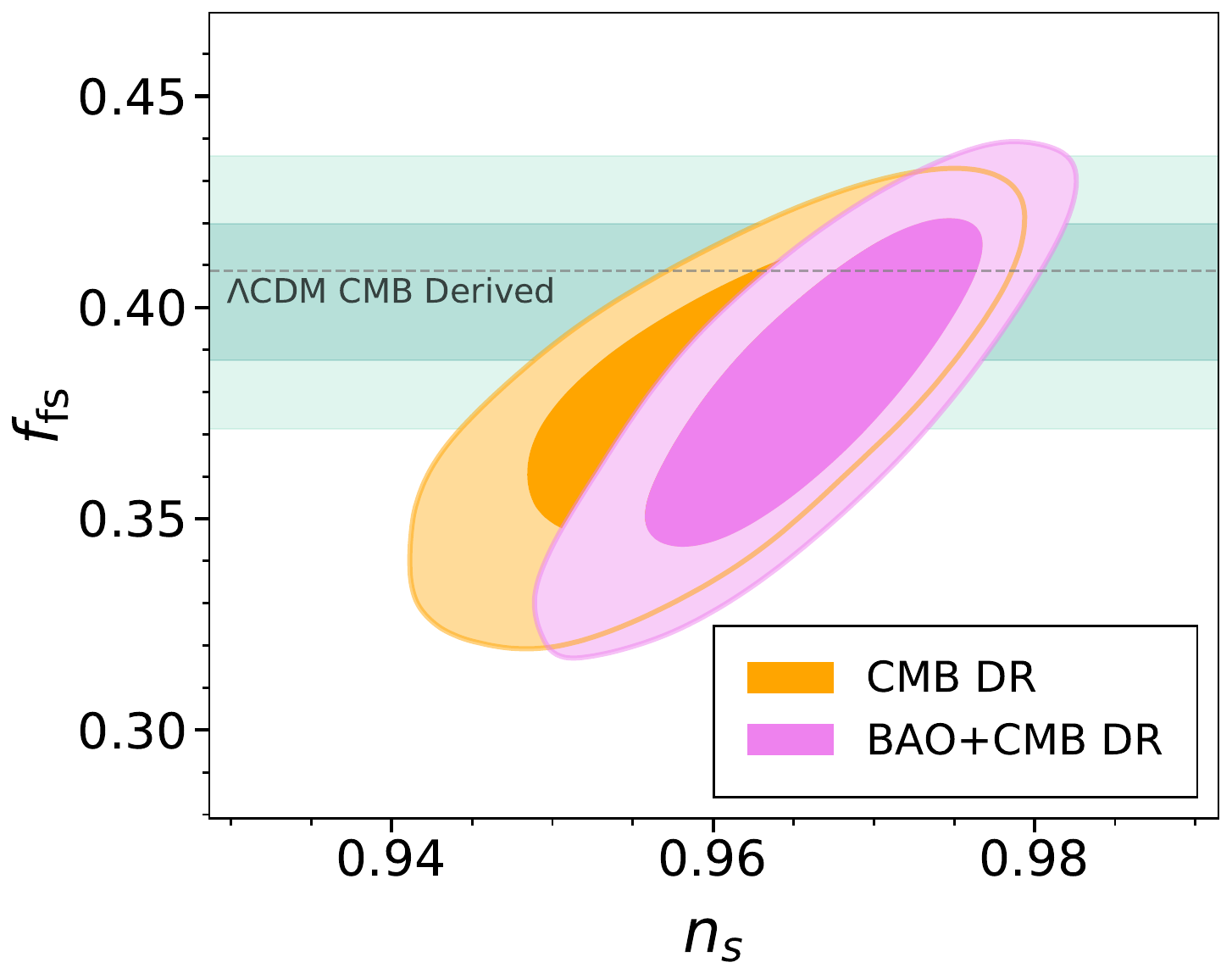}
\hfill
\includegraphics[width=.32\textwidth]{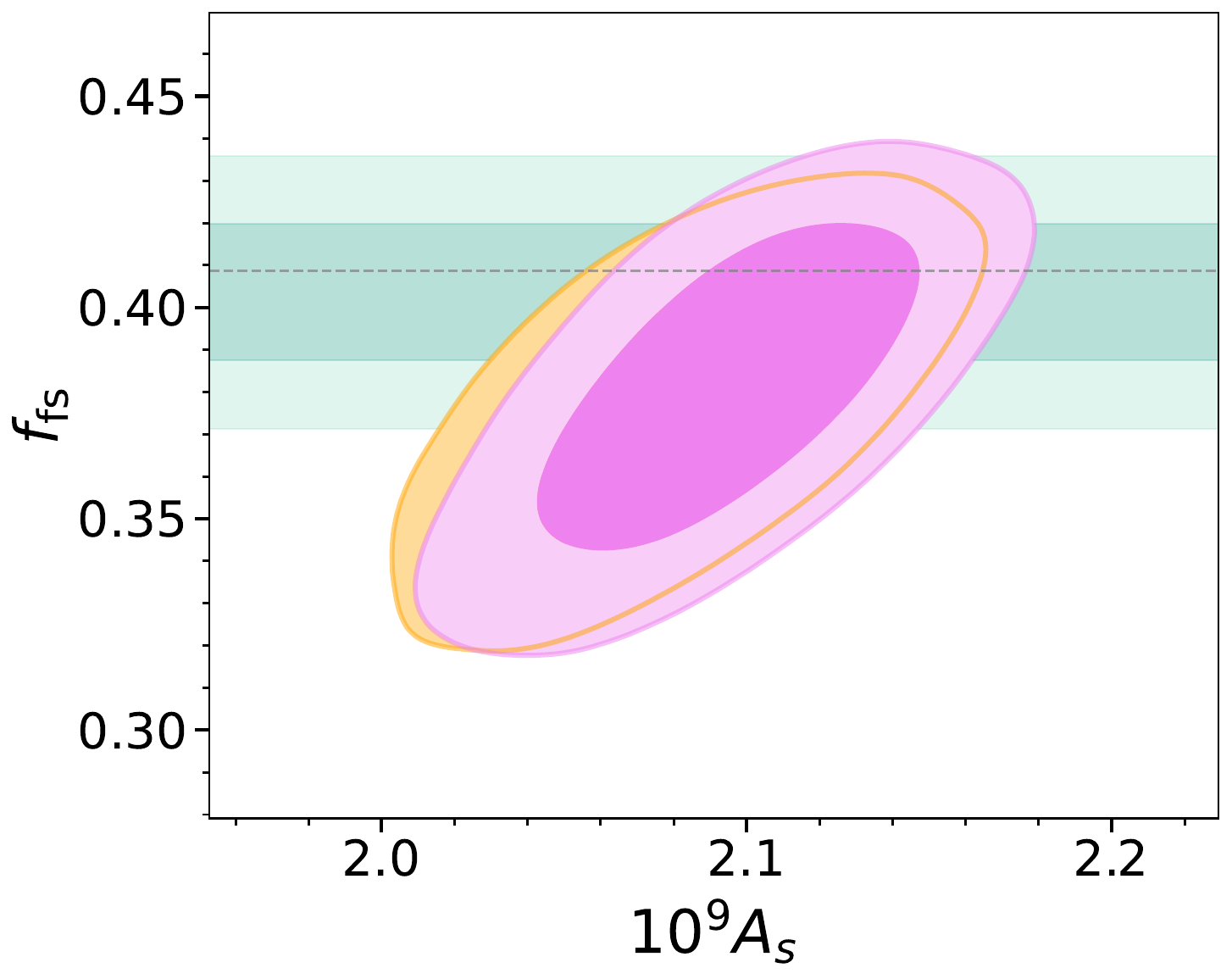}
\hfill
\includegraphics[width=.32\textwidth]{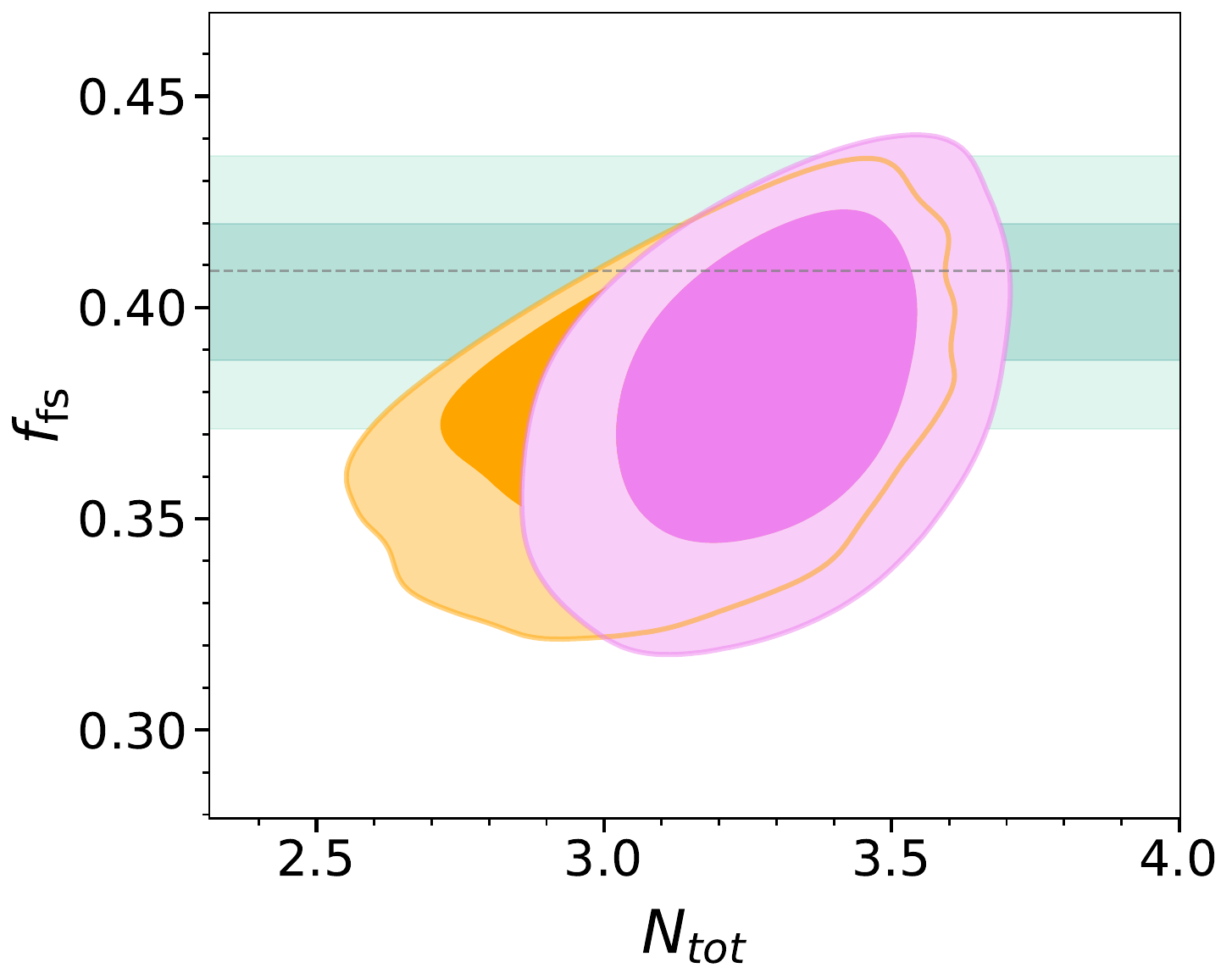}
\caption{Marginalised 2D posterior distributions (68\% and 95\% credible intervals) on $n_s$ and $f_{\rm fs}$ (left panel), on $A_s$ and $f_{\rm fs}$ (central panel) and on $N_{\rm tot}$ and $f_{\rm fs}$ (right panel), from CMB only (orange contours) and CMB+BAO (magenta contours), both computed under the assumption of a DR cosmology model. The turquoise band shows the value $f_{\rm fs}$ derived from CMB assuming $\Lambda$CDM cosmology (following equation \ref{Ntot LCDM}). For reference, the grey dashed line shows the Standard Model prediction, $\left.f_{\rm fs}\right|_{\rm SM}=0.4087$.} \label{fig:ns_PR4}
\end{figure}

We start by examining the impact of DR on the primordial parameters.
As we mentioned before, for simplicity, we consider a primordial power spectrum, 
\begin{equation}
    \mathcal{P}_{\mathcal{R}}(k) = A_s \left( \frac{k}{k_p} \right)^{n_s - 1},
\end{equation}
where $A_s$ is the primordial amplitude of the spectrum, and $n_s$ the primordial index. The pivot scale $k_p$ is conventionally fixed to $k_p=0.05$ Mpc$^{-1}$. Figure \ref{fig:ns_PR4} displays the 2D posterior distributions for these two primordial parameters and the total number of relativistic species, and the fraction of free streaming particles, $f_{\mathrm{fs}}$.  

\vspace{0.2cm}

In addition, we also display the result inferred from CMB data assuming $\Lambda$CDM, $\left.f_{\rm fs}\right|_{\rm \Lambda CDM}=0.402\pm0.015$ in turquoise horizontal bands. These results show an excellent consistency with the DR framework values. They are also in agreement with the standard model prediction shown in an horizontal dashed line, $\left.f_{\rm fs}\right|_{\rm SM}=0.4087$ (by assuming $\left.N_{\rm tot}\right|_{\rm SM}=\left.N_{\rm fs}\right|_{\rm SM}=3.044)$.
We conclude that the DR framework features are not strongly sensitive to the primordial power spectrum features. In particular the values for the spectral index under the CMB+BAO datasets,
\begin{equation}
    \left.n_s\right|_{\rm BAO+CMB}=
\begin{cases}
        0.9672\pm 0.0034 \quad(\Lambda{\rm CDM})\\
        0.9659\pm 0.0069 \quad({\rm DR})
\end{cases}
\end{equation}
are not significantly modified when introducing the DR parameters although the statistical precision decreases by a factor of two.

\vspace{0.2cm}

\begin{table}[htbp]
\centering
\renewcommand{\arraystretch}{1.4} 

\resizebox{\textwidth}{!}{%
\begin{tabular}{|l c c c c c c c| }
\hline 
Model/Dataset & $\Omega_m$ & $H_0$ & $r_d$ & $N_{\rm fs}$ & $N_{\rm fld}$ & $\Delta N_{\rm tot}$ & $f_{\rm fs}$ \\
 & & [km s$^{-1}$ Mpc$^{-1}$] & [Mpc] & & & & \\
\hline\hline

\multicolumn{8}{|l|}{\textbf{$\Lambda$CDM}} \\  
 CMB & $0.3126\pm 0.0063$ & $67.33\pm 0.46$ & $147.42\pm 0.24$  & --- & --- & --- & ---\\
 BAO & $0.2968\pm 0.0086 $ & --- &  --- & --- & --- & --- & ---\\
CMB + BAO  & $0.2989\pm 0.0036 $ & $ 68.52\pm 0.29$ & $147.72\pm 0.19 $ & --- & --- & --- & ---\\

\hline 

\multicolumn{8}{|l|}{\textbf{Dark Radiation}} \\ 
 BAO \quad & $0.2952 \pm 0.0086$ & --- & --- & $3.27^{+0.47}_{-1.2}$ & --- & $0.23^{+0.47}_{-1.2}$ & ---  \\
 CMB & $0.3088\pm0.0092$& $68.1\pm 1.6$& $146.7\pm 2.2$ & $2.85\pm 0.22$& $<0.66$ & $0.07\pm{+0.23}$ & $0.378^{+0.025}_{-0.020}$ \\
 BAO + CMB & $0.2977\pm 0.0041$ & $69.8\pm 1.1$ & $145.0\pm 1.8$ & $2.93\pm 0.23$ & $0.36^{+0.16}_{-0.21}$ & $0.25\pm 0.18$ & $0.381^{+0.028}_{-0.024}$ \\
 CMB + Pantheon$+$ & $0.3134\pm 0.0082$ & $67.4^{+1.3}_{-1.5}$ & $147.4\pm 2.0$ & $2.81\pm 0.21$ & $<0.60$ & $-0.01\pm0.21$ & $0.378^{+0.025}_{-0.020}$ \\
 CMB + DESY5 & $ 0.3141\pm 0.0080$ & $67.2\pm 1.4$ & $147.6\pm 2.0$ & $2.79\pm 0.21$ & $<0.59$ & $-0.03\pm 0.20$ & $0.376^{+0.025}_{-0.020}$\\
 CMB + BAO + Pantheon$+$ & $0.2994\pm 0.0041$ & $69.5\pm 1.1$ & $145.4^{+1.8}_{-1.9}$ & $2.91\pm 0.23$ & $0.35^{+0.16}_{-0.21}$ & $0.21 \pm 0.19$ & $0.379^{+0.027}_{-0.024}$ \\
 CMB + BAO + DESY5 & $0.3000\pm 0.0040$ & $69.4\pm 1.0$ & $145.5\pm 1.8$ & $2.88\pm 0.23$ & $0.36^{+0.16}_{-0.20}$ & $0.20\pm 0.18$ & $0.377^{+0.026}_{-0.023}$ \\

  CMB + BAO + DESY5 w. SH0ES & $0.2949\pm 0.0034$ & $71.25\pm 0.73$ & $142.6^{+1.3}_{-1.2}$ & $3.09\pm 0.22$ & $ 0.46\pm 0.20$ & $0.50^{+0.12}_{-0.14}$ & $0.389\pm 0.026$ \\

 CMB + BAO + Pantheon$+$ w. SH0ES & $0.2933\pm 0.0034$ & $71.80\pm 0.75$ & $141.8^{+1.3}_{-1.2}$ & $3.15\pm 0.23$ & $ 0.48\pm 0.20$ & $0.59^{+0.13}_{-0.15}$ & $0.392\pm 0.026$ \\

\hline 

\multicolumn{8}{|l|}{$\bm{w_0w_a}$\textbf{CDM}} \\ 
CMB + BAO + Pantheon$+$ & $0.3096\pm 0.0056$ & $67.48\pm 0.59$ & $147.58\pm 0.21$ &  --- & --- & --- & --- \\
CMB + BAO + DESY5 & $0.3110\pm 0.0053$ & $67.36\pm 0.55$ & $147.57\pm 0.21$ &  --- & --- & --- & --- \\
\hline
\end{tabular}%
}

\caption{Summary of the main cosmological parameters for different cosmological models: standard $\Lambda$CDM, the Dark Radiation framework considered in this work and a CPL parametrisation (see eq~\ref{eq:CPL}) of an evolving dark energy model ($w_0w_a$CDM). For each of these cosmologies, we provide the results when fitting a combination of CMB, BAO and SNe datasets (DESY5 and Pantheon+) as described in Section~\ref{models}. Moreover, for the DR framework, we provide the results when adding the calibrated SH0ES distances. Results quoted for all parameters are the marginalised posterior means and 68\% credible intervals in each case where two-sided constraints are possible, or the 95\% upper limits when only the upper bound constraints are possible. Recall we define, $N_{\rm tot}\equiv\Delta N_{\rm tot}+3.044$.}
\label{tab:cosmo_results_PR4}. 
\end{table}

\vspace{0.2cm}

As previously discussed, the increased expansion rate $H(z)$ introduced by DR reduces the efficiency of Thomson scattering prior to recombination. This happens because there is less time for Thomson scattering to occur before the universe becomes `too sparse'. Since there are fewer scattering events per unit of expansion, the photons can travel (or diffuse) over larger distances, which would naturally lead to an enhanced suppression (a damping) at high-$\ell$. To satisfy observational constraints and keep the high-$\ell$ CMB spectrum invariant, the angular diffusion scale $\theta_D$ must remain nearly fixed (see  Appendix~\ref{damping}). A physical mechanism to maintain a constant $\theta_D$ in the presence of DR is to allow for a variation in the free electron number density, $n_e$. This parameter depends on the baryon abundance and the primordial helium fraction, $Y_{\rm He}$. Specifically, the electron density is given by $n_e(a)=x_e(a)(1-Y_{\rm He})\rho_b/m_{\rm H}$, where $x_e$ is the hydrogen ionisation fraction, $\rho_b$ is the baryon density, and $m_{\rm H}$ is the mass of the hydrogen atom. By adjusting $Y_{\rm He}$ or $\omega_b$, the model compensates for the enhanced diffusion caused by the higher radiation density.

\vspace{0.2cm}

\begin{figure}[htbp]
\centering

\includegraphics[width=.49\textwidth]{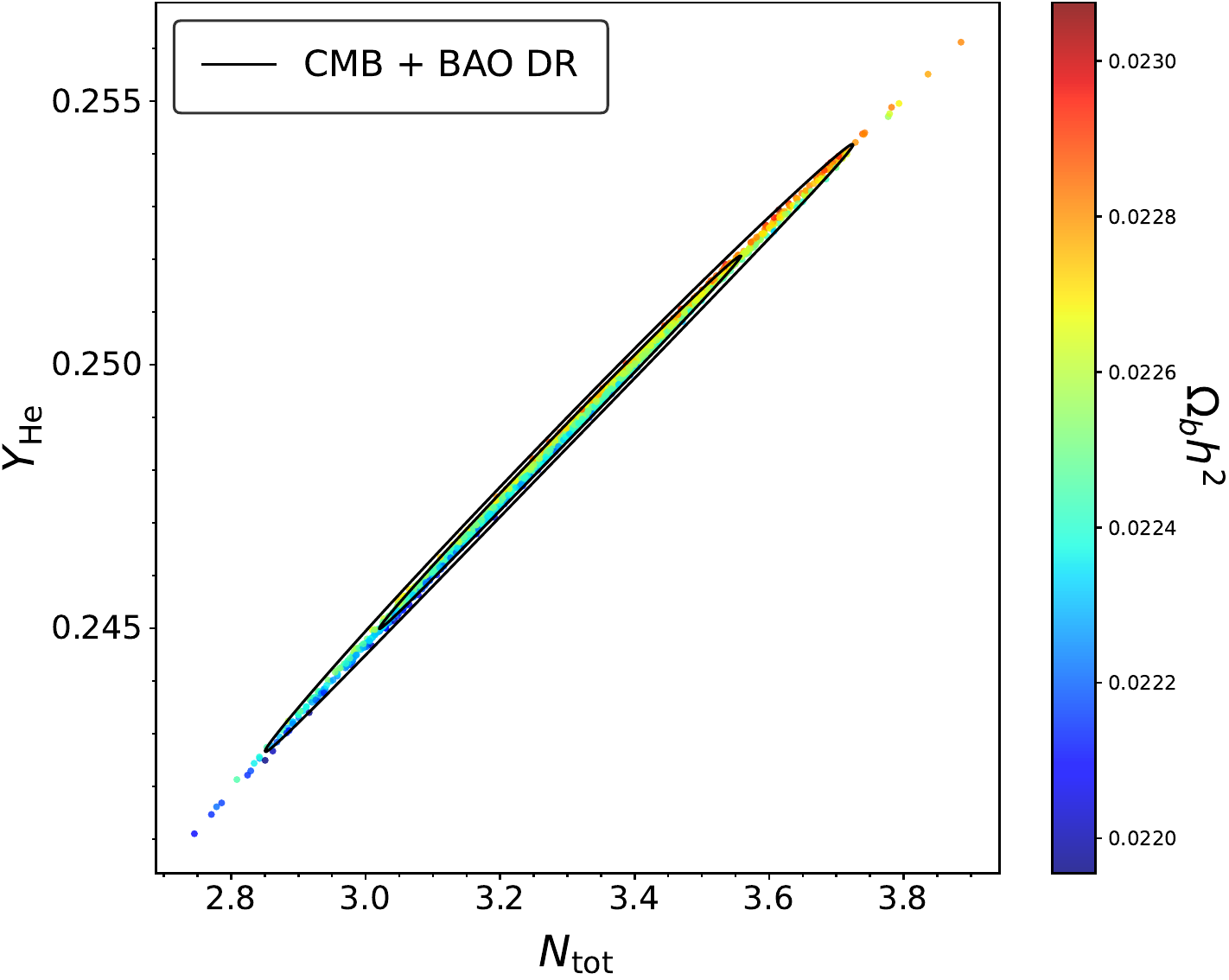}
\includegraphics[width=.49\textwidth]{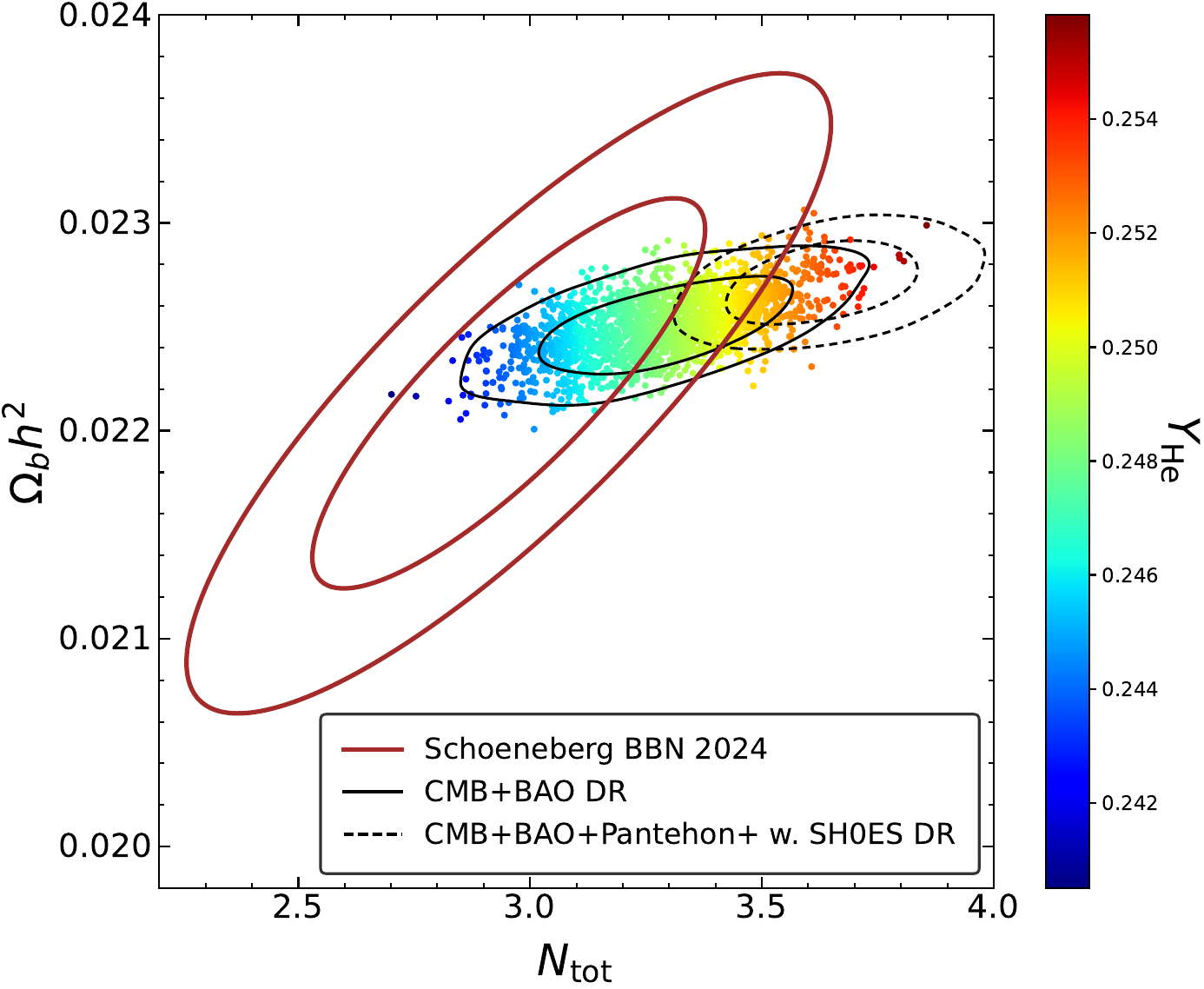}
\caption{Marginalised 2D posterior distributions (68\% and 95\% credible intervals) for $Y_{\rm He}$, $\Omega_bh^2$, and $N_{\rm tot}$ from CMB+BAO, computed under the assumption of a DR cosmology. The left and right panels display the same consistency relation, projected onto two of the three variables for complementary visualisation, while the third variable is represented by the colour scale. The burgundy elipses in the right panel shows the 2D marginalised posteriors from \cite{schoeneberg24}, derived from light-element abundances only, and is generally in good agreement with the CMB+BAO posterior under DR. The dashed black lines correspond to the DR best-fitting model to CMB+BAO+Pantheon+ SNe with SH0ES callibration (see Section~\ref{sec:EDE_vs_DR})} \label{fig:Yhe_PR4}
\end{figure}

Thus, while treating $Y_{\rm He}$ as a free parameter can help suppress the excess radiation diffusion caused by DR without altering the baryon content \cite{Hou_2013_Neutrinos}, we instead enforce BBN consistency in the present work. In this framework, the primordial helium abundance is not varied independently, but is determined by the baryon density and total radiation content through the standard BBN evolution, where the expansion rate controls the weak-interaction freezeout and the subsequent nuclear reaction network that converts the surviving neutrons into $^4$He. In particular, additional radiation increases the early-time Hubble expansion rate, causing proton-neutron freezeout to occur earlier and leaving behind a larger neutron fraction. This in turn enhances the primordial helium production, yielding a larger $Y_{\rm He}$. Consequently, $Y_{\rm He}$, $N_{\rm tot}$, and $\Omega_bh^2$ become tightly coupled through the BBN consistency relation. As illustrated in the left panel of Figure~\ref{fig:Yhe_PR4}, an increase in the number of light relics leads to a corresponding increase in the physical baryon density, while simultaneously increasing the helium fraction. Note that there is a small region of uncertainty between these three parameters (the plotted ellipses as the 68 and 95\% confidence levels), which is more evident if we display instead the number of relativistic species versus the baryon density, as done in the right panel. This happens because the BBN consistency relation does not uniquely fix $Y_{\rm He}$, $N_{\rm tot}$ and $\Omega_bh^2$ but instead predicts a correlated region with finite width arising from theoretical uncertainties in the BBN calculation, primarily associated with nuclear reaction rates and the neutron lifetime. As a result, different combinations of relativistic energy density and baryon abundance can produce similar primordial helium abundances within the allowed confidence region. For reference, in the right panel, we add as a burgundy contours the `BBN prior' presented in Ref.~\cite{schoeneberg24} derived from fitting primordial abundances, which traces those combinations of $N_{\rm tot}-\Omega_bh^2$ that preserve the light-element abundances. The difference between these two arises because the BBN-consistency contours are obtained from the full CMB likelihood, where changes in $N_{\rm tot}$, $\Omega_bh^2$, and the induced variation in $Y_{\rm He}$ simultaneously affect the recombination history and diffusion damping scale. In contrast, the degeneracy reported in \cite{schoeneberg24} solely corresponds to the pure BBN abundance relation inferred from preserving the primordial light-element yields. We see that both are in relative good agreement, as we know that BBN predictions from light-element abundances are in general good agreement with CMB observations.

\vspace{0.2cm}

As discussed previously, this shift in the baryon density is a key effect required to suppress the additional small-scale anisotropies induced by the presence of DR.
The best fit values obtained for the physical density of baryons within the DR framework (or Dark Radiation in Table~\ref{tab:cosmo_results_PR4}) are given by,

\begin{align}
\label{eq:besfit_omegab}    \left.\Omega_bh^2\right|_{\rm DR}= \begin{cases}
            0.02230 \pm 0.00022\qquad ({\rm CMB})\\
            0.02251 \pm 0.00016 \qquad ({\rm CMB+ BAO}),
           \end{cases}           
\end{align}
Compared to the $\Lambda$CDM value from \textit{Planck}, $ \left.\Omega_bh^2\right|_{\Lambda{\rm CDM}}=0.02237\pm0.00015$, the value inferred from CMB data in the DR framework is slightly lower, even though it corresponds to a value of $N_{\rm tot}$ higher than the Standard Model prediction (see Table \ref{tab:cosmo_results_PR4}). This might seem in contradiction with the previous statement that higher values of $N_{\rm tot}$ correspond to higher values of $\Omega_bh^2$. However, since the relationship between these two parameters also depends on $Y_{\rm He}$ and the value of $N_{\rm tot}$ is consistent with the Standard Model prediction, the value of $\Omega_bh^2$ can be somewhat lower than the expected one. Within the DR framework, the addition of BAO data reduces the uncertainty on $\Omega_bh^2$ by approximately 30\%. This significant improvement arises because BAO data provides a low-redshift anchor for the expansion history and the total matter density $\Omega_m$. By independently constraining the late-time geometry, BAO breaks the degeneracies between the extra radiation and the baryon-to-photon ratio, effectively pinning down the baryon density required to match the observed CMB power spectrum.

\FloatBarrier  
\subsection{Fluid-like and free-streaming fractions}

\label{fluidlike vs fs content}

We now focus on the content of fluid-like and free-streaming DR in our universe, both from CMB data alone, and with BAO. The best-fit values for $N_{\rm fs}$ are shown in table \ref{tab:cosmo_results_PR4}, which lead  to the following values for the free-streaming fraction, $f_{\rm fs}$,

\begin{align}
\label{eq:eq:best_fit_ffs}    \left.f_{\mathrm{fs}}\right|_{\rm DR}= \begin{cases}
            0.378^{+0.025}_{-0.020}\qquad ({\rm CMB})\\
            0.381^{+0.028}_{-0.024} \qquad ({\rm CMB+ BAO}).
           \end{cases}
\end{align}

\vspace{0.2cm}

The values obtained in the considered CMB and CMB+BAO models are consistent within their respective uncertainties compared to the standard model prediction value for this parameter: $\left.f_{\mathrm{fs}}\right|_{{\rm SM}}=0.4087$. 

\vspace{0.2cm}

\begin{figure}[htbp]
\centering
\includegraphics[scale=0.4]{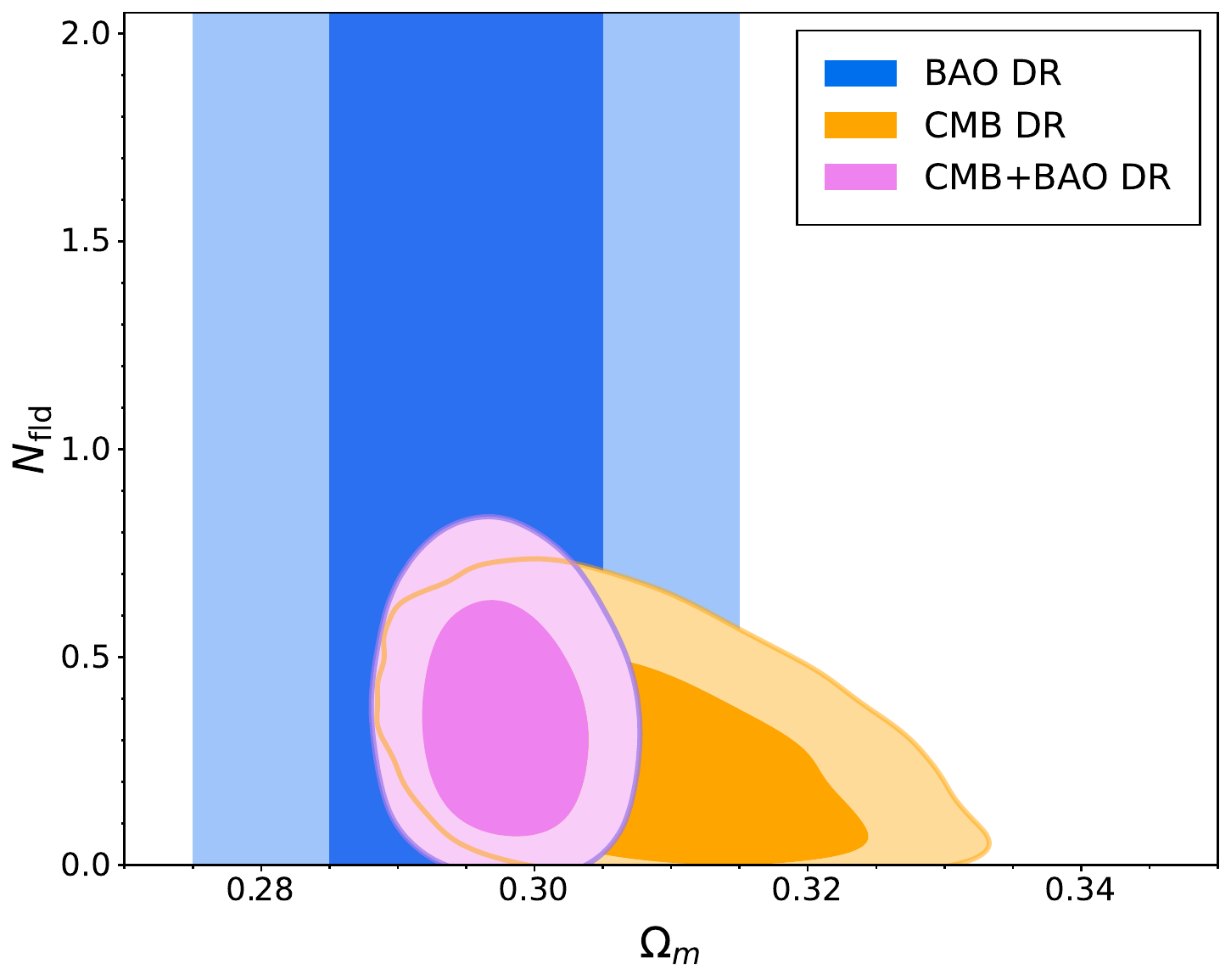}
\qquad
\caption{Marginalised 2D posterior distributions (68\% and 95\%) on $N_{\rm fld}$ and $\Omega_m$ from BAO only (blue bands), CMB only (orange contours) and CMB combined with BAO (magenta contours), all for a DR cosmology framework. \label{fig:N_fld 1D}}
\end{figure}

In Figure~\ref{fig:N_fld 1D} we display the 2D marginalised posterior distributions for $N_{\rm fld}$ and $\Omega_m$ for the DR framework, using CMB, BAO and CMB+BAO data. The inferred values for these parameters are reported in Table~\ref{tab:cosmo_results_PR4}. BAO data alone is not sensitive to $N_{\rm fld}$ as it involves data from redshifts where radiation barely has an effect on the expansion history of the universe\footnote{At $z\simeq 4$, $\Omega_r(z)\simeq 10^{-3}$, and this value decreases down to $\simeq 10^{-5}$ as $z\rightarrow0$.}. This effect can also be seen in the left panel of Figure \ref{fig:H(z)}. The total content of radiation inferred within the DR framework is,

\begin{align}
    \left. N_{\rm tot}\right|_{\rm DR}= \begin{cases}
            3.11\pm 0.23 \qquad ({\rm CMB})\\
            3.29\pm 0.18 \qquad ({\rm CMB+ BAO}).
           \end{cases}
\end{align}
These results are consistent with the standard model prediction, $\left.N_{\rm tot}\right|_{\rm SM}=3.044$, although the CMB+BAO shows an interesting (though not statistically significant) higher value. For reference, these values within the $\Lambda$CDM framework are

\begin{align}
    \left. N_{\rm tot}\right|_{\Lambda{\rm CDM}}= \begin{cases}
            2.98\pm 0.20 \qquad ({\rm CMB})\\
            3.22\pm 0.18 \qquad ({\rm CMB+ BAO}),
           \end{cases}
\label{Ntot LCDM}
\end{align}
which are also in good agreement with both the Standard Model prediction and the DR framework results, given their large uncertainties. It is interesting to note that, even in $\Lambda$CDM, CMB+BAO also yields a higher $N_{\rm tot}$ value, with respect to the standard model prediction.  In the DR framework, the high value of $N_{\rm tot}$ is partially caused by the $N_{\rm fld}$ parameter (=0 for SM) which takes $N_{\rm fld}=0.36^{+0.16}_{-0.21}$ for the CMB+BAO datasets, while the free streaming content, $N_{\rm fs}=2.93\pm0.23$, is more consistent with the SM prediction of $\left.N_{\rm tot}\right|_{\rm SM}=\left.N_{\rm fs}\right|_{\rm SM}=3.044$ (see Table \ref{tab:cosmo_results_PR4}).

\vspace{0.2cm}

\vspace{0.2cm}

\FloatBarrier  

\section{Dark Radiation and the Hubble tension}
\label{sec: DR_tensions}

In the previous sections, we have discussed the impact that DR has on modifying the confidence regions of certain parameters with respect to the standard $\Lambda$CDM model, in light of CMB and BAO data. Consequently, DR is a compelling candidate for alleviating tensions within $\Lambda$CDM. In particular, we focus on the well-known Hubble tension \cite{Planck_2018, Verde_2024_H0, Riess_2022_SH0ES}, but we also explore the recent (although weak) discrepancy in $\Omega_m$ values among BAO, CMB, and uncalibrated SNe data, which has triggered discussion about whether dark energy could be evolving with time \cite{DESI_DR2} This discrepancy can be addressed not only by our framework, but also by alternative frameworks such as the previously mentioned DAO \cite{Garny_2025_DAO} or early dark energy frameworks \cite{Poulin_2019_EDE, Chaussidon_2025_DESI_EDE}.

\vspace{0.2cm}

\begin{figure}[htbp]
\centering
\includegraphics[width=.49\textwidth]{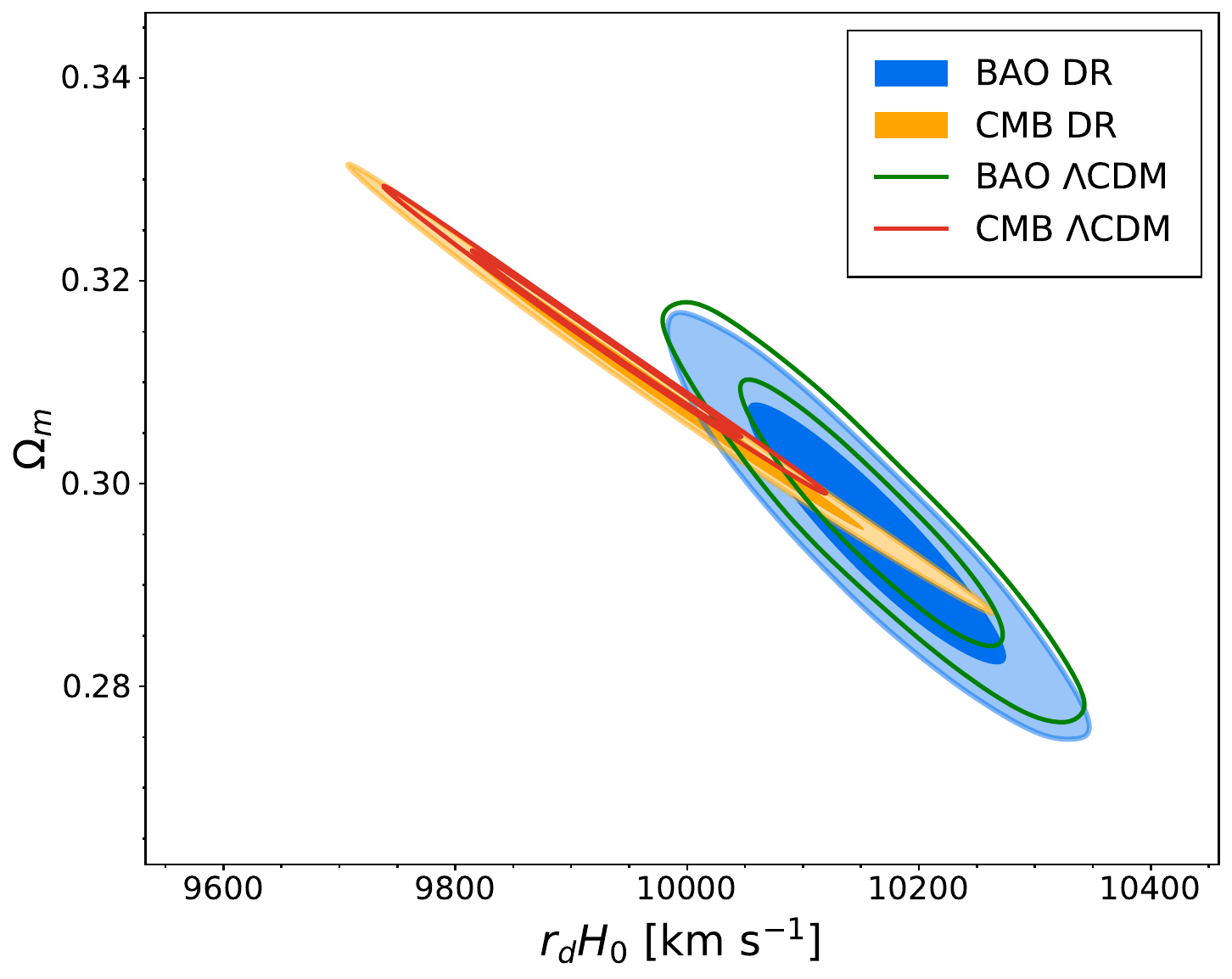}
\includegraphics[width=.49\textwidth]{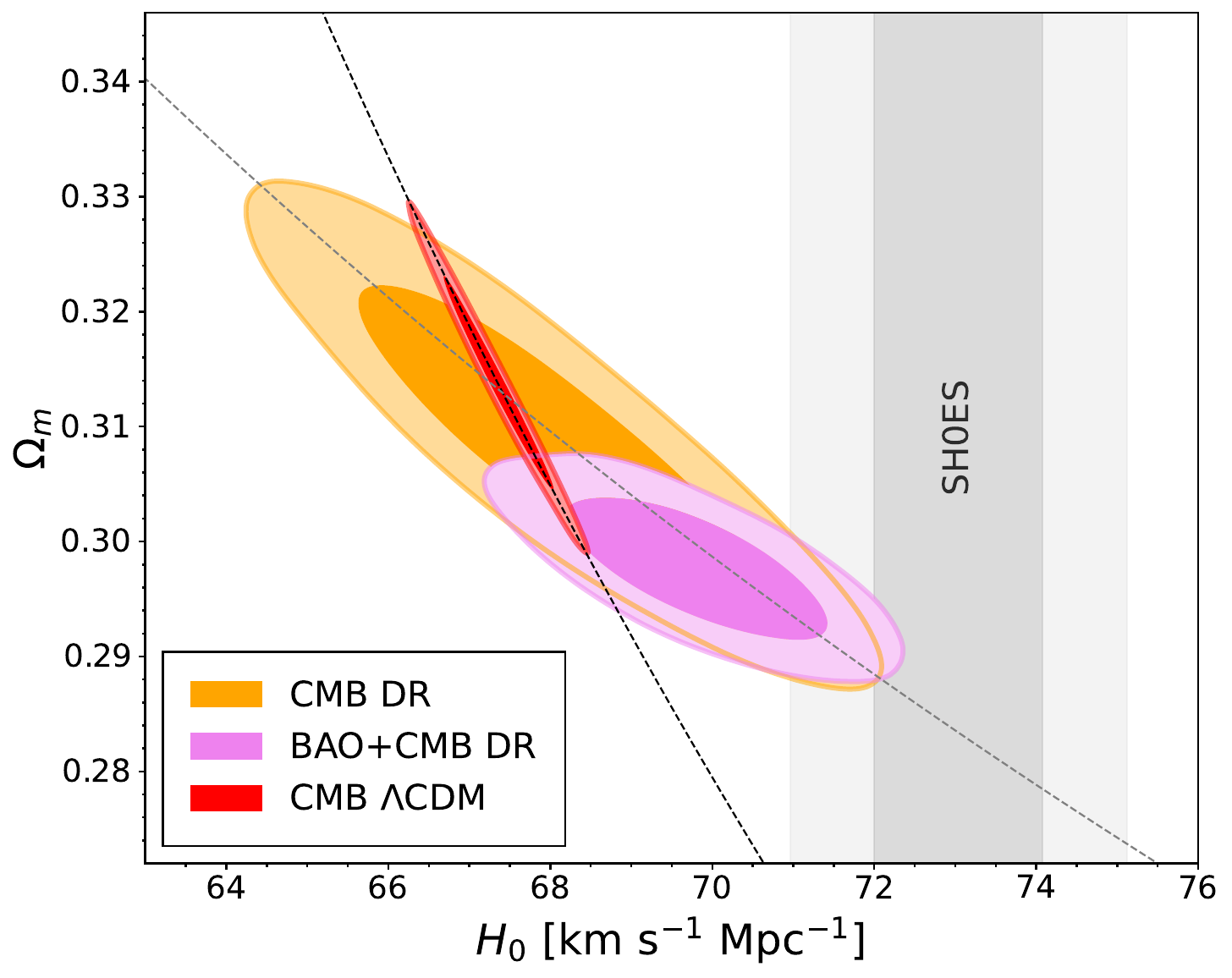}
\qquad
\caption{{\it Left Panel}: Marginalised 2D posterior distributions (68\% and 95\% credible intervals) on $\Omega_m$ and $r_dH_0$, from BAO and CMB datasets within the $\Lambda$CDM model (green and red contours, respectively) and within the DR framework (blue and orange, respectively). Within the DR framework, the two datasets display a higher degree of consistency (see text). {\it Right Panel}: Same for $\Omega_m$ and $H_0$ from the CMB (orange) and CMB+BAO (magenta) datasets for the DR cosmology. For reference, the result from CMB alone on the $\Lambda$CDM cosmology is also shown as red contours. The black dashed lines show the expected degeneracy direction for the $\Lambda$CDM model, $\Omega_m \propto h^{-3}$. The grey dashed line shows the new degeneracy direction for the DR cosmology, which is $\Omega_m\propto h^{-1.09}$. The grey vertical band displays the direct $H_0$ measurement from SH0ES \cite{Riess_2022_SH0ES}. }

\label{fig:omega_m_and_H0}
\end{figure}

\begin{figure}[htbp]
\centering
\includegraphics[width=.49\textwidth]{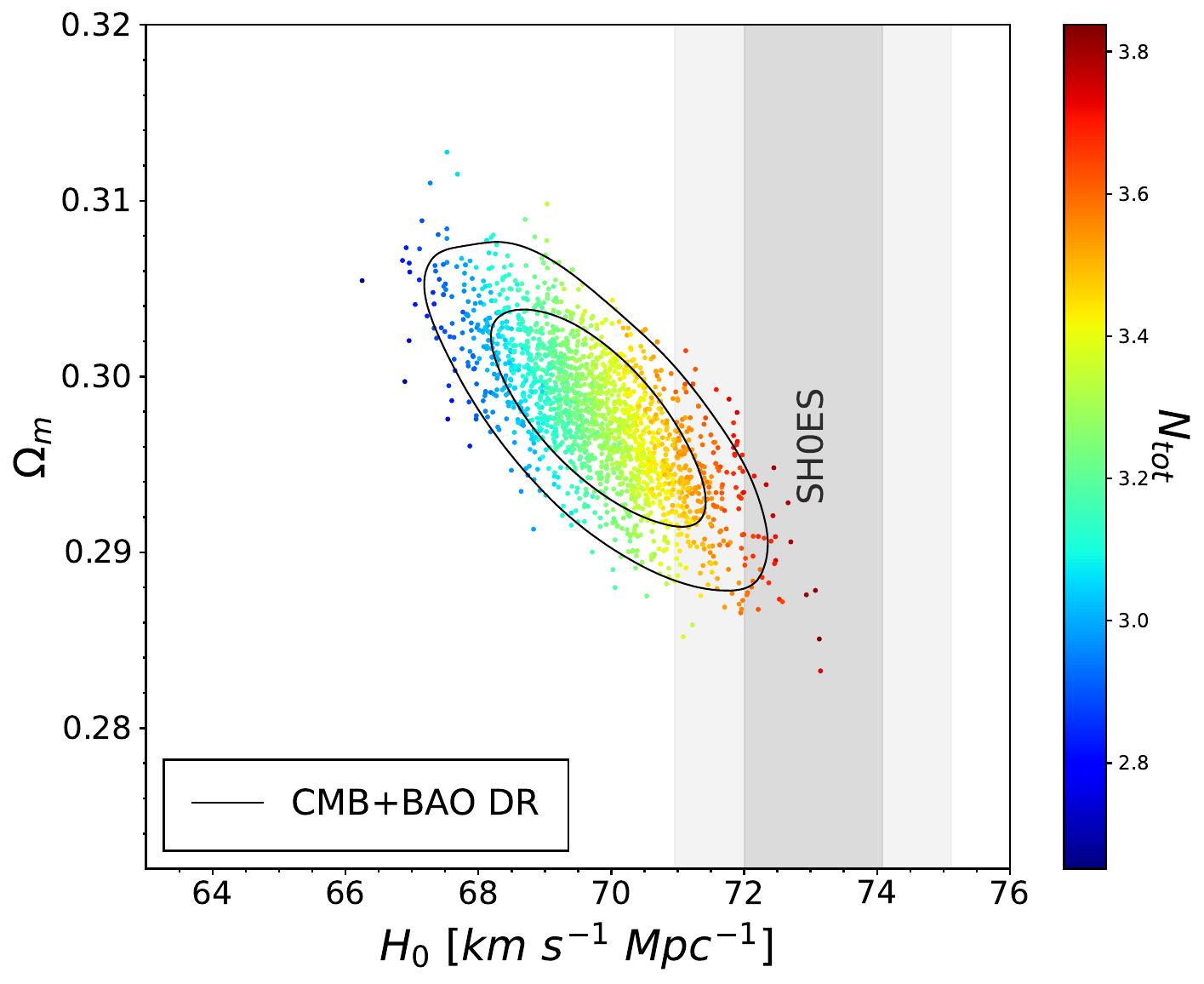}
\includegraphics[width=.49\textwidth]{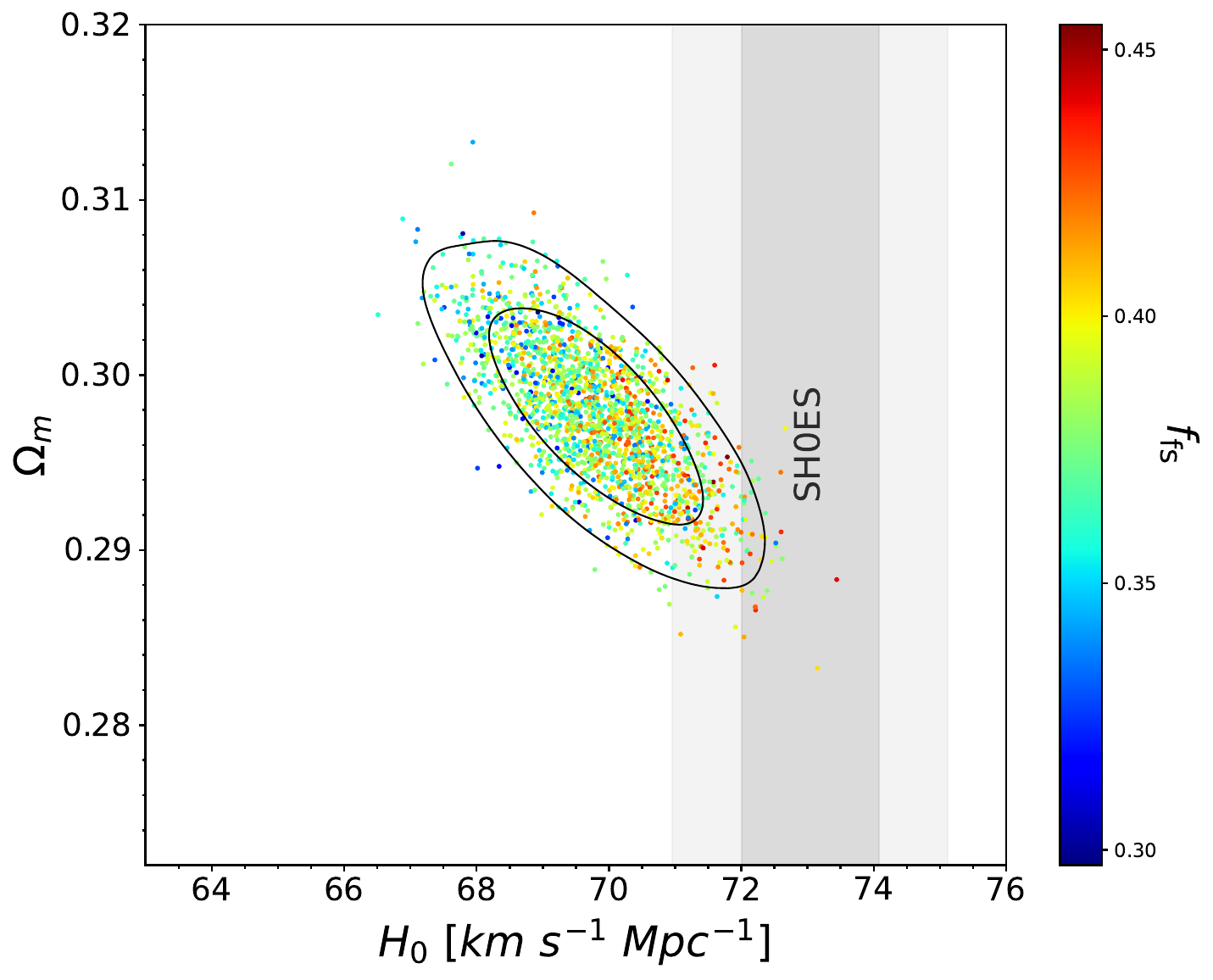}
\qquad
\caption{Marginalised 2D posterior distributions (68\% and 95\% credible intervals) on $\Omega_m$ and $H_0$ from the CMB+BAO datasets for the DR cosmology, as presented in pink contours in the right panel of Figure~\ref{fig:omega_m_and_H0}. The colour gradient represents a change on $N_{\rm tot}$ and $f_{\rm fs}$ in the left and right panels, respectively, which highlights how the dark radiation parameters change as we move along the 2D posteriors. As $H_0$ increases $N_{\rm tot}$ also increases, whereas there is no strong observed trend for $f_{\rm fs}$.}
\label{fig:scatterH0}
\end{figure}

These two discrepancies are exemplified in the two panels of Figure~\ref{fig:omega_m_and_H0}. The right panel displays how the DR framework (orange and pink contours for CMB and CMB+BAO, respectively) can relax the tension in $H_0$ between the direct measurements by SH0ES (grey band) and the $\Lambda$CDM measurements (red contours for CMB).
We clearly see how the extra degrees of freedom introduced by the DR framework (through $N_{\rm fld}$ and $N_{\rm fs}$, or through $N_{\rm tot}$ and $f_{\rm fs}$) are able to relax the tension by broadening credible interval contours in the line where $\Omega_mh^{1.09}$ remains constant (see caption of Figure~\ref{fig:omega_m_and_H0}), but also slightly shifting the maximum towards higher values of $H_0$. The left panel shows the results of BAO and CMB experiments for both $\Lambda$CDM and the DR framework, in terms of $\Omega_m$ and the Hubble parameter in units of the sound horizon scale at drag epoch, $r_dH_0$. Although not strong, there is a slight discrepancy in the 2D-parameter space between CMB and BAO within $\Lambda$CDM, which is alleviated within DR. In Figure~\ref{fig:scatterH0} we display the posteriors of $N_{\rm tot}$ and $f_{\rm fs}$ values in the $\Omega_m-H_0$ space, for the BAO+CMB case for the DR framework. We see a clear gradient for $N_{\rm tot}$ values in the degeneracy direction, exemplifying the relationship between high values of $H_0$ and high values of $N_{\rm tot}$. In contrast, for $f_{\rm fs}$ values, we do not observe a significant gradient in the degeneracy direction. This indicates that increasing the free-streaming radiation content is mostly what leads to higher $H_0$ values, while $f_{\rm fs}$ helps to accommodate this higher $N_{\rm tot}$ value on CMB data. Also, as described in Figure~\ref{fig:theta_star} the value of the sound horizon scale at drag epoch is reduced down to $r_d=145.0\pm1.8\,{\rm Mpc}$ as a consequence of the increase of $N_{\rm tot}$, as reported in Table~\ref{tab:cosmo_results_PR4}.

\vspace{0.2cm}

We can quantify the degree of tension between two given datasets by employing the $\mathcal{T}$-statistic (see eq 4 of Ref~\cite{Verde_2013_Tension}),
\begin{equation}
    \mathcal{T}\equiv \frac{\bar{\varepsilon}|_{\rm maxA=maxB}}{\varepsilon},
\end{equation}
where $A$ and $B$ denote two experiments (such as CMB and SH0ES for example), whose posterior distributions are $P_{A,B}(\theta|D_{A,B})$, where $\theta$ are the parameters of a given model, $D_i$ is the data obtained for the $i$-experiment, and $\varepsilon$ is the unnormalised evidence, $\int P_AP_Bd\theta=\varepsilon$. In order to avoid dealing with the evidence normalisation factor, Ref~\cite{Verde_2013_Tension} proposes to shift the distributions in $\theta$ and define the shifted distributions, $\bar{P}_{A,B}$. This translation allows to shift of the location of the maximum
without changing the shape or the width of the distribution. When the maxima of the two distributions coincide, we find $\int \bar{P}_A\bar{P}_Bd\theta=\bar{\varepsilon}|_{\rm maxA=maxB}$, corresponding to the null hypothesis. Thus, the ratio of the unnormalised evidences between this null hypothesis and any other situation can be related to the Jeffrey's scale (see table 1 of Ref~\cite{Verde_2013_Tension}), establishing a degree of preference for one model over the other within a Bayesian framework. 

\vspace{0.2cm}

In Table~\ref{tab:T-stat} we report the results for $\ln(\mathcal{T})$ associated with $H_0$ (i.e., replacing $\theta\rightarrow H_0$), comparing different datasets within both $\Lambda$CDM and the Dark Radiation framework (DR). For completeness, we also include results for an evolving Dark Energy model ($w_0w_a$CDM), which will be discussed in more detail in Section~\ref{sec:EDE_vs_DR}. There, we further analyse model selection for both DR and evolving Dark Energy, with and without the SH0ES calibration, and assess their relative ability to alleviate the Hubble tension.

\vspace{0.2cm}

Here, we highlight that the inclusion of DR significantly reduces the $H_0$ tension, bringing it from a ``highly significant" discrepancy between SH0ES and CMB alone to a ``substantial event" (odds of roughly 1:3), and further down to ``not worth a bare mention" (approximately 1:1 odds) when CMB is combined with BAO data. For completeness, we have also computed the $\ln(\mathcal{T})$ for $\Omega_m$ (i.e., replacing $\theta\rightarrow\Omega_m$) between CMB and BAO, as by eye the left panel of Figure~\ref{fig:omega_m_and_H0} could cast some doubts about the consistency of these two datasets for $\Lambda$CDM. We find $\ln(\mathcal{T}(\Omega_m))|_{\Lambda {\rm CDM}}=0.4$ and $\ln(\mathcal{T}(\Omega_m))|_{\Lambda {\rm CDM+DR}}=0.14$, which in both cases qualfies as `not a bare mention' according to the scale for interpreting $\mathcal{T}$, with odds 1:1.5 and 1:1.15, respectively. Therefore, BAO and CMB data can undoubtedly be combined within both $\Lambda$CDM and DR frameworks.

\begin{table}[htbp]
\centering
\renewcommand{\arraystretch}{1.4} 

\begin{tabular}{|lcc|}
\hline 
Model/Datasets & $\ln(\mathcal{T}(H_0))$ & Strenght of Tension (odds) \\
\hline\hline

\multicolumn{3}{|l|}{\textbf{$\Lambda$CDM}} \\ 
 CMB vs. SH0ES & $9.9$ & Highly Significant (1:20.000) \\
 CMB+BAO vs. SH0ES  & $6.1$ & Highly Significant (1:445) \\

\hline 
\multicolumn{3}{|l|}{\textbf{Dark Radiation}} \\ 
 CMB vs. SH0ES  & $1.2$ & Substantial (1:3) \\
 CMB+BAO vs. SH0ES  & $0.02$ & Not worth a bare mention (1:1) \\

\hline 
\multicolumn{3}{|l|}{\textbf{$w_0w_a$CDM}} \\
CMB+BAO+Pantheon$+$ vs. SH0ES  & $8.2$ & Highly Significant (1:3600) \\
CMB+BAO+DESY5 vs. SH0ES  & $9.3$ & Highly Significant (1:11.000) \\
\hline 

\end{tabular}%

\caption{Values of the $\ln(\mathcal{T})$ statistic used to evaluate the tension in $H_0$ between different datasets, assuming the baseline $\Lambda$CDM model (first rows), the dark radiation framework (second rows) and the evolving dark energy framework through the CPL parametrisation of Eq.~\ref{eq:CPL} (third rows). The third column reports the interpretation of $\ln(\mathcal{T})$ in terms of a modified Jeffrey's scale to establish the strength of the discrepancy and the corresponding odds.}
\label{tab:T-stat} 
\end{table}

\vspace{0.2cm}

\vspace{0.2cm}

We may wonder how DR modifies the (uncallibrated) expansion history of the universe to alleviate the discrepancy on $H_0$ via reducing the $\Omega_m$ value, as shown in the panels of Figure~\ref{fig:omega_m_and_H0}. At the level of expansion history, the extra component of DR behaves exactly like standard radiation. Along with matter and dark energy, the expansion history is given by:
\begin{equation}
    H(z)=H_0\left[\Omega_m(1+z)^3+\Omega_r(1+z)^4+ \Omega_\Lambda\right]^{1/2},
\end{equation}
where we have assumed a flat universe such that $\Omega_m+\Omega_r+\Omega_\Lambda=1$. We consider the uncallibrated expansion history $E(z)\equiv H(z)/H_0$ to represent this evolution independently of its calibration (or the exact $H_0$ value).

\vspace{0.2cm}

\begin{figure}[htbp]
\centering
\includegraphics[width=.49\textwidth]{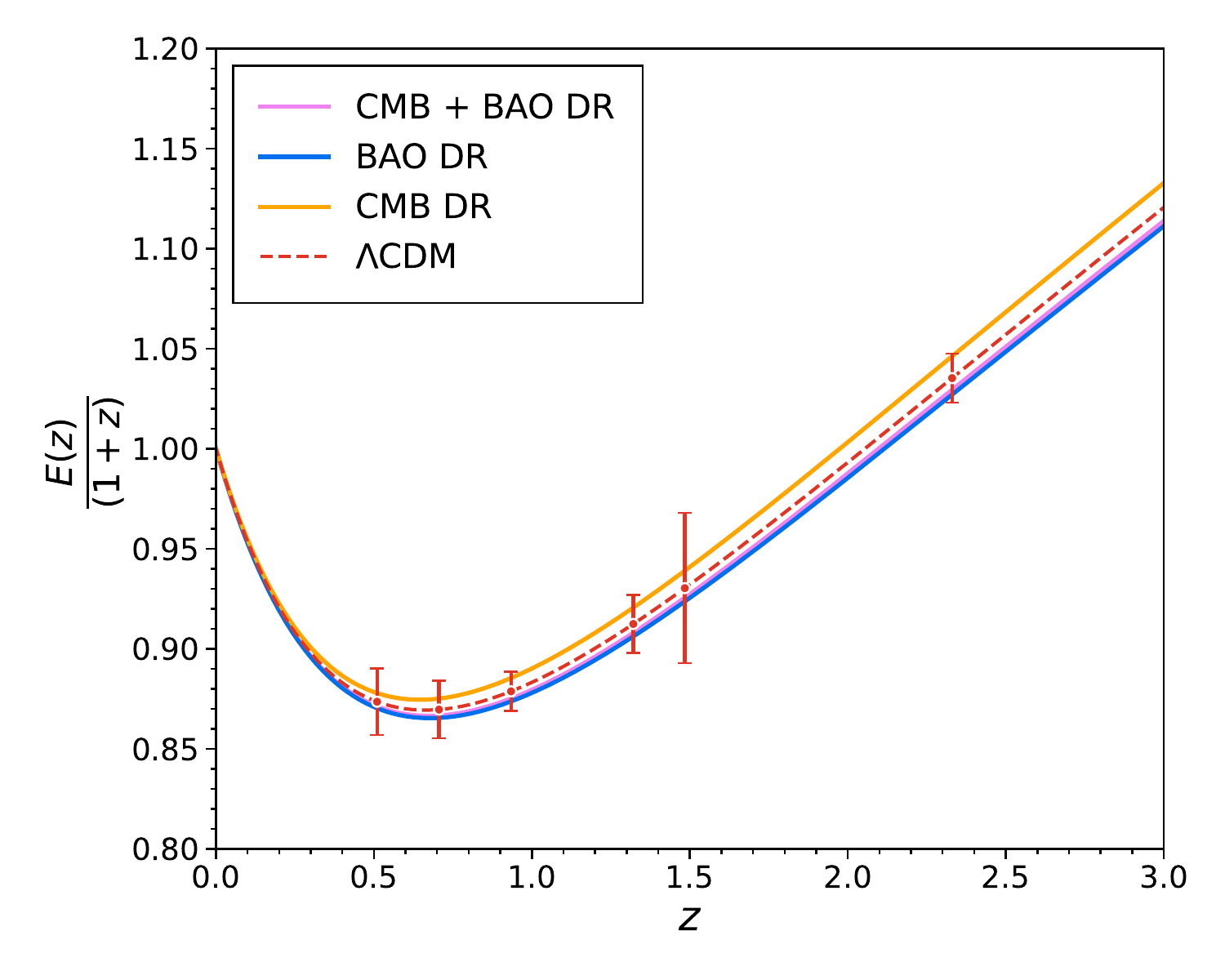}
\hfill
\includegraphics[width=.49\textwidth]{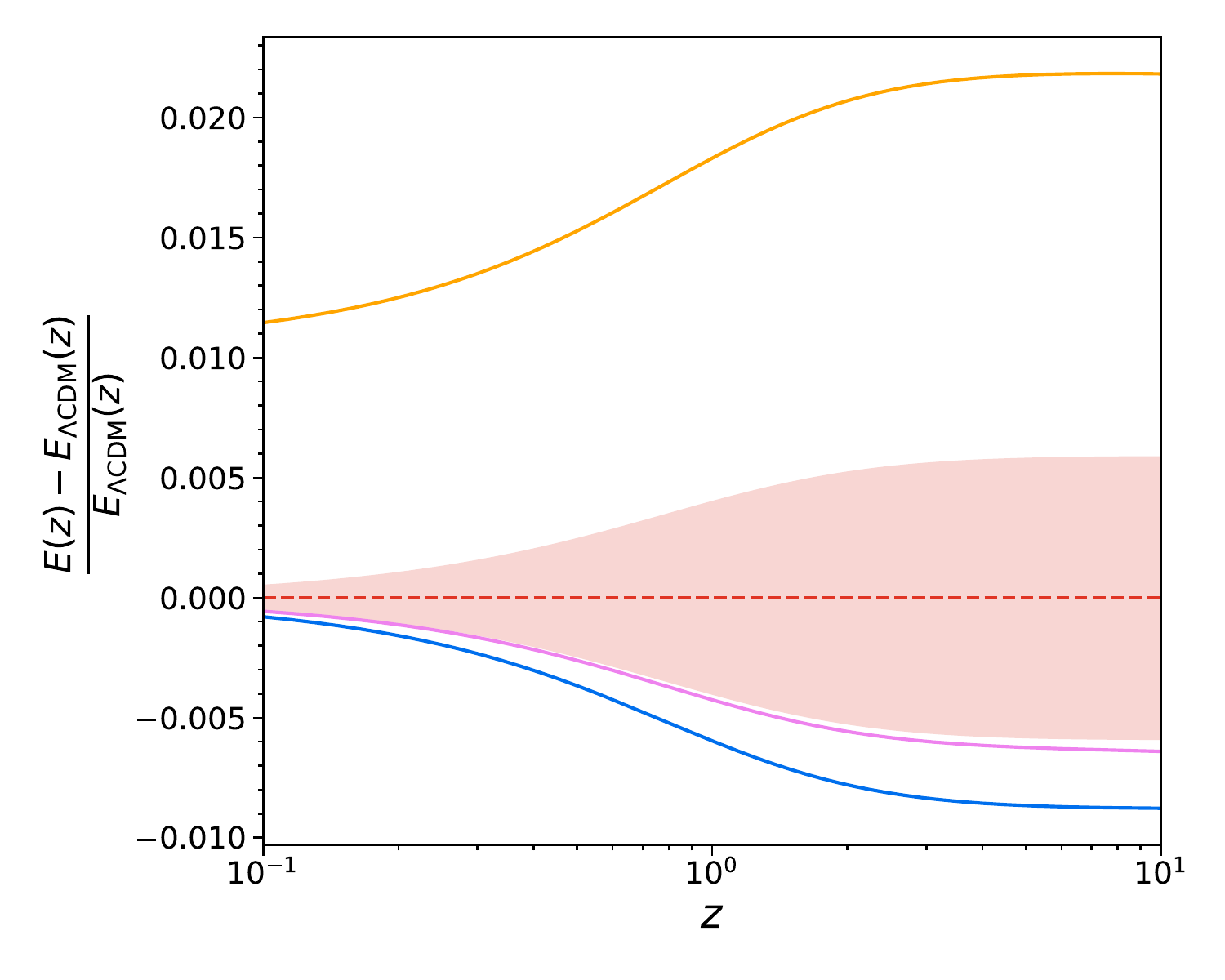}
\caption{Uncallibrated expansion history for different models considered in this work: $\Lambda$CDM model with CMB+BAO best-fitting values as reference (red dashed lines), DR framework with CMB best-fitting values (orange solid lines), DR with BAO best-fitting values (blue solid lines), DR with BAO+CMB best-fitting values (magenta solid lines). The left panel displays the uncallibrated expansion history, $E(z)$ scaled by $1/(1+z)$ for late-times ($z<3$). Errorbars, shown for guidance, are estimated from DESI DR2 $D_H(z)/r_d$ measurements. They use radial BAO information only, omit transverse BAO constraints, and rely on assumptions based on a $\Lambda$CDM model (see text).
The right panel displays the relative change of the uncallibrated expansion history relative to the $\Lambda$CDM with CMB best-fitting values for $z<10$. The red solid coloured band denotes the 68\% confidence interval. }
\label{fig:H(z)}
\end{figure}

The left panel of Figure~\ref{fig:H(z)} displays the uncallibrated expansion history for different frameworks: CMB+BAO best-fitting $\Lambda$CDM (red dashed lines), and DR one using the CMB best-fitting parameters (orange), BAO best-fitting parameters (blue), and CMB+BAO best-fitting parameters (magenta lines). Bear in mind that all these DR frameworks keep $\theta_s$ relatively close to \textit{Planck}'s $\Lambda$CDM one.
For reference, the red points with error bars provide an estimate of the DESI DR2 BAO precision in discriminating among these models; these points are displayed on top of the reference $\Lambda$CDM model for convenience. Since the BAO technique does not directly infer $E(z)$, we have transformed the radial BAO distance errorbars into $E(z)$ error bars by choosing a fiducial $\Lambda$CDM value for $c/[H_0r_d]$, since $D_H(z)/r_d\equiv c/[H_0 r_d E(z)]$. Note that DESI also measures the transverse BAO signal through $D_M(z)/r_d$, which depends on the integrated value of $1/E(z')$ from $z'=0$ up to the effective redshift of the galaxy sample ($z=0.295$, 0.510, 0.706, 0.934, 1.321, 1.484 and 2.330 for the seven DESI $z$-bins), but for simplicity, we ignore this contribution in the errorbars representation of Figure~\ref{fig:H(z)}.

\vspace{0.2cm}

The right panel of Figure~\ref{fig:H(z)} represents the fractional difference of the uncallibrated expansion history with respect to CMB+BAO $\Lambda$CDM best-fit. Here, we focus on a wider redshift range ($0.1<z<10$) displayed on a logarithmic scale, where the red band represents the uncertainty of CMB+BAO $\Lambda$CDM best-fitting parameters.

\vspace{0.2cm}

We see that the DR framework studied here produces changes on the $E(z)$ for $0.5<z<2.5$ of order of $\sim1\sigma$ for LRG3+ELG1 and Ly-$\alpha$, $\sim0.5\sigma$ for LRG2 and ELG2, and $<0.25\sigma$ for LRG1 and QSO. In the right panel of Figure \ref{fig:H(z)}, the relative difference in $E(z)$ shows the early time modification induced by DR in CMB data, and the redshift-dependent modification induced by BAO data. All the models induce modifications of at most a 2\% from $\Lambda$CDM.

\vspace{0.2cm}

In conclusion, find that the DR framework can alleviate the Hubble tension by increasing the effective number of relativistic species, $N_{\rm tot}$, while a non-zero $f_{\rm fs}$ ensures that the resulting CMB power spectrum remains consistent with observational constraints, as well as isotropic and anisotropic BAO data. As the tension is reduced, it becomes feasible to jointly fit CMB and BAO data together with the calibrated SH0ES distances within this DR scenario. We present these results in the next section, along with a comparison to evolving dark energy models.

\FloatBarrier
\section{Dark Radiation and  Evolving Dark Energy}
\label{sec:EDE_vs_DR}

We now compare Dark Radiation with Evolving Dark Energy (EvoDE) models, which has become a hot topic in cosmology following the results presented in Refs.~\cite{DESI_DR1_1,DESI_DR2}, as these models may show a preference over $\Lambda$CDM when DESI BAO data is combined with CMB and SNe data. EvoDE is based on considering a time variation of the dark energy equation of state, $w(a)$. To examine EvoDE we adopt the Chevalier-Polarski-Linder (CPL, \cite{Chevallier_2001_CPL, Linder_2003_CPL}) parametrisation as it is widely used in the literature (for the performance of alternative parametrisations see \cite{Lodhaetal2025}), 
\begin{equation}
    w(a)=w_0+w_a(1-a),
    \label{eq:CPL}
\end{equation}
where $w_0$ is the equation of state parameter today, and $w_a$ parametrises the change with the scale factor $a$. This yields the so-called $w_0w_a$CDM model. In this parametrisation, the energy density of dark energy, $\rho_{\rm DE}$, evolves with time as,
\begin{equation}
    \frac{\rho_{\rm DE}(a)}{\rho_{\rm DE,0}}=a^{-3(1+w_0+w_a)} \cdot e^{-3w_a(1-a)}.
\end{equation}

By taking $w_0=-1$ dark energy behaves as a cosmological constant today, and by taking $w_a=0$ dark energy does not evolve with time, and we recover $\Lambda$CDM. In this work we apply uniform and wide priors over both EvoDE parameters, as well as the usual prior $w_0+w_a\leq 0$, which ensures matter domination at early times (see Appendix~\ref{sec:priors}). The $w_0w_a$CDM parametrisation of EvoDE is a general and flexible framework capable of fitting observational data. One of the issues with the $w_0w_a$CDM model is the excess of freedom in the $w_0-w_a$ parameter space, which can lead to the so-called ``phantom crossing" phenomenon, where the value of $w(a)$ can allow for phantom dark energy at any epoch $a$ ($w(a)<1$). This may lead to the violation of the null energy condition and the interpretation of dark energy as a single scalar field becomes challenging, and more complex models for dark energy (such as multiple fields models) are introduced \cite{Caldwell_2002_Phantom, Carroll_2003_Phantom, Vikman_2005_Phantom, Hu_2005_Phantom, Koussour_2023_Phantom, Wolf_2025_Phantom, Clifton_2012_ModifiedGravity, Ishak_2019_TestingGR}. To better constrain $w_0w_a$CDM models, low redshift data ($z\lesssim1$) is needed. For this reason, uncalibrated SNe data are implemented in this section (see Section~\ref{models} for details).

\vspace{0.2cm}

\begin{figure}[htbp]
\centering
\includegraphics[width=.49\textwidth]{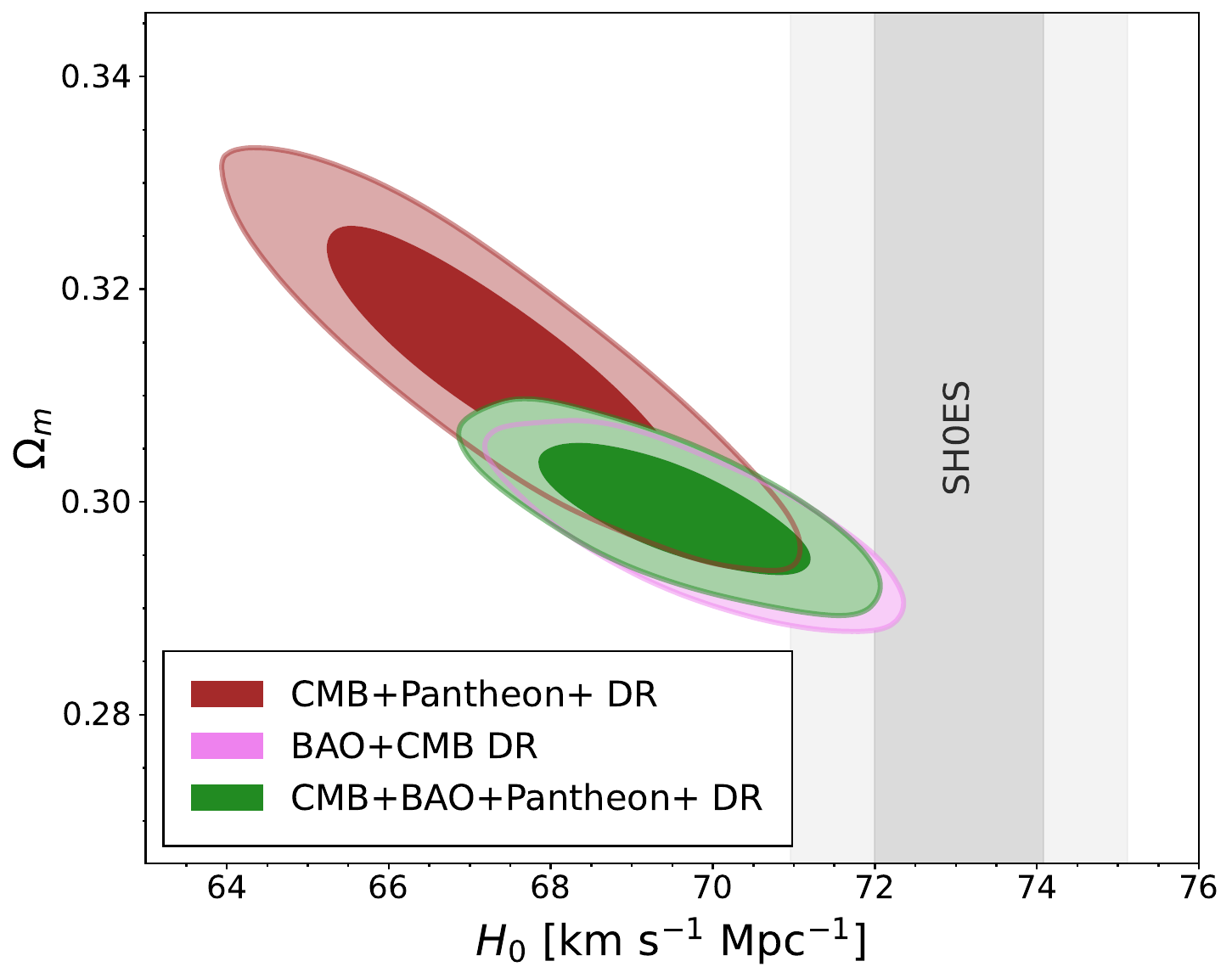}
\includegraphics[width=.49\textwidth]{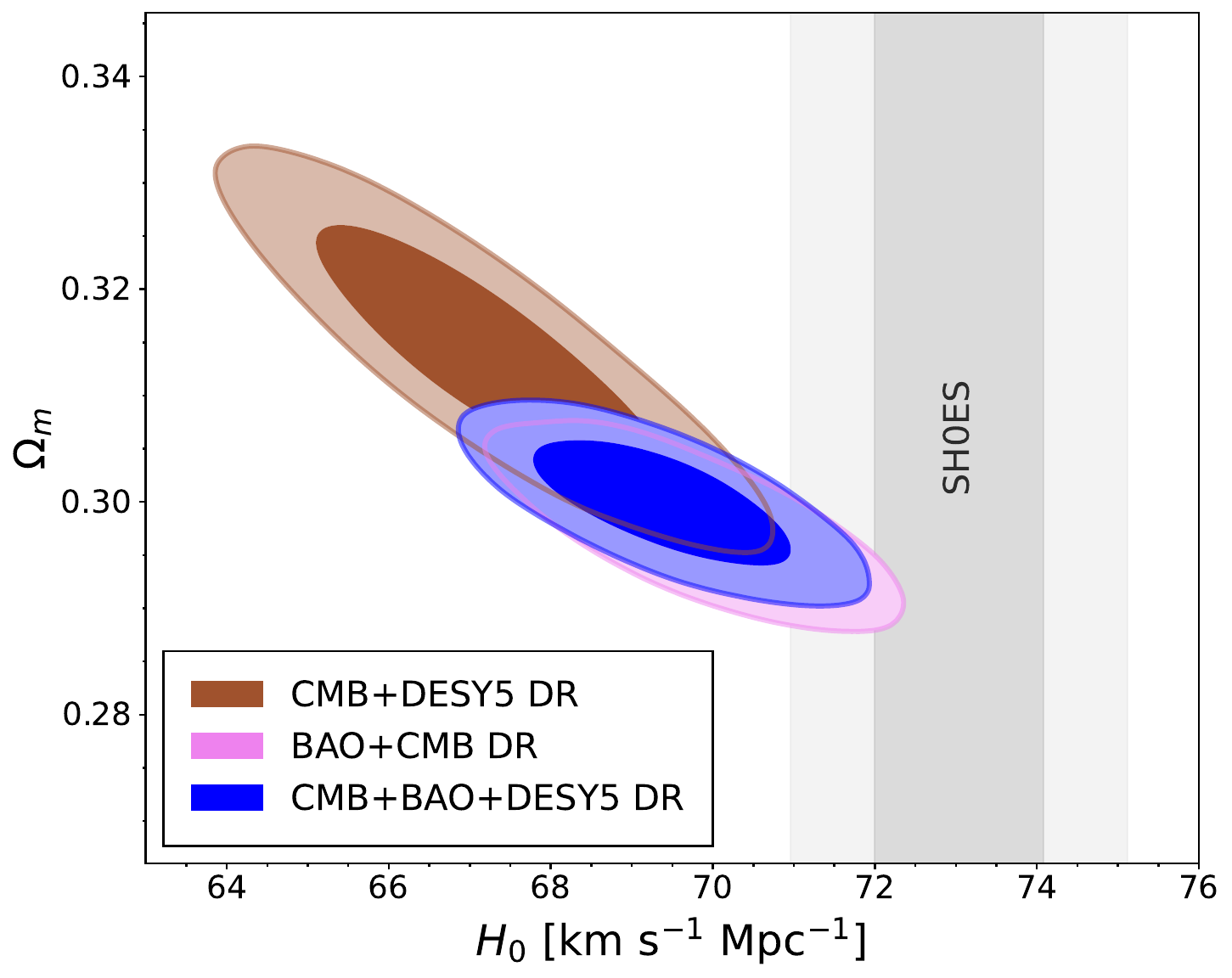}
\qquad
\caption{{\it Left Panel}: Marginalised 2D posterior distributions (68\% and 95\% credible intervals) on $\Omega_m$ and $H_0$, from CMB+BAO, CMB+Pantheon$+$, and CMB+BAO+Pantheon$+$ datasets within the DR framework (magenta, maroon, and green, respectively). {\it Right Panel}: Same for for the CMB+BAO, CMB+DESY5, and CMB+BAO+DESY5 datasets (magenta, brown and blue respectively). The grey vertical band displays the direct $H_0$ measurement from SH0ES \cite{Riess_2022_SH0ES}.}
\label{Sne_DR}
\end{figure}

We start by analysing in Figure~\ref{Sne_DR} the effect of adding SNe data (Pantheon+ in the left panel and DES-Dovekie in the right panel) to the CMB and BAO data within the DR framework. When these SNe datasets are added we see a small shift towards higher $\Omega_m$ values compared to the results without SNe, as SNe-only datasets prefer higher $\Omega_m$ values (see also Section \ref{sec: DR_tensions} and Table \ref{tab:cosmo_results_PR4}). The large overlap between CMB+BAO and CMB+SNe allows us to safely combine the CMB, BAO, and SNe datasets within the DR framework\footnote{Whereas in $\Lambda$CDM, the discrepancies among datasets are more pronounced, making it difficult to combine all three}, resulting in the green and blue contours in the left and right panels for Pantheon+ and DES-Dovekie, respectively. For reference, we display as a vertical grey band the direct ladder $H_0$ measurement by SH0ES \cite{Riess_2022_SH0ES}, as done before in the right panel of Figure \ref{fig:omega_m_and_H0}. Within the DR framework, we find that the inclusion of uncalibrated SNe does not significantly alter the contours obtained from BAO+CMB alone. This is because SNe provide little information about DR-induced modifications to the expansion history, as they probe only relatively low redshifts (which are important for EvoDE models, though). Furthermore, SNe data alone do not represent a competitive constraint for $\Omega_m$, as they retrieve a precision of approximately $\pm 0.02$ at 68\% confidence, and do not substantially improve upon the precision already achieved by BAO+CMB, namely $\pm 0.004$, about five times tighter. The strong reason for including the SNe in this section is to make an equal comparison between DR and EvoDE models. 

\vspace{0.2cm}

\begin{figure}[htbp]
\centering
\includegraphics[width=.49\textwidth]{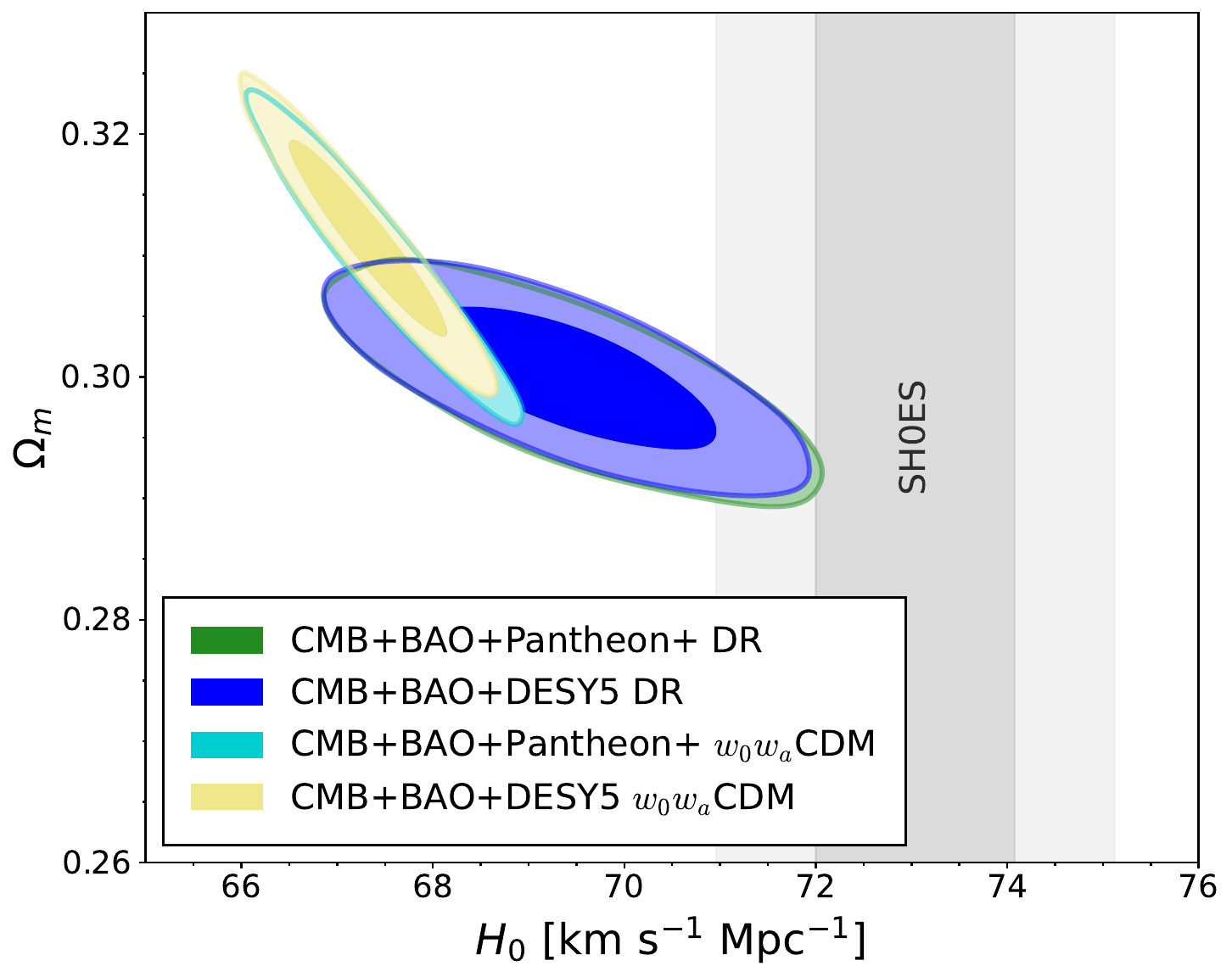}
\includegraphics[width=.49\textwidth]{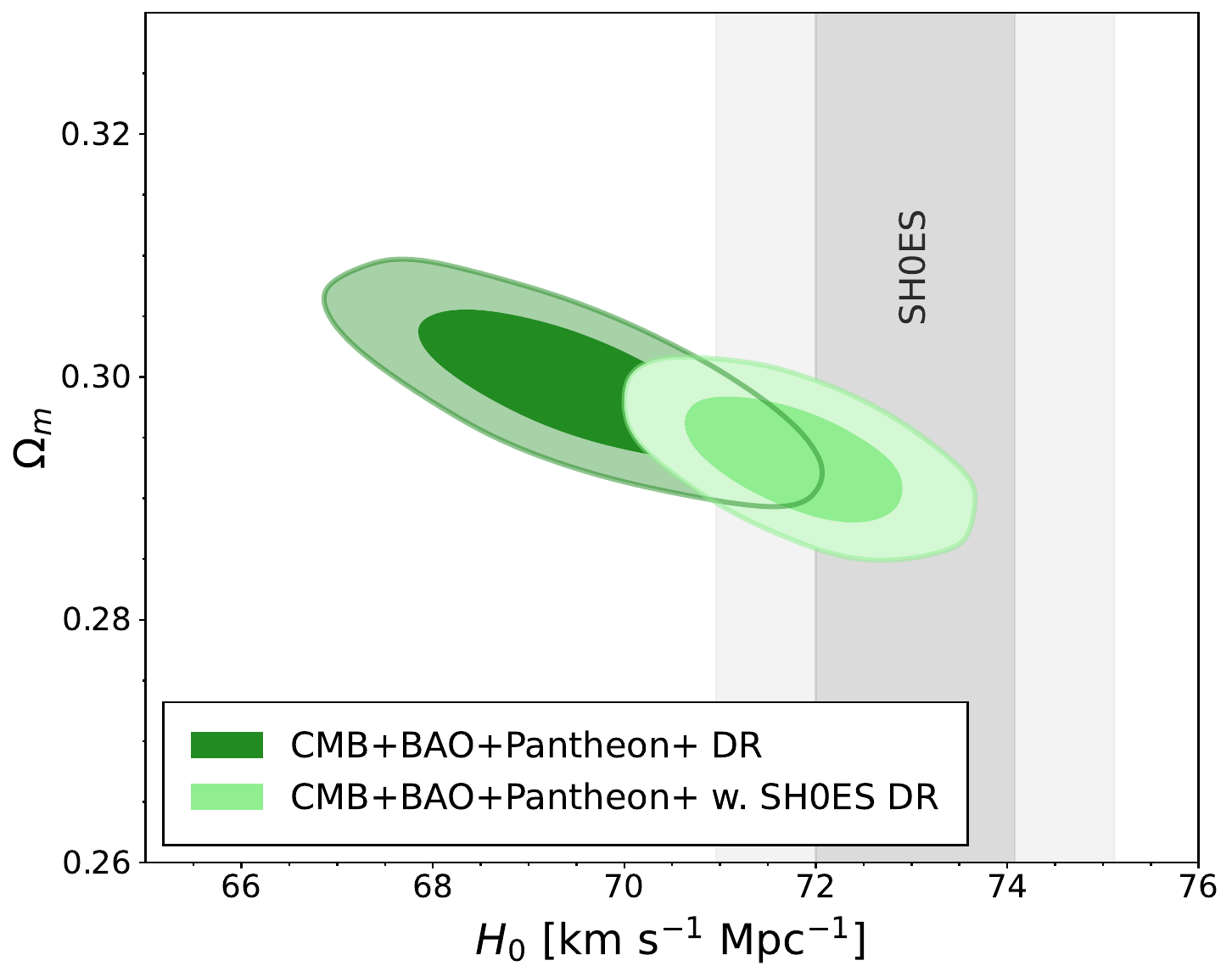}
\qquad
\caption{{\it Left Panel}: Marginalised 2D posterior distributions (68\% and 95\% credible intervals) on $\Omega_m$ and $H_0$, from the CMB+BAO+Pantheon$+$ and CMB+BAO+DESY5 datasets within the DR framework (green and blue respectively) and the $w_0w_a$CDM model (cyan and khaki respectively). {\it Right Panel}: Same for CMB+BAO+Pantheon$+$ and CMB+BAO+Pantheon$+$ w. SH0ES datasets within the DR framework. The grey vertical band displays the direct $H_0$ measurement from SH0ES \cite{Riess_2022_SH0ES}.}
\label{fig:wowa}
\end{figure}

The left panel of Figure \ref{fig:wowa} displays the performance of the DR framework alongside the $w_0w_a$CDM one when CMB, BAO and SNe data are considered. Qualitatively, we see that {\it i)} the ellipses of the DR framework are larger than those of the EvoDE in this particular projection, even though both models have the same (extra) degrees of freedom; {\it ii)} the EvoDE model is unable to alleviate the tension with SH0ES data significantly.

\vspace{0.2cm}

In order to quantitatively compare both models, we also employ several model selection tools. First, we make use of the Deviance Information Criterion (DIC) \cite{Spiegelhalter_2014_DIC, Kass_1995_Bayes, Trotta_2008_Bayes, Grandis_2016_Tension, Gelman_2014_WAIC}. DIC is a frequentist model selection tool based on the goodness of fit and the complexity of a model. The complexity term in the DIC can be estimated using the Gelman or Spiegelhalter criterion. The former estimates the complexity term using the variance of the posterior log-likelihoods, while the latter estimates it using the difference in maximum a posteriori $\chi^2$ ($\Delta\chi^2_{\rm MAP}$) between the models compared. We compute the results with both criteria for comparison, for both DR and EvoDE, relative to $\Lambda$CDM, where both models have the same number of parameters (two plus the $\Lambda$CDM ones). As a rule of thumb, values of $|\Delta \rm DIC|$ lower than $2$ indicate a ``negligible'' preference, values between $2-5$ indicate a ``substantial" preference, values between than $5-10$ indicate a ``strong" preference, and values above 10 ``decisive", for the tested model with respect to $\Lambda$CDM or the oposite depending on the sign of $\Delta \rm DIC$: positive standing for $\Lambda$CDM and negative for DR/$w_0w_a$CDM in this case \cite{Grandis_2016_Tension}. Second, we exploit the fact that both $w_0w_a$CDM and DR are nested models of $\Lambda$CDM with two extra parameters by using the Savage-Dickey-Density-Ratio (SDDR) \cite{Trotta_2007_Applications, Trotta_2008_Bayes} to estimate the Bayes factor between $\Lambda$CDM and the above-mentioned extended models. In other words, with this criterion we compute an estimate of $\ln(B)=\ln(\mathcal{Z_{\rm ext}})-\ln(\mathcal{Z}_{\Lambda \rm CDM})$ where $\mathcal{Z_{\rm ext}}$ is the evidence for the extended model and $\mathcal{Z}_{\Lambda \rm CDM}$ the evidence for $\Lambda$CDM. SDDR estimates $B$ as,

\begin{equation}
    B^{-1} \equiv \left. \frac{P_{\rm ext}(\theta \mid \mathcal{D}_i)}{\pi_{\rm ext}(\theta )} \right|_{\theta = \theta_0}.
\end{equation}

Here, $\theta$ represents the additional parameters introduced by the extended models (in our case $w_0$ and $w_a$ for $w_0w_a$CDM and $N_{\rm fs}$ and $N_{\rm fld}$ for the DR framework). The posterior $P_{\mathrm{ext}}(\theta \mid \mathcal{D}_i)$ and the prior $\pi_{\mathrm{ext}}(\theta)$ of the extended model are evaluated at the value of $\theta$ that recovers the $\Lambda$CDM model, $\theta_0$. Note that the SDDR is directly dependent on the prior chosen, a characteristic shared by all evidence-based Bayesian model selection frameworks. We consider $|\ln(B)|<1$ values as ``weak'' preference,  $1 \leq |\ln B| < 2.5$ values as ``substantial'' preference, $2.5 \leq |\ln B| < 5$ values as ``strong'' preference and $5 \leq |\ln B|$ values as ``decisive'' preference for the $\Lambda$CDM model (negative values) or the extended model (positive values). 

\vspace{0.2cm}

For completeness, in this work we report both the Bayes factor $\ln B$ and the DIC statistics. The DIC offers a useful indication of the balance between fit quality and model complexity, but it does not quantify relative model evidence in the Bayesian sense. Since the Bayes factor is directly related to the marginal likelihood of the model, we regard $\ln B$ as the more robust indicator for model selection, and therefore assign greater weight to its conclusions.

\vspace{0.2cm}

Table \ref{tab:SDDR} shows the results for the three considered model selection tools. Readers should note that while Spiegelhalter’s criterion is widely used for hierarchical frameworks, Gelman’s definition often imposes a more stringent penalty on parameter-heavy models. Consequently, the Spiegelhalter $\Delta$DIC may lean toward more complex architectures if they provide a sufficiently improved fit to the observed data, whereas the Gelman criterion and the $\ln B$ thresholds provide a more conservative penalisation against over-parameterisation. By presenting both, we ensure that our conclusions remain robust against different definitions of ``effective'' model complexity\footnote{The Spiegelhalter criterion is the one adopted by the DESI collaboration. Our results are in agreement with those reported in Table IV of Ref.~\cite{DESI_DR2}, with minor differences arising from the choice of CMB likelihoods and the Boltzmann solver.}. As discussed above, only $\ln B$ constitutes a true Bayesian model-comparison statistic, since it is directly related to the relative evidence of competing models given the data. In contrast, $\Delta$DIC quantifies the balance between goodness-of-fit and effective model complexity, and is therefore more closely related to predictive performance than to relative model probabilities. For this reason, we regard $\ln B$ as the primary indicator for model selection, while the DIC results are reported mainly as a complementary cross-validation diagnostic.

\vspace{0.2cm}

\begin{figure}[htbp]
\centering
\includegraphics[width=1\textwidth]{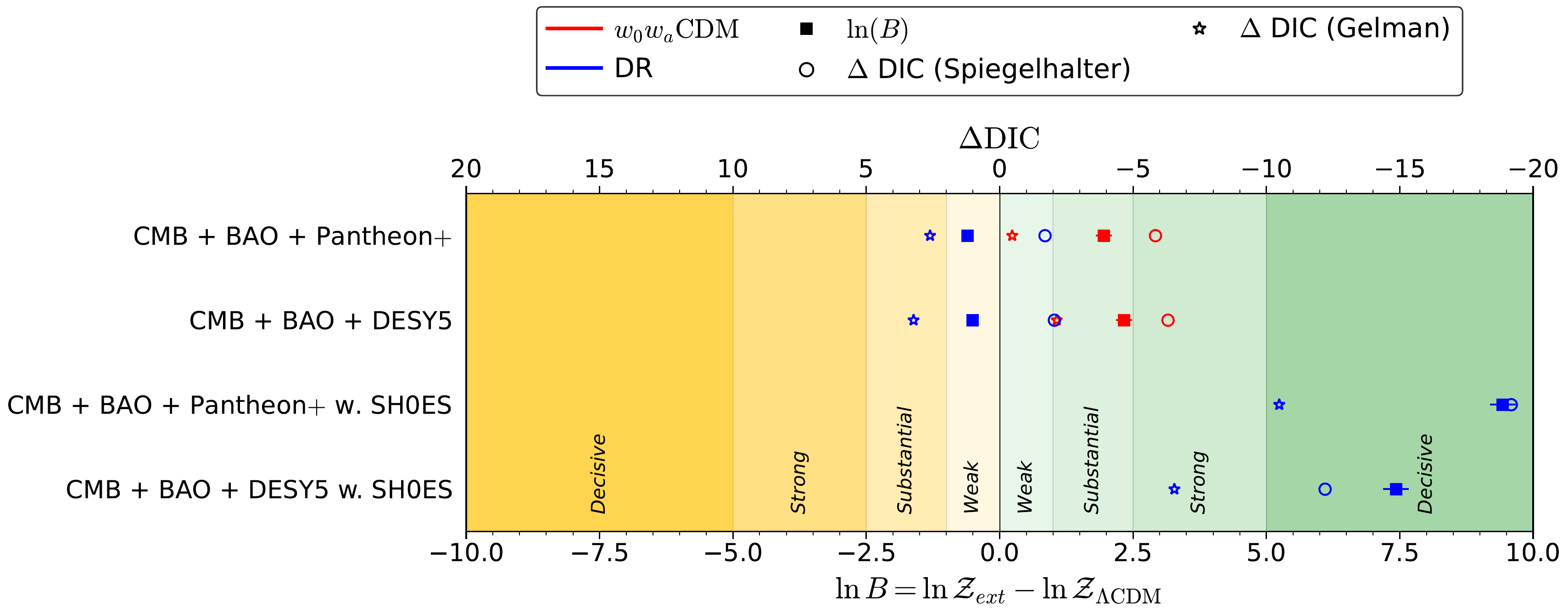}
\caption{Visualisation of the results presented in Table~\ref{tab:SDDR}. Values for $\ln(B)$ (solid square markers) and $\Delta$DIC (non-solid markers, with stars referring to the Gelman criterion and circles to the Spegelhalter criterion) for $w_0w_a$CDM (in red) and the DR framework (in blue). Confidence bands show Jeffrey's scale, with green colours favouring the test-model ($\ln B>0$ and $\Delta{\rm DIC}<0$), and yellow colours favouring $\Lambda$CDM ($\ln B<0$ and $\Delta{\rm DIC}>0$). The upper $x$-axis is aligned with the bottom $x$-axis by taking the approximate relation $\Delta{\rm DIC} =-2\ln(B)$, just for visualisation purposes.}
\label{fig:SDDR_DIC_plot}
\end{figure}

For a clearer visualisation of these results, Figure~\ref{fig:SDDR_DIC_plot} displays a whisker plot with the values reported in Table \ref{tab:SDDR}. Using Jeffrey's scale for $\ln(B)$ (bottom $x$-axis and coloured bands), we also represent the $\Delta$DIC results (upper $x$-axis, with $\Delta{\rm DIC}\simeq -2\ln B$ chosen only for visual comparison purposes). We find that, when CMB+BAO+SNe data are included, the overall results do not indicate decisive evidence either in favour of $\Lambda$CDM or of the tested extended models ($w_0w_a$CDM and DR). Nevertheless, the different statistics show mild trends, with $\ln B$ and the Gelman criterion generally preferring $\Lambda$CDM over DR, while favouring $w_0w_a$CDM over $\Lambda$CDM. In summary, we find that {\it i}) for $\Lambda$CDM versus DR, the evidence ranges from substantial-to-weak support for $\Lambda$CDM to weak support for DR depending on the dataset combination and criterion adopted; {\it ii}) for $w_0w_a$CDM, the results range from weak to substantial support for the extended model, reaching moderately strong support only according to the Spiegelhalter criterion.

\vspace{0.2cm}

A remarkable feature is that the area enclosed by the 2D contours in the left panel of Figure~\ref{fig:wowa} is significantly larger for the DR framework than for the $w_0w_a$CDM one. This induces a larger complexity penalty in the model-selection statistics for the two additional parameters of the DR framework relative to $\Lambda$CDM than for the corresponding extension in $w_0w_a$CDM. Consequently, DR is typically less favoured by the statistical indicators than $w_0w_a$CDM, despite both models introducing the same number of additional parameters. Only the Spiegelhalter criterion yields a moderately strong preference for $w_0w_a$CDM over $\Lambda$CDM; however, as discussed above, this estimator generally penalises parameter volume less aggressively whenever an improved fit is achieved. Even in this case, the result remains well below the strong-to-decisive regime. On the other hand, when SH0ES data are included, the $\ln B$ criterion shifts decisively towards the DR framework over $\Lambda$CDM, consistently with the behaviour observed in Figure~\ref{fig:scatterH0}. The Gelman and Spiegelhalter $\Delta$DIC statistics point in the same direction, indicating a strong-to-decisive improvement in predictive performance for the DR framework relative to $\Lambda$CDM. We therefore conclude that the DR framework remains decisively favoured by the Bayesian evidence when its ability to alleviate the Hubble tension is taken into account (see also Table~\ref{tab:T-stat}).

\vspace{0.2cm}

Finally, it is worth noting that, after the Dovekie DESY5 recalibration \cite{Popovic_2025_DES_Evolving, Popovic_2025_Dovekie_Calib}, the results inferred from Pantheon$+$ and DESY5 became highly consistent with each other, leading to a much lower preference for EvoDE than that reported in DESI DR2 \cite{DESI_DR2} (see also \cite{ong_yallup_handley26}). Consequently, the values derived from both datasets are in excellent agreement, as shown in Table~\ref{tab:cosmo_results_PR4}.

\vspace{0.2cm}

In the right panel of Figure~\ref{fig:wowa}, we compare the 2D marginalised posterior distributions for $\Omega_m$ and $H_0$ for the CMB+BAO+Pantheon$+$ sample under the DR framework, with and without the SH0ES direct distance ladder calibration. The SH0ES-calibrated and SH0ES-free results are mutually consistent. However, as expected, including SH0ES data shifts the results towards lower values of $\Omega_m$, higher values of $H_0$, a larger number of relativistic species, with $N_{\rm tot}=3.63^{+0.13}_{-0.15}$, lower values of the sound horizon scale, $r_d=141.8^{+1.3}_{-1.2}$ Mpc, and higher values of the helium fraction, $Y_{\rm He}= 0.2530\pm 0.0017$. We already anticipated this behaviour in the right panel of Figure~\ref{fig:Yhe_PR4}, where we showed the abundances-only measurements from \cite{schoeneberg24} together with the CMB+BAO+Pantheon$+$ analysis including SH0ES calibration as dashed contours. Despite these trends, both the abundances-only and DR results remain highly consistent within the current error bars.

\vspace{0.2cm}

\begin{table}[htbp]
\centering
\renewcommand{\arraystretch}{1.4} 

\resizebox{\textwidth}{!}{%
\begin{tabular}{|lccc|}
\hline 
Model/Dataset &  $\ln(B)$ (SDDR) & $\Delta$DIC (Spiegelhalter)&$\Delta$DIC (Gelman)\\
\hline\hline

\multicolumn{4}{|l|}{\textbf{Dark Radiation}}  \\ 
 CMB + BAO + Pantheon$+$ & $-0.60 \pm 0.02$ & $-1.70$   & $2.61$ \\
 CMB + BAO + DESY5 & $-0.51\pm0.02$ & $-2.05$   & $3.23$ \\

 CMB + BAO + Pantheon$+$ w. SH0ES & $ 9.43 \pm 0.24 $ & $-19.18$   & $-10.48$ \\

  CMB + BAO + DESY5 w. SH0ES & $7.47 \pm 0.23$ & $-12.20$   & $-6.55$ \\
\hline 

\multicolumn{4}{|l|}{$\bm{w_0w_a}$\textbf{CDM}}  \\ 
CMB + BAO + Pantheon$+$ & $ 1.95 \pm 0.30$ & $-5.84$  & $-0.47$\\
CMB + BAO + DESY5 & $2.33 \pm 0.30$  & $ -6.31$  &$-2.14$\\
\hline
\end{tabular}%
 }
\caption{Values of $\ln(B)=\ln(\mathcal{Z_{\rm ext}})-\ln(\mathcal{Z}_{\Lambda \rm CDM})$ computed with SDDR estimation for different nested models. Negative values show preference for the $\Lambda$CDM model, while positive values show preference for the extended model. We also provide values for $\Delta \rm DIC=DIC_{ext}-DIC_{\Lambda CDM}$ following the Gelman and Spiegelhalter criteria (see text). These results are also displayed in Figure~\ref{fig:SDDR_DIC_plot}.}
\label{tab:SDDR} 
\end{table}

In Table~\ref{tab:helium} we present the predicted primordial helium fraction for the different cosmological models and datasets considered in this work, and compare them with direct helium abundance determinations.

\begin{table}[]
    \centering
    \resizebox{\textwidth}{!}{%
\
    \begin{tabular}{|c|c|c|}
    \hline
        Method & $Y_{\rm He}$ & Reference  \\
        \hline\hline
        CMB ($N_{\rm tot}\Lambda$CDM) & $0.2443\pm 0.0028$ & This work\\
        CMB (DR) & $0.2460\pm 0.0031$ & This work\\
        CMB+BAO (DR) & $0.2486\pm 0.0024$ & This work\\

        \multirow{2}{*}{CMB+BAO+SNe w. SH0ES (DR)}
            & $0.2530\pm0.0017$ (P+) & \multirow{2}{*}{This work} \\
            & $0.2519^{+0.0015}_{-0.0017}$ (DESY5) & \\
        \hline
        CMB ($\Lambda$CDM) & $0.24672\pm0.00031$ & \textit{Planck} Collaboration 2018 \cite{Planck_2018}\\
        \hline
        Metal-poor H\,{\sc ii} regions (EMPRESS) & $0.2402\pm0.0040$ & Yanagisawa et al. 2026 \cite{Yanagisawa26} \\
        Metal-poor H\,{\sc ii} regions (Leo P) & $0.2453 \pm 0.0034$ & Aver et al. 2021 \cite{Averetal2021} \\
        Extragalactic H\,{\sc ii} regions (NGC 346) & $0.2451\pm0.0026$ & Valerdi et al. 2019 \cite{Valerdi19} \\
        Intergalactic absorption systems & $0.250_{-0.025}^{+0.033}$ & Cook \& Fumagalli 2018 \cite{Cooke_Fumagalli18} \\
\hline
    \end{tabular}
     }
% I found this arXiv:2506.24050v2, it should be empress with Y_he=0.2402 \pm 0.0040
     
%Maybe add https://arxiv.org/abs/2601.22238 ?? which is low wrt to EMPRESS
    \caption{Summary of helium fraction with 68\% confidence levels reported direct determinations (bottom rows) and comparison with model predictions in this work (upper rows). For reference, we also include the \textit{Planck} collaboration derived parameter using the BBN internal consistency within $\Lambda$CDM, as done in this work for the reported models and datasets.} 
    \label{tab:helium}
\end{table}

\vspace{0.2cm}

The first block of rows displays the inferred values of $Y_{\rm He}$ for the different cosmological models and dataset combinations analysed in this work, including the DR model with CMB+BAO+SNe data calibrated with SH0ES for both supernova compilations. We find that the first three cases are generally consistent with the official \textit{Planck} Collaboration result obtained under $\Lambda$CDM with BBN consistency, $Y_{\rm He}\simeq 0.246$ \cite{Planck_2018}. However, when the SH0ES calibration is included, the preferred value increases to approximately $Y_{\rm He}\simeq 0.252$. This shift is associated with the alleviation of the Hubble tension within the DR framework, which favours a modified early-time expansion history and consequently a larger primordial helium fraction relative to the standard $\Lambda$CDM + SM prediction.

\vspace{0.2cm}

We compare these inferred values with direct abundance determinations from the literature. In particular, we consider:
\begin{itemize}
\item the recent EMPRESS determination of the primordial helium abundance by Yanagisawa et al. \cite{Yanagisawa26}, based on Subaru near-infrared spectroscopy of a large sample of extremely metal-poor star-forming galaxies. This analysis combines helium and hydrogen emission-line measurements with detailed photoionization modelling and stringent sample-selection criteria.

\item the helium abundance determination of Aver et al. \cite{Averetal2021}, based on a detailed reanalysis of the extremely metal-poor galaxy Leo~P. This study incorporates high-resolution optical and near-infrared spectroscopy, including the density-sensitive He\,{\sc i} emission line.

\item the determination of the primordial helium abundance by Valerdi et al. \cite{Valerdi19} is based on an extensive spectroscopic study of NGC~346, the brightest H\,{\sc ii} region in the Small Magellanic Cloud. This analysis employs long-slit optical observations covering a large fraction of the nebula, focusing on a single nearby, chemically homogeneous, low-metallicity H\,{\sc ii} region which minimizes uncertainties associated with extrapolations from heterogeneous galaxy samples.

\item the intergalactic absorption-system determination by Cooke \& Fumagalli \cite{Cooke_Fumagalli18}, which uses helium and hydrogen absorption features from a near-pristine intergalactic gas cloud observed against a background quasar. This method is fundamentally different from the emission-line analyses above, and is therefore largely independent from the associated nebular-emission systematic uncertainties. Although its current precision is significantly weaker, it provides an important complementary cross-check of the standard abundance determinations.
\end{itemize}

Overall, we find that the DR+SH0ES determination lies at the level of approximately $\sim2.9\sigma$ above the central value reported by Yanagisawa et al. \cite{Yanagisawa26} (EMPRESS), $\sim 2.5\sigma$ above the central value reported by Valerdi et al. \cite{Valerdi19} (NGC 346), and at the level of $\sim 2\sigma$ with respect to the Aver et al. determination \cite{Averetal2021} (LeoP), when uncertainties are combined in quadrature. It remains fully consistent with the Cooke \& Fumagalli \cite{Cooke_Fumagalli18} determination once its substantially larger uncertainty is taken into account. If we look at the $\mathcal{T}$-statistic for these measurements, we find $\ln{(\mathcal{T}(Y_{\rm He})})=2.38, 0.04, 0.95, 0.06$ respectively. This indicates that the value of $Y_{\rm He}$ coming from the DR+SH0ES combination is in a ``substantial" tension with the EMPRESS measurement, and a ``not worth a bare mention" tension with the other three. We therefore conclude that, although the larger helium fraction preferred by the DR framework in the presence of SH0ES data mildly increases the tension with some direct abundance measurements, the results remain statistically compatible within current observational uncertainties.

\vspace{0.2cm}

Finally, the performance of the best-fitting model within the DR framework using CMB, BAO, and Pantheon$+$ SNe data with the SH0ES calibration is presented and compared with the best-fitting EvoDE model obtained from CMB, BAO, and Pantheon$+$ SNe and a CMB and CMB+BAO $\Lambda$CDM model in Appendix~\ref{sec:model_residuals}. The flexibility of the EvoDE model allows for a slightly better fit to the data than $\Lambda$CDM, although it does not satisfactorily resolve the Hubble tension. By contrast, the DR framework is able to significantly alleviate the tension with SH0ES while only requiring a moderate decrease in $\Omega_m$ (and $r_d$) relative to the $w_0w_a$CDM and $\Lambda$CDM models. This leads to a region in the \textit{Planck} $TT$ spectrum, approximately within $500<\ell<1000$, where the DR model slightly overpredicts the measured $C_{\ell}^{TT}$ data points. However, the statistical significance of this feature cannot be assessed from the plots alone and instead requires a proper Bayesian model selection analysis, such as the one performed earlier in this section. These results should therefore be regarded as visually informative examples of how the DR framework can alleviate the Hubble tension while inducing only mild discrepancies in limited regions of the CMB spectrum.

\vspace{0.2cm}

\FloatBarrier  
\section{Discussion and Conclusions}
\label{discussion and conclus}

\vspace{0.2cm}

In this paper, we explore phenomenological modifications of the dark radiation (DR) sector and their impact on cosmology. We describe these in terms of two broad classes — free-streaming and fluid-like radiation — providing a model-independent framework to test deviations from the Standard Model with cosmological data. Both components are necessary to account for potential cancellations in the phase shifts of the CMB acoustic peaks, from which we derive the corresponding observational constraints on such scenarios. We perform a battery of studies on the performance of this two-class Dark Radiation framework.

\begin{itemize}

\item We first present the constraints inferred from the DR framework using CMB and BAO data. First, we find that the DR framework keeps the primordial power spectrum parameters $n_s$ and $A_s$ in excellent agreement with the $\Lambda$CDM inferred values. Second, we study the content of free-streaming and fluid-like radiation derived within the DR framework. The free-streaming radiation content is in excellent agreement with the Standard Model and $\Lambda$CDM predictions, as shown by the values reported in Equation~\ref{eq:eq:best_fit_ffs}. While BAO data are not sensitive to radiation and CMB data alone cannot tightly constrain the fluid-like component, their combination yields $N_{\rm fld}=0.36^{+0.16}_{-0.21}$.

\item Second, we examine whether the additional radiation degrees of freedom introduced in the DR framework can alleviate the Hubble tension. We find that the tension is significantly relaxed: the presence of DR both shifts the inferred value of $H_0$ to slightly higher values and increases its uncertainty in CMB and CMB+BAO analyses (see Table~\ref{tab:cosmo_results_PR4}). Quantitatively, using the $\mathcal{T}$-statistic (see Table~\ref{tab:T-stat}), the tension is reduced from highly significant in $\Lambda$CDM to substantial for CMB data alone, and to statistically non-significant when CMB and BAO data are combined within the DR framework. Finally, we characterise the impact of DR on the expansion history of the Universe and find deviations from $\Lambda$CDM remain at the level of $\lesssim 2\%$, and are therefore consistent with current late-time expansion constraints.

\item Finally, we compare the dark radiation (DR) framework with an evolving dark energy model (EvoDE), parametrised through the CPL form ($w_0w_a$CDM), including supernova (SNe) data from Pantheon$+$ and DES-Y5 from the Dovekie reanalysis. Both scenarios introduce new dark sector physics via two additional parameters relative to $\Lambda$CDM ($w_0$, $w_a$ for $w_0w_a$CDM, and $N_{\rm fs}$, $N_{\rm fld}$ for DR). To assess their performance, we employ two model selection criteria relative to $\Lambda$CDM: the Deviance Information Criterion (DIC) and the Savage--Dickey Density Ratio (SDDR), expressed in terms of $\ln B$. Since the Bayes factor is directly related to the marginal likelihood of the model, we regard $\ln B$ as a more robust indicator for model selection.
When CMB+BAO+SNe data are used, we find no decisive preference for any model over $\Lambda$CDM. However, $w_0w_a$CDM is moderately to strongly favoured, while the DR model is weakly disfavoured with respect to $\Lambda$CDM, leading overall to inconclusive evidence.
In contrast, when SH0ES data are included, we find that the DR model shows decisive preference over $\Lambda$CDM, whereas for the $w_0w_a$CDM model the inclusion of SH0ES is not statistically consistent due to its strong tension with early-Universe calibrated data.
\end{itemize}

We conclude that DR is a compelling framework to provide a potential solution to the Hubble tension by achieving novel results to quantify the tension with the $\mathcal{T}$-statistic and to check whether data allows for the DR framework with the model selection analysis. When we examine more closely how this tension is resolved, we find that it occurs at the expense of increasing the effective number of relativistic species to $N_{\rm tot}\simeq 3.6$, as well as increasing the predicted helium fraction from $Y_{\rm He}\simeq 0.246$ up to $Y_{\rm He}\simeq 0.252$ (see Tables~\ref{tab:cosmo_results_PR4} and \ref{tab:helium} for specific values depending on the dataset choice). When we compare this increased helium fraction with abundance determinations, we find that it implies a $\sim 2$–$2.9\sigma$ discrepancy with respect to measurements based on metal-poor H\,{\sc ii} regions, and a much milder difference with respect to works based on helium and hydrogen absorption features from near-pristine intergalactic gas clouds. Therefore, although predicting larger primordial helium fractions than those inferred from abundance measurements, these results remain statistically compatible within current observational uncertainties.

\vspace{0.2cm}

To definitively rule out or confirm the dark radiation framework, further work is essential to precisely constrain $N_{\rm tot}$. One promising avenue is to incorporate the full-shape information from galaxy clustering \cite{DESI2024_V,DESI2024_VII,Elbersetal25} alongside the Lyman-$\alpha$ 1D power spectrum data \cite{Chaves_Montero_2026_DESILya}. The inclusion of these datasets is particularly powerful due to their sensitivity to a broad range of $k$-scales; while full-shape analysis captures the broadband suppressions in the matter power spectrum, Lyman-$\alpha$ probes the small-scale (high-$k$) regime where the specific clustering properties of fluid-like versus free-streaming relics become most distinct. This information is highly complementary to early-time CMB and late-time BAO measurements and can provide critical insights into potential new physics within the relativistic species sector. Another avenue is the study of other CMB measurements, such as the ones provided by the ACT collaboration \cite{ACT}. These measurements probe smaller scales compared to \textit{Planck}, which have led to a substantial change in bounds on DR (towards lower contents of DR). however, this issue appears to be solved when allowing for a change in the running and running of the running of the primordial power spectrum index \cite{Garny_2026_ACT_DR6}. Ultimately, upcoming experiments, such as LiteBird \cite{LiteBird} and the Simons Observatory \cite{SO}, are expected to provide the definitive sensitivity required to probe these dark radiation scenarios with unprecedented precision

\FloatBarrier  
\acknowledgments

We thank Licia Verde for useful suggestions and comments throughout this work. MM acknowledges support by the Spanish Ministry of Education and Vocational Training (Ministerio de Educación y Formación Profesional) through a `Beca de Colaboración en Departamentos Universitarios' (Grant No. 25C01/003680). HGM acknowledges support through the Consolidación Investigadora (CNS2023-144605) of the Spanish Ministry of Science and Innovation. Plots and summary statistics are obtained with \texttt{getdist} \cite{Lewis_2025_GetDist}. All the calculations presented in this paper were done with the \texttt{Aganice} cluster at the Institut de Ciències del Cosmos at the University of Barcelona.

\appendix
\section{Dark Radiation effects on the $C_\ell$ of the CMB power spectrum}
\label{sec:Cl_on_CMB}
\begin{figure}[htbp]
\centering
\includegraphics[width=.32\textwidth]{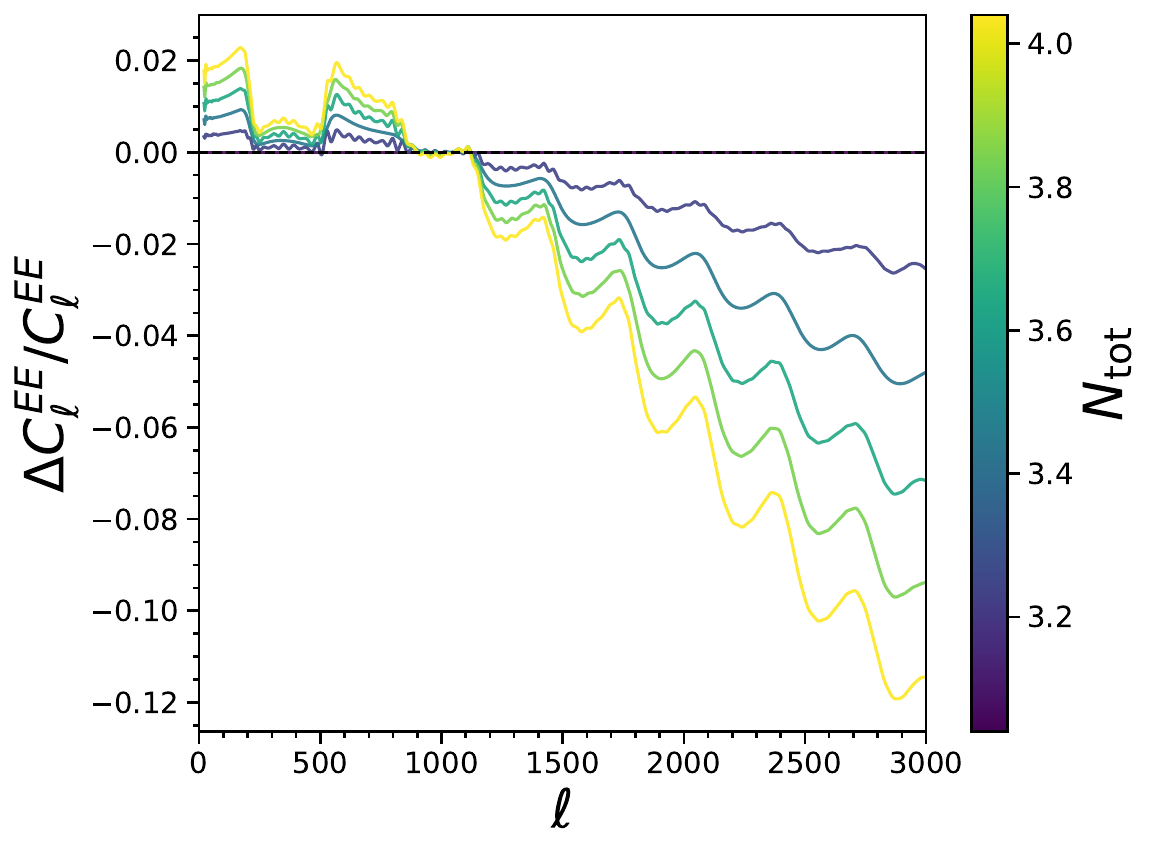}
\includegraphics[width=.32\textwidth]{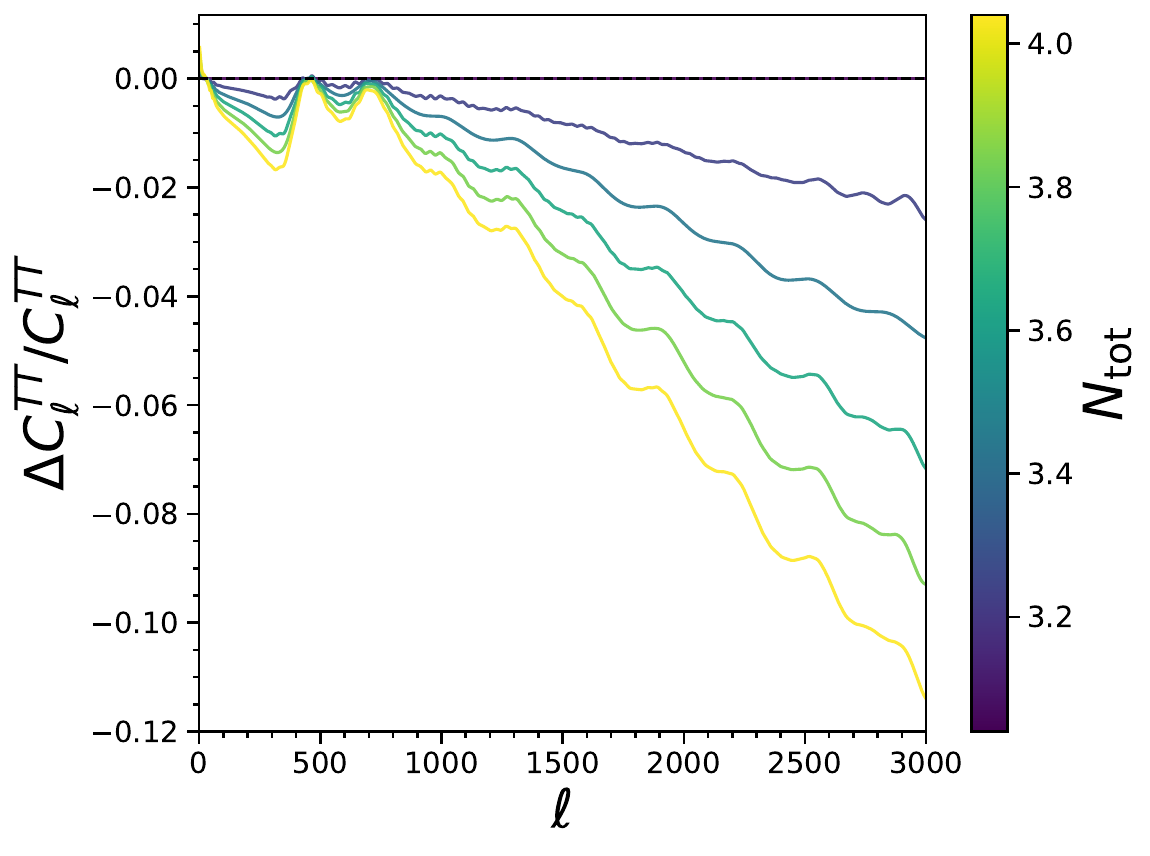}
\includegraphics[width=.32\textwidth]{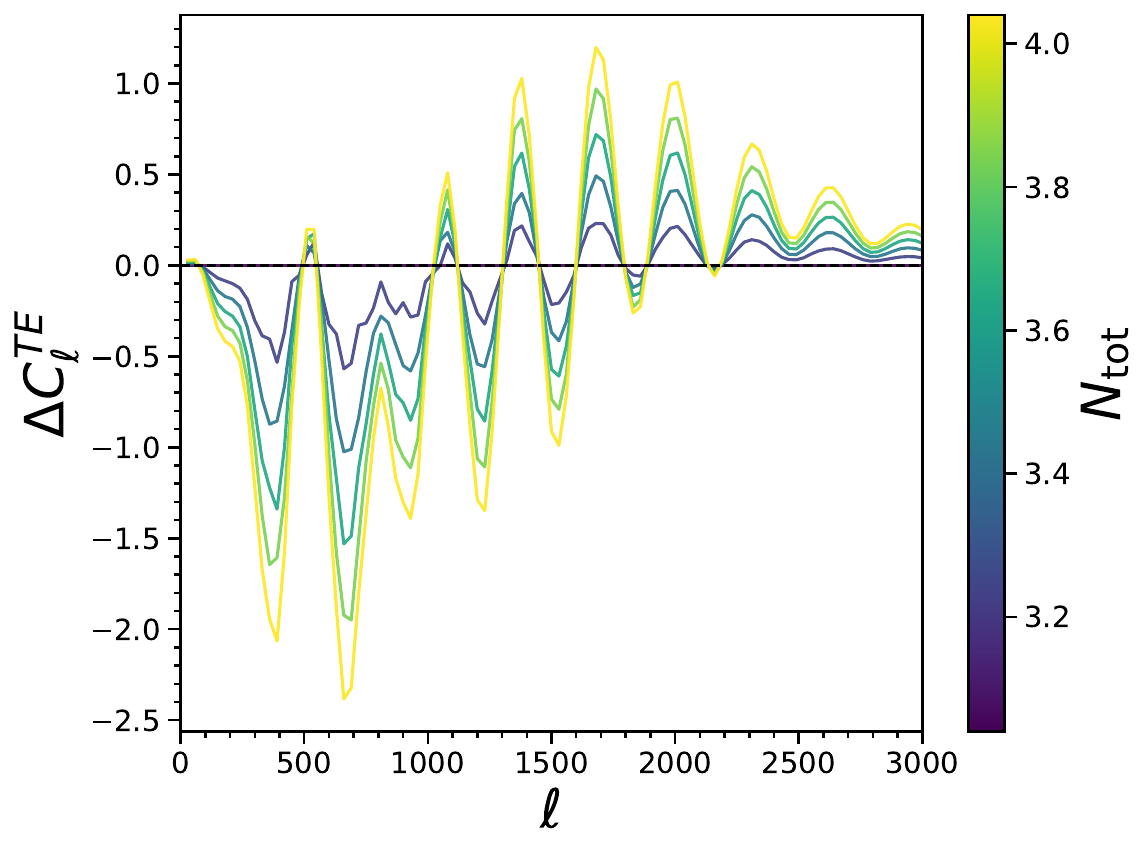}
\caption{Effect of changing the effective number of relativistic species on the different $C_\ell$ of CMB, at fixed $f_{\mathrm{fs}}$, $\theta_s$ and $a_{\rm eq}$. These quantities are fixed by modifying $\Lambda$CDM parameters. Results are shown with relative differences with a $\Lambda$CDM cosmology ($N_{\rm tot}=N_{\rm eff}=3.044$).\label{fig:Cl_CMB}}
\end{figure}

Introducing DR has effects on the $C_\ell$ of the CMB power spectrum. Focusing on the effects at fixed $f_{\mathrm{fs}}$ while accomplishing the observational constraints explained in Section~\ref{sec:constraints_DR}, we recover the results found in \cite{DR_CMB}. Results are shown in Figure~\ref{fig:Cl_CMB}. The variations in $C_\ell$ are not large, suggesting that DR induces a subtle change in the cosmological parameters.

\FloatBarrier

\section{The Damping tail constraints on Dark Radiation}
\label{damping}

\begin{figure}[htbp]
\centering
\includegraphics[width=.49\textwidth]{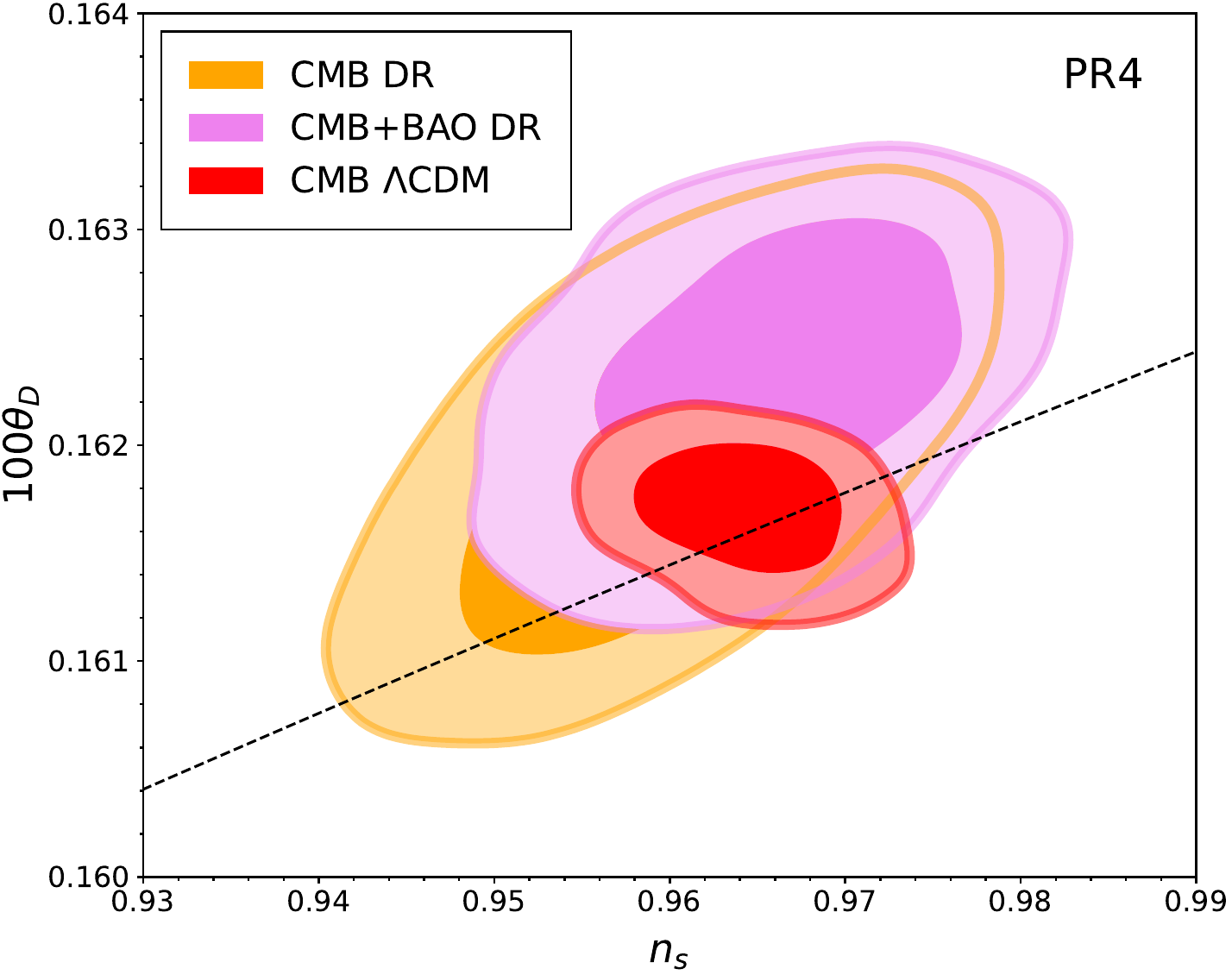}
\hfill
\includegraphics[width=.49\textwidth]{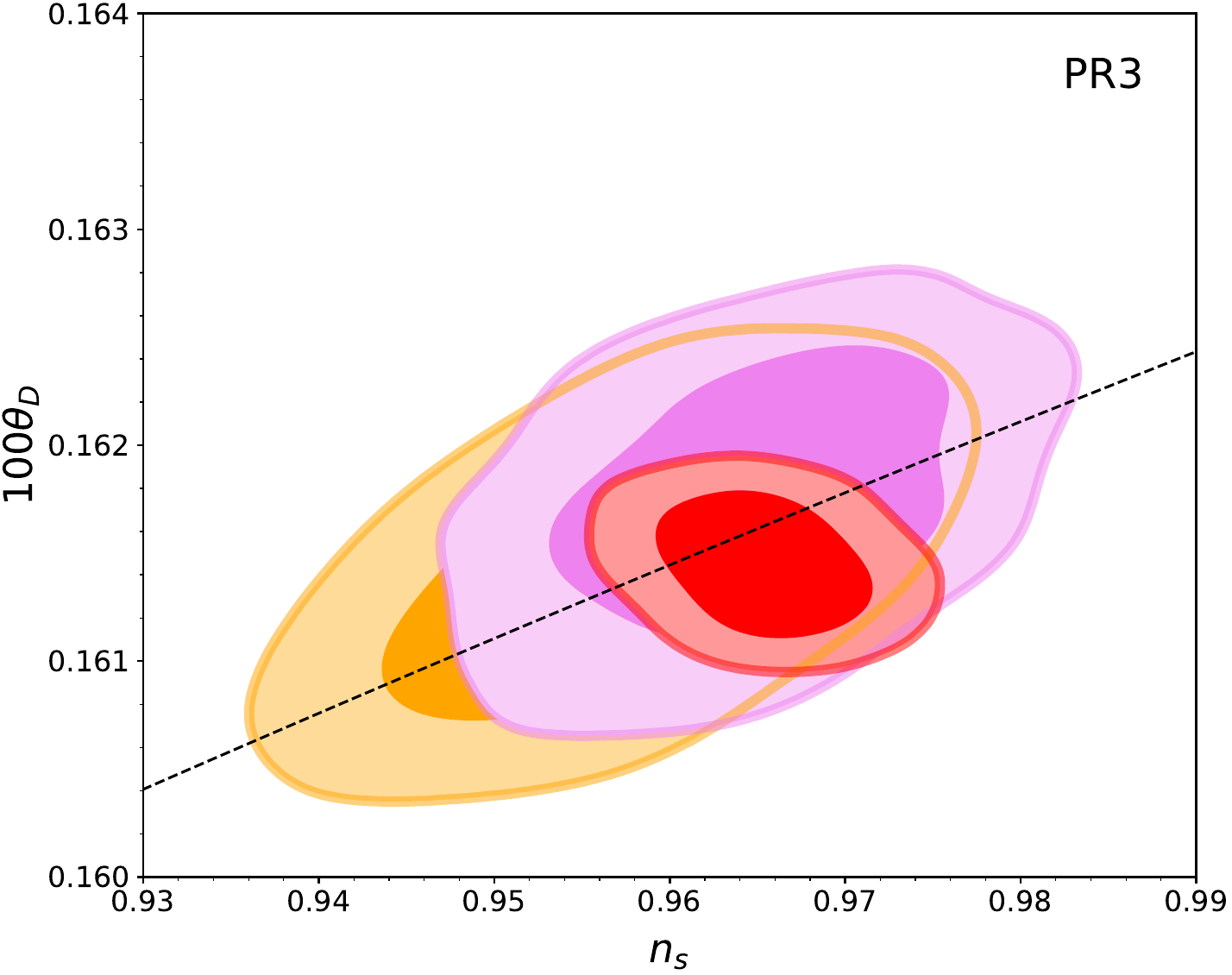}
\caption{Marginalised 2D posterior distributions (68\% and 95\% credible intervals) for the damping scale $\theta_D$ and the scalar spectral index $n_s$. The left panel shows constraints using \textit{Planck} PR4, while the right panel utilises \textit{Planck} PR3. In both panels, orange contours represent the dark radiation (DR) model constrained by CMB data alone, while magenta contours show the effect of combining \textit{Planck} with DESI DR2 BAO data. For comparison, the red contours indicate the standard $\Lambda$CDM constraints (without DR) using the respective \textit{Planck} datasets. The black dashed line corresponds to the theoretical relationship defined by Eq.~\ref{eq:damping tail}.\label{fig:theta_D_PR4}}
\end{figure}

As discussed in Sections~\ref{sec:constraints_DR} and \ref{sec:primordial i damping}, the angular diffusion scale at recombination $\theta_D$ must be kept relatively fixed to maintain the observational constraints on the small-scale CMB power spectrum. We show quantitatively if we can maintain the shape of the damping tail when introducing DR. $\theta_D$ is computed according to \textit{Planck}'s collaboration conventions \cite{Planck_2018,Smith_2024_EDE}:
\begin{align}
\label{eq:theta_D}   &\theta_D(z_\star) \equiv \frac{\pi}{k_D(z_\star)D_M(z_\star)}; \\
     &k_D^{-2} \equiv -\frac{1}{6} \int_{z_\star}^\infty \frac{cdz'}{\dot{\kappa}H(z')} \frac{R^2 + 16(1 + R)/15}{(1 + R)^2};
\end{align}

where, $k_D \equiv \pi/r_D$,  $\dot{\kappa}\equiv\frac{d\kappa}{d\eta}=n_e\sigma_T a$ is the rate of change of the photon's optical depth, $\sigma_T$ is the Thomson scattering cross section, $a$ the scale factor, and $\eta$ the conformal time $\eta(t)=\int\frac{cdt}{a}$. The following relation must hold in order to maintain the damping tail relatively unchanged \cite{Smith_2024_EDE}, 
\begin{equation}
    \label{eq:damping tail}
    \frac{\delta \theta_D}{\theta_D}\approx+0.2\frac{\delta n_s}{n_s},
\end{equation}
where $\delta \theta_D=\theta_D-\theta_D^{\Lambda {\rm CDM}}$ and $\delta n_s=n_s-n_s^{\Lambda \rm{CDM}}$. The DR framework shows notable differences when studying this equation between \textit{Planck}'s PR4 and PR3 datasets. PR4 prefers higher values of $\theta_D$ compared to PR3, as shown in Figure~\ref{fig:theta_D_PR4}. For this reason, PR3 is in slightly better agreement with Eq.~\ref{eq:damping tail} than PR4. Higher values of $\theta_D$ are related to the preference of PR4 for larger values of $Y_{\rm He}$ (as shown in \cite{DR_CMB}). An increase in $Y_{\rm He}$, leads to a decrease in $n_e$, which according to Eq.~\ref{eq:damping tail} leads to higher values of $\theta_D$

\section{Residual Analysis of the Best-Fitting Dark Radiation Model}
\label{sec:model_residuals}

This appendix provides a detailed visualisation of the residuals for the best-fitting model within the dark radiation framework. By isolating the deviations between the observed data and the best-fitting model, we can assess the goodness-of-fit across the diverse cosmological datasets used in this study.

\vspace{0.2cm}

We present the residuals of the best-fitting model relative to the primary cosmological probes: the Cosmic Microwave Background (CMB), Baryon Acoustic Oscillations (BAO), Type Ia Supernovae (SNe), and the local Hubble constant ($H_0$) from SH0ES. For simplicity, we only take Pantheon+ data as the SNe data, as the results are very similar for those with the DESY5 SNe sample. For comparison, we also display the EvoDE best-fitting model to BAO, CMB and Pantheon+ SNe data, and we take the $\Lambda$CDM best-fitting model to CMB data (from \textit{Planck} 2018 \cite{Planck_2018}) as fiducial reference cosmology. 

\vspace{0.2cm}

\begin{figure}[htb!]
    \centering
    \includegraphics[width=0.32\linewidth]{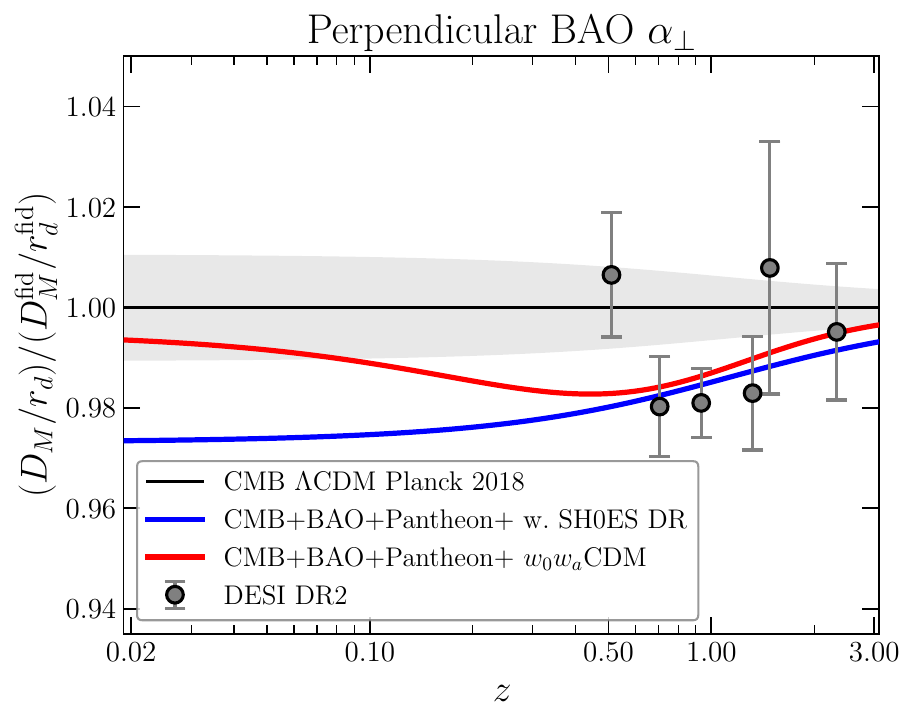}
    \includegraphics[width=0.32\linewidth]{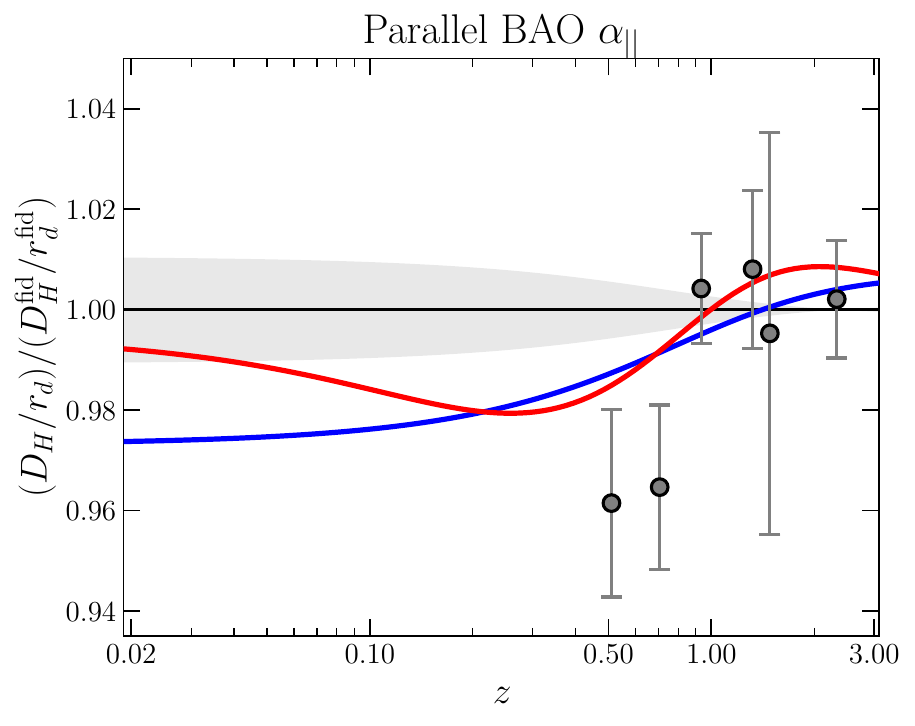}
    \includegraphics[width=0.32\linewidth]{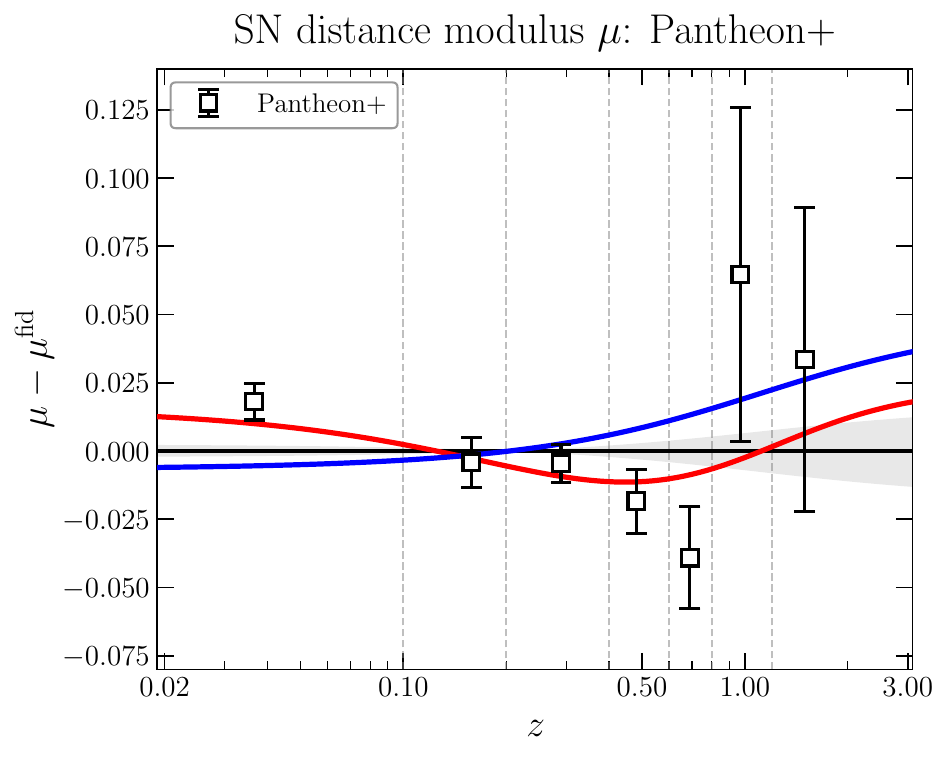}

    \includegraphics[width=0.32\linewidth]{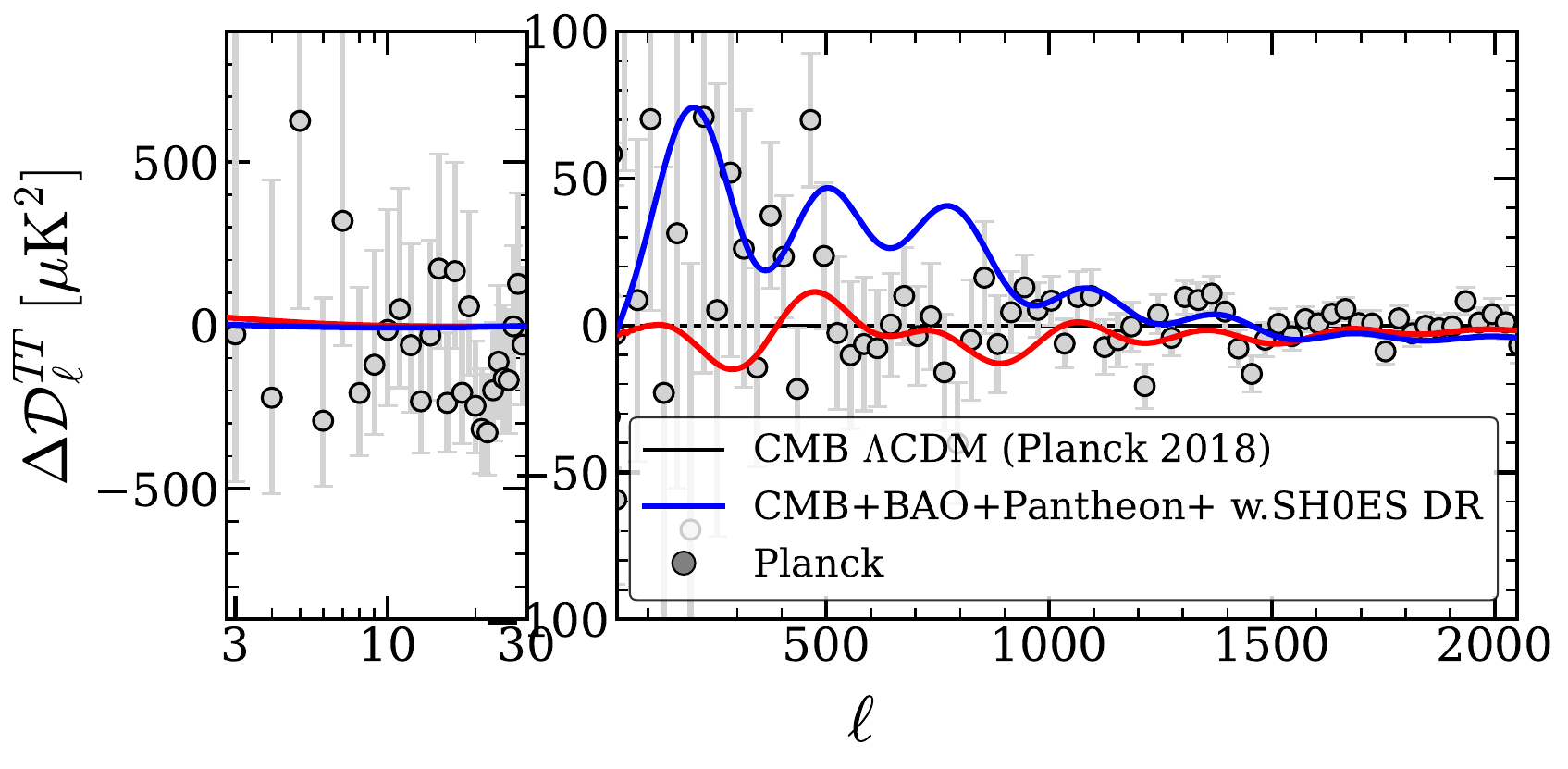}
    \includegraphics[width=0.32\linewidth]{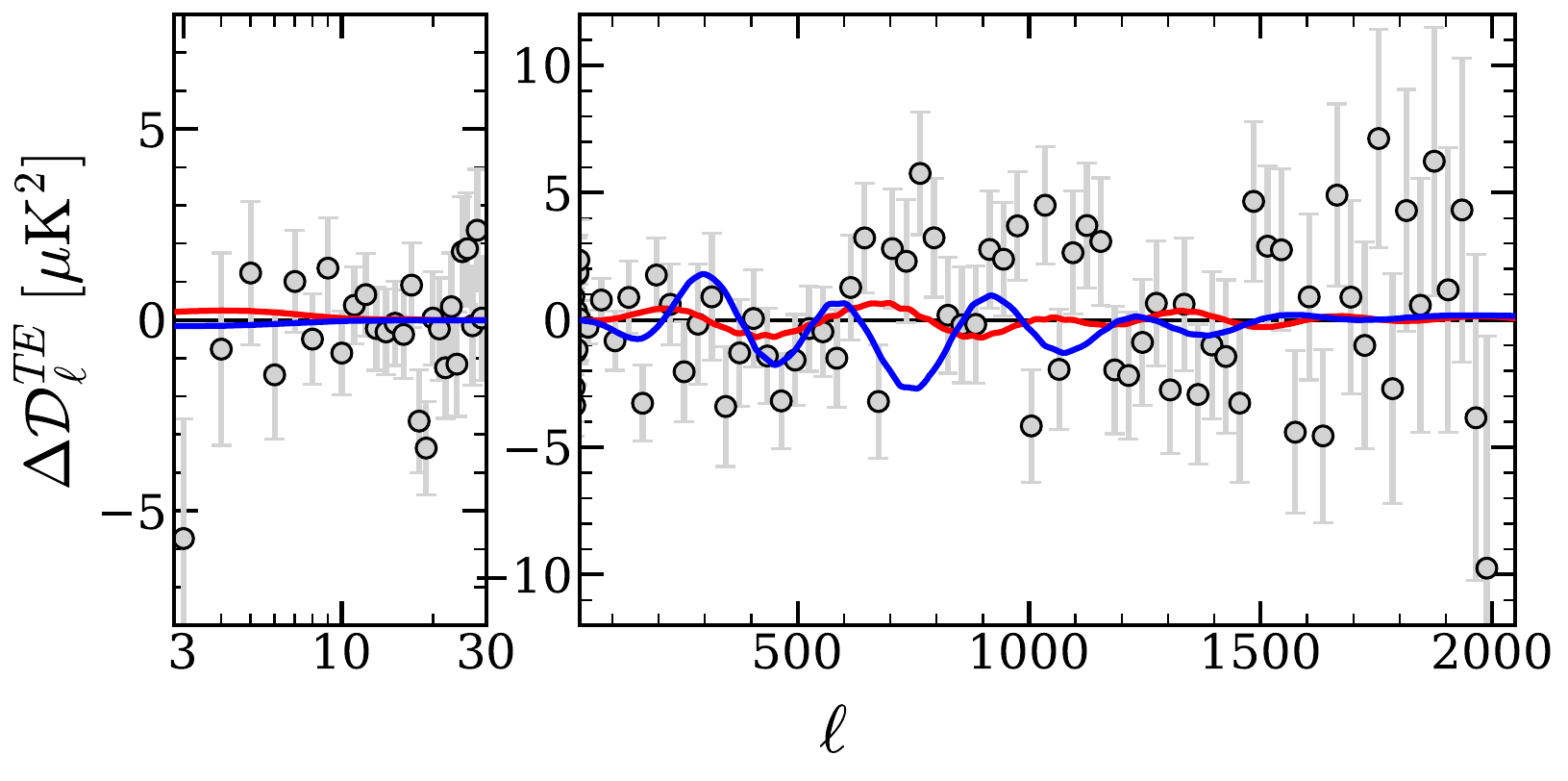}
    \includegraphics[width=0.32\linewidth]{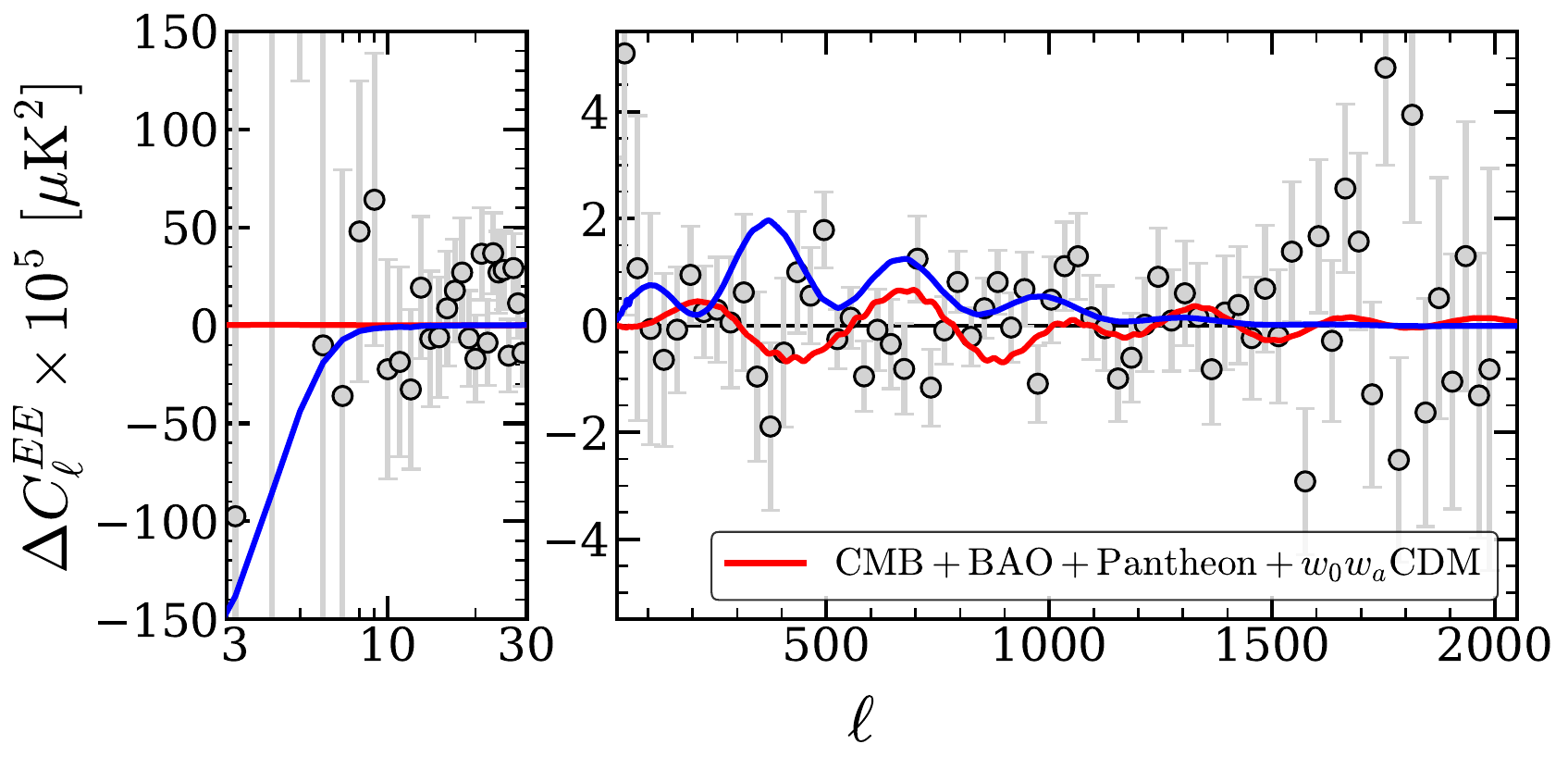}    
    \caption{{\it Top panels}: Hubble diagrams showing comparisons of DESI DR2 BAO (radial and angular distances) and Pantheon$+$ SNe distance modulus data to the best-fitting models: Dark Radiation to CMB+BAO+Pantheon+ SNe with SH0ES (blue lines); and $w_0w_a$CDM model to CMB+BAO+Pantheon$+$ without SH0ES (red line). For simplicity, we choose to represent only the Pantheon$+$ SNe, and have re-binned the 1701 SNe samples into just 7 $\mu(z)$-bins. The solid grey band denotes the 68\% confidence interval of the baseline reference model. {\it Bottom panels:} \textit{Planck} CMB data temperature and polarisation spectra, $D_\ell\equiv\frac{\ell(\ell+1)}{2\pi}C_\ell$, relative to the best-fitting CMB $\Lambda$CDM model (black line), and the best-fitting models: Dark Radiation model fit to CMB+BAO+Pantheon+ SNe with SH0ES data (blue lines); and $w_0w_a$CDM model to CMB+BAO+Pantheon+ SNe without SH0ES data (red lines).  }
    \label{fig:BAOmures}
\end{figure}

Top panels of Figure \ref{fig:BAOmures} provide a three-panel summary of the late-time expansion history residuals. The first two panels show the residuals for the transverse and longitudinal DESI DR2 BAO measurements, which constrain the angular diameter and the Hubble distance at various redshifts in units of the sound horizon scale. The third panel displays the residuals for the distance modulus ($\mu$) from the Pantheon$+$ SNe, where $\mu$ has been re-binned accordingly for a better visualisation.

\vspace{0.2cm}

Bottom panels of Figure \ref{fig:BAOmures} display the residuals for the CMB temperature ($TT$), cross-polarisation ($TE$), and $E$-mode polarisation ($EE$) power spectra, for the same models presented before. The aim of this representation is to illustrate the model's ability to capture the acoustic peak structure and damping tail while accounting for the additional relativistic degrees of freedom introduced by the dark radiation component and SH0ES data. We see that while the $w_0w_a$CDM model has a very similar structure as the vanilla $\Lambda$CDM model, the Dark Radiation framework thend to predict a slightly higher signal for the $TT$ component in the $500<\ell<1000$ range, although it is compatible with the CMB data. 

\vspace{0.2cm}

Together, these figures demonstrate that the best-fitting dark radiation model maintains high consistency across both high-redshift early-universe physics and low-redshift expansion markers, and how they compare to popular models of the literature, $w_0w_a$CDM to CMB+BAO+Pantehon$+$, and $\Lambda$CDM to CMB.

\vspace{0.2cm}

\begin{figure}
    \centering
    \includegraphics[width=0.6\linewidth]{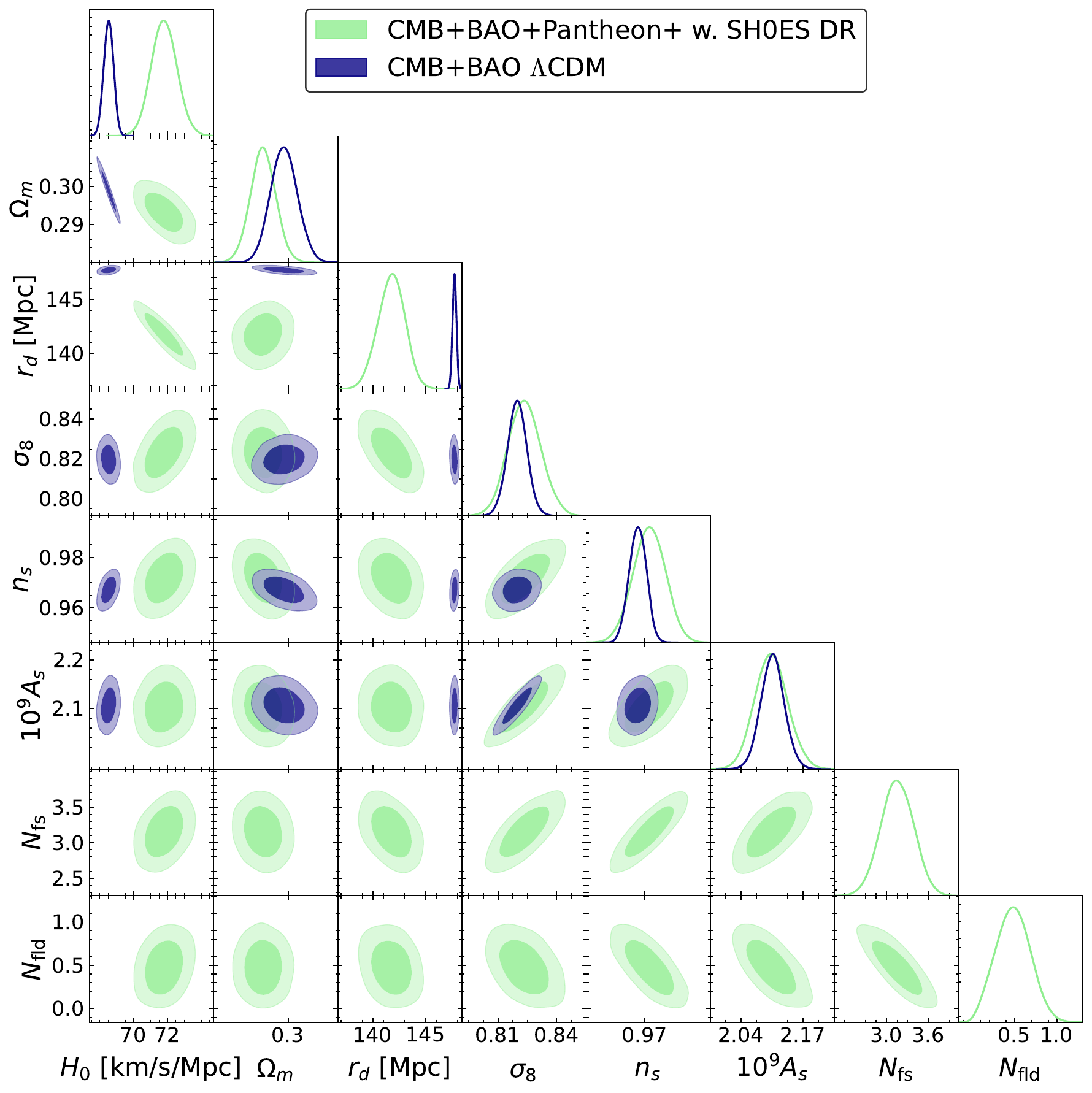}
    \caption{Marginalised 2D posterior distributions (68\% and 95\% credible intervals) for the cosmological parameters obtained when considering the Dark Radiation best-fitting model CMB+BAO+Pantheon+ SNe with SH0ES data (navy contours) and a vanilla $\Lambda$CDM model fit to CMB+BAO data (light green contours).}
    \label{fig:cosmo}
\end{figure}

Finally, in Figure~\ref{fig:cosmo} we present a full corner plot with the final cosmological parameters, so the correlations among parameters are displayed. For clarity we also display the derived parameters, $\sigma_8$ and $r_d$. We see a highly consistent agreement between DR and $\Lambda$CDM model in all parameters, except on $H_0$, where DR predicts a higher value as widely discussed in Section~\ref{sec:EDE_vs_DR}.

\FloatBarrier  
In Table \ref{tab:chi2B} we display the best-fitting $\chi^2$ for the different frameworks studied in this work, EvoDE ($w_ow_a$CDM) and Dark Radiation, taking into account CMB, BAO, SNe data (Pantheon$+$ in this case), with and without the SH0ES callibration. For reference, we also display for each case the $\Delta\chi^2\equiv \chi^2_{\rm model}-\chi^2_{\Lambda \rm CDM}$. 

\begin{table}[htb]
\centering
\label{tab:chi2}
\begin{tabular}{|lc|cc|cc|}
\hline
Dataset Combination 
& $\Lambda$CDM 
& $w_0w_a$CDM & $\Delta\chi^2$ 
& Dark Radiation & $\Delta\chi^2$ \\
\hline\hline
CMB + BAO + SNe       & $12028.5$ & $12020.9$ & $-7.6$ & $12025.5$ & $-3.0$ \\
CMB + BAO + SNe + SH0ES  & $12110.0$ & $12106.5$ & $-3.5$ & $12081.8$ &  $-28.2$ \\
\hline
\end{tabular}
\caption{Best-fit $\chi^2$ values for the different cosmological models, EvoDE and DR frameworks,  using two of the dataset combinations considered in this work, as indicated. For the SNe data, we take Pantheon$+$, as done for the rest of the figures of this appendix.  We also report the $\chi^2$-difference with respect to the $\Lambda$CDM model. The total number of elements in the data vector is 9982 (for CMB) + 13 (for BAO) + 1701 (for SNe) + 1 (when SH0ES is included). The number of free parameters (including cosmological and nuisance) is 15 for $\Lambda$CDM, and 17 for DR and EvoDE.}
    \label{tab:chi2B}
\end{table}

\section{Priors}
\label{sec:priors}

We display in Table~\ref{tab:priors} the priors used for both cosmological and nuisance parameters in our study. For the nuisance parameters we employ the internal notation suggested in \texttt{Cobaya}. $\mathcal{U}[x_{\rm min},\,x_{\rm max}]$ stands for a uniform prior between $x_{\rm min}$ and $x_{\rm max}$, whereas $\mathcal{N}(\mu,\,\sigma^2)$ is Gaussian prior with mean $\mu$ and variance, $\sigma^2$. As commented in the main text, we additionally apply the usual prior $w_0+w_a\leq 0$, which ensures matter domination at early times.

\begin{table}[htbp]
\centering
\renewcommand{\arraystretch}{1.3}

\begin{tabular}{|lc||lc|}
\hline

\multicolumn{4}{|l|}{\textbf{Cosmological parameters}} \\
\hline
$\Omega_b h^2$   & $\mathcal{U}[0.019, 0.025]$ 
& $n_s$           & $\mathcal{U}[0.9, 1.1]$ \\

$\Omega_c h^2$   & $\mathcal{U}[0.09, 0.15]$ 
& $N_{\rm fld}$   & $\mathcal{U}[0, 2]$ \\

$H_0$ [km/s/Mpc]            & $\mathcal{U}[50, 90]$ 
& $N_{\rm ur}$    & $\mathcal{U}[0.5, 4]$ \\

$\tau$           & $\mathcal{U}[0.01, 0.2]$ 
& $w_0$           & $\mathcal{U}[-3, 1]$ \\

$10^9\cdot A_s$            & $\mathcal{U}[1,5]$ 
& $w_a$           & $\mathcal{U}[-3,2]$ \\

\hline
\hline
\multicolumn{4}{|l|}{\textbf{Nuisance parameters}} \\
\hline
$y_\mathrm{cal}$ & $\mathcal{N}(1, 0.0025^2)$ 
& $c_{TE}$ & $\mathcal{N}(1, 0.01^2)$ \\

$A^{\rm power}_{143}$ & $\mathcal{U}[0, 50]$ 
& $c_{EE}$ & $\mathcal{N}(1, 0.01^2)$ \\

$A^{\rm power}_{217}$ & $\mathcal{U}[0, 50]$ 
& $\gamma^{\rm power}_{143}$ & $\mathcal{U}[0, 5]$ \\

$A^{\rm power}_{143\times217}$ & $\mathcal{U}[0, 50]$ 
& $\gamma^{\rm power}_{217}$ & $\mathcal{U}[0, 5]$ \\

& &   $\gamma^{\rm power}_{143\times217}$ & $\mathcal{U}[0, 5]$ \\

\hline
\end{tabular}

\caption{Prior limits for all the parameters utilised in this work. Recall that we have used the definition $N_{\rm fs}\equiv N_{\rm ur}+1.$}
\label{tab:priors}
\end{table}

\bibliographystyle{JHEP}
\bibliography{biblio}

@ARTICLE{Valerdi19,
       author = {{Valerdi}, Mabel and {Peimbert}, Antonio and {Peimbert}, Manuel and {Sixtos}, Andr{\'e}s},
        title = "{Determination of the Primordial Helium Abundance Based on NGC 346, an H II Region of the Small Magellanic Cloud}",
      journal = {\apj},
         year = 2019,
        month = may,
       volume = {876},
       number = {2},
          eid = {98},
        pages = {98},
          doi = {10.3847/1538-4357/ab14e4},
archivePrefix = {arXiv},
       eprint = {1904.01594},
 primaryClass = {astro-ph.GA},
       adsurl = {https://ui.adsabs.harvard.edu/abs/2019ApJ...876...98V}
}

@ARTICLE{Yanagisawa26,
       author = {{Yanagisawa}, Hiroto and {Ouchi}, Masami and {Matsumoto}, Akinori and {Kawasaki}, Masahiro and {Murai}, Kai and {Nakajima}, Kimihiko and {Kohri}, Kazunori and {Sugahara}, Yuma and {Nagamine}, Kentaro and {Tanaka}, Ichi and {Kim}, Ji Hoon and {Ono}, Yoshiaki and {Nakane}, Minami and {Fukushima}, Keita and {Harikane}, Yuichi and {Hirai}, Yutaka and {Isobe}, Yuki and {Kusakabe}, Haruka and {Onodera}, Masato and {Rauch}, Michael and {Yajima}, Hidenobu},
        title = "{EMPRESS. XV. A New Determination of the Primordial Helium Abundance Suggesting a Moderately Low $Y_\mathrm{P}$ Value}",
      journal = {arXiv e-prints},
         year = 2025,
        month = jun,
          eid = {arXiv:2506.24050},
        pages = {arXiv:2506.24050},
          doi = {10.48550/arXiv.2506.24050},
archivePrefix = {arXiv},
       eprint = {2506.24050},
 primaryClass = {astro-ph.GA},
       adsurl = {https://ui.adsabs.harvard.edu/abs/2025arXiv250624050Y}
}

@ARTICLE{Averetal2021,
       author = {{Aver}, Erik and {Berg}, Danielle A. and {Olive}, Keith A. and {Pogge}, Richard W. and {Salzer}, John J. and {Skillman}, Evan D.},
        title = "{Improving helium abundance determinations with Leo P as a case study}",
      journal = {\jcap},
         year = 2021,
        month = mar,
       volume = {2021},
       number = {3},
          eid = {027},
        pages = {027},
          doi = {10.1088/1475-7516/2021/03/027},
archivePrefix = {arXiv},
       eprint = {2010.04180},
 primaryClass = {astro-ph.CO},
       adsurl = {https://ui.adsabs.harvard.edu/abs/2021JCAP...03..027A}
}

@ARTICLE{Cooke_Fumagalli18,
       author = {{Cooke}, Ryan J. and {Fumagalli}, Michele},
        title = "{Measurement of the primordial helium abundance from the intergalactic medium}",
      journal = {Nature Astronomy},
         year = 2018,
        month = oct,
       volume = {2},
        pages = {957-961},
          doi = {10.1038/s41550-018-0584-z},
archivePrefix = {arXiv},
       eprint = {1810.06561},
 primaryClass = {astro-ph.CO},
       adsurl = {https://ui.adsabs.harvard.edu/abs/2018NatAs...2..957C}
}

@ARTICLE{ong_yallup_handley26,
       author = {{Ong}, Dily Duan Yi and {Yallup}, David and {Handley}, Will},
        title = "{A Bayesian Perspective on Evidence for Evolving Dark Energy}",
      journal = {arXiv e-prints},
         year = 2025,
        month = nov,
          eid = {arXiv:2511.10631},
        pages = {arXiv:2511.10631},
          doi = {10.48550/arXiv.2511.10631},
archivePrefix = {arXiv},
       eprint = {2511.10631},
 primaryClass = {astro-ph.CO},
       adsurl = {https://ui.adsabs.harvard.edu/abs/2025arXiv251110631O}
}

@ARTICLE{schoeneberg24,
       author = {{Sch{\"o}neberg}, Nils},
        title = "{The 2024 BBN baryon abundance update}",
      journal = {\jcap},
         year = 2024,
        month = jun,
       volume = {2024},
       number = {6},
          eid = {006},
        pages = {006},
          doi = {10.1088/1475-7516/2024/06/006},
archivePrefix = {arXiv},
       eprint = {2401.15054},
 primaryClass = {astro-ph.CO},
       adsurl = {https://ui.adsabs.harvard.edu/abs/2024JCAP...06..006S}
}

@ARTICLE{Elbersetal25,
       author = {{Elbers}, W. and {Aviles}, A. and {Noriega}, H.~E. and {Chebat}, D. and {Menegas}, A. and {Frenk}, C.~S. and {Garcia-Quintero}, C. and {Gonzalez}, D. and {Ishak}, M. and {Lahav}, O. and {Naidoo}, K. and {Niz}, G. and {Y{\`e}che}, C. and {Abdul-Karim}, M. and {Ahlen}, S. and {Alves}, O. and {Andrade}, U. and {Armengaud}, E. and {Behera}, J. and {BenZvi}, S. and {Bianchi}, D. and {Brieden}, S. and {Brodzeller}, A. and {Brooks}, D. and {Burtin}, E. and {Calderon}, R. and {Canning}, R. and {Carnero Rosell}, A. and {Casas}, L. and {Castander}, F.~J. and {Charles}, M. and {Chaussidon}, E. and {Chaves-Montero}, J. and {Claybaugh}, T. and {Cole}, S. and {Cooper}, A.~P. and {Cuceu}, A. and {Dawson}, K.~S. and {de la Macorra}, A. and {de Mattia}, A. and {Deiosso}, N. and {Dey}, A. and {Dey}, B. and {Ding}, Z. and {Doel}, P. and {Eisenstein}, D.~J. and {Ferraro}, S. and {Font-Ribera}, A. and {Forero-Romero}, J.~E. and {Garrison}, L.~H. and {Gazta{\~n}aga}, E. and {Gil-Mar{\'\i}n}, H. and {Gontcho}, S. Gontcho A. and {Gonzalez-Morales}, A.~X. and {Gutierrez}, G. and {He}, S. and {Herbold}, M. and {Herrera-Alcantar}, H.~K. and {Howlett}, C. and {Huterer}, D. and {Juneau}, S. and {Kehoe}, R. and {Kirkby}, D. and {Kisner}, T. and {Kremin}, A. and {Lamman}, C. and {Landriau}, M. and {Le Guillou}, L. and {Leauthaud}, A. and {Levi}, M.~E. and {Li}, Q. and {Lodha}, K. and {Magneville}, C. and {Manera}, M. and {Martini}, P. and {Matthewson}, W.~L. and {Meisner}, A. and {Mena-Fern{\'a}ndez}, J. and {Miquel}, R. and {Moustakas}, J. and {Nadathur}, S. and {Newman}, J.~A. and {Paillas}, E. and {Palanque-Delabrouille}, N. and {Percival}, W.~J. and {Pieri}, M.~M. and {Poppett}, C. and {Prada}, F. and {P{\'e}rez-R{\`a}fols}, I. and {Rabinowitz}, D. and {Ram{\'\i}rez-P{\'e}rez}, C. and {Rashkovetskyi}, M. and {Ravoux}, C. and {Rivera-Morales}, H. and {Rohlf}, J. and {Ross}, A.~J. and {Rossi}, G. and {Ruhlmann-Kleider}, V. and {Samushia}, L. and {Sanchez}, E. and {Schlegel}, D. and {Schubnell}, M. and {Seo}, H. and {Sinigaglia}, F. and {Sprayberry}, D. and {Tan}, T. and {Tarl{\'e}}, G. and {Taylor}, P. and {Turner}, W. and {Vargas-Maga{\~n}a}, M. and {Verde}, L. and {Walther}, M. and {Weaver}, B.~A. and {Whitford}, A. and {Wolfson}, M. and {Zarrouk}, P. and {Zhao}, C. and {Zhou}, R. and {Zou}, H. and {DESI Collaboration}},
        title = "{Constraints on neutrino physics from DESI DR2 BAO and DR1 full shape}",
      journal = {\prd},
         year = 2025,
        month = oct,
       volume = {112},
       number = {8},
          eid = {083513},
        pages = {083513},
          doi = {10.1103/w9pk-xsk7},
archivePrefix = {arXiv},
       eprint = {2503.14744},
 primaryClass = {astro-ph.CO},
       adsurl = {https://ui.adsabs.harvard.edu/abs/2025PhRvD.112h3513E}
}

@ARTICLE{DESI2024_VII,
       author = {{Adame}, A.~G. and {Aguilar}, J. and {Ahlen}, S. and {Alam}, S. and {Alexander}, D.~M. and {Allende Prieto}, C. and {Alvarez}, M. and {Alves}, O. and {Anand}, A. and {Andrade}, U. and {Armengaud}, E. and {Avila}, S. and {Aviles}, A. and {Awan}, H. and {Bahr-Kalus}, B. and {Bailey}, S. and {Baltay}, C. and {Bault}, A. and {Behera}, J. and {BenZvi}, S. and {Beutler}, F. and {Bianchi}, D. and {Blake}, C. and {Blum}, R. and {Bonici}, M. and {Brieden}, S. and {Brodzeller}, A. and {Brooks}, D. and {Buckley-Geer}, E. and {Burtin}, E. and {Calderon}, R. and {Canning}, R. and {Carnero Rosell}, A. and {Cereskaite}, R. and {Cervantes-Cota}, J.~L. and {Chabanier}, S. and {Chaussidon}, E. and {Chaves-Montero}, J. and {Chebat}, D. and {Chen}, S. and {Chen}, X. and {Claybaugh}, T. and {Cole}, S. and {Cuceu}, A. and {Davis}, T.~M. and {Dawson}, K. and {de la Macorra}, A. and {de Mattia}, A. and {Deiosso}, N. and {Dey}, A. and {Dey}, B. and {Ding}, Z. and {Doel}, P. and {Edelstein}, J. and {Eftekharzadeh}, S. and {Eisenstein}, D.~J. and {Elbers}, W. and {Elliott}, A. and {Fagrelius}, P. and {Fanning}, K. and {Ferraro}, S. and {Ereza}, J. and {Findlay}, N. and {Flaugher}, B. and {Font-Ribera}, A. and {Forero-S{\'a}nchez}, D. and {Forero-Romero}, J.~E. and {Frenk}, C.~S. and {Garcia-Quintero}, C. and {Garrison}, L.~H. and {Gazta{\~n}aga}, E. and {Gil-Mar{\'\i}n}, H. and {Gontcho}, S. Gontcho A. and {Gonzalez-Morales}, A.~X. and {Gonzalez-Perez}, V. and {Gordon}, C. and {Green}, D. and {Gruen}, D. and {Gsponer}, R. and {Gutierrez}, G. and {Guy}, J. and {Hadzhiyska}, B. and {Hahn}, C. and {Hanif}, M.~M.~S. and {Herrera-Alcantar}, H.~K. and {Honscheid}, K. and {Howlett}, C. and {Huterer}, D. and {Ir{\v{s}}i{\v{c}}}, V. and {Ishak}, M. and {Joyce}, R. and {Juneau}, S. and {Kara{\c{c}}ayl{\i}}, N.~G. and {Kehoe}, R. and {Kent}, S. and {Kirkby}, D. and {Kong}, H. and {Koposov}, S.~E. and {Kremin}, A. and {Krolewski}, A. and {Lahav}, O. and {Lai}, Y. and {Lan}, T.-W. and {Landriau}, M. and {Lang}, D. and {Lasker}, J. and {Le Goff}, J.~M. and {Le Guillou}, L. and {Leauthaud}, A. and {Levi}, M.~E. and {Li}, T.~S. and {Lodha}, K. and {Magneville}, C. and {Manera}, M. and {Margala}, D. and {Martini}, P. and {Matthewson}, W. and {Maus}, M. and {McDonald}, P. and {Medina-Varela}, L. and {Meisner}, A. and {Mena-Fern{\'a}ndez}, J. and {Miquel}, R. and {Moon}, J. and {Moore}, S. and {Moustakas}, J. and {Mudur}, N. and {Mueller}, E. and {Mu{\~n}oz-Guti{\'e}rrez}, A. and {Myers}, A.~D. and {Nadathur}, S. and {Napolitano}, L. and {Neveux}, R. and {Newman}, J.~A. and {Nguyen}, N.~M. and {Nie}, J. and {Niz}, G. and {Noriega}, H.~E. and {Padmanabhan}, N. and {Paillas}, E. and {Palanque-Delabrouille}, N. and {Pan}, J. and {Penmetsa}, S. and {Percival}, W.~J. and {Pieri}, M.~M. and {Pinon}, M. and {Poppett}, C. and {Porredon}, A. and {Prada}, F. and {P{\'e}rez-Fern{\'a}ndez}, A. and {P{\'e}rez-R{\`a}fols}, I. and {Rabinowitz}, D. and {Raichoor}, A. and {Ram{\'\i}rez-P{\'e}rez}, C. and {Ramirez-Solano}, S. and {Rashkovetskyi}, M. and {Ravoux}, C. and {Rezaie}, M. and {Rich}, J. and {Rocher}, A. and {Rockosi}, C. and {Roe}, N.~A. and {Rosado-Marin}, A. and {Ross}, A.~J. and {Rossi}, G. and {Ruggeri}, R. and {Ruhlmann-Kleider}, V. and {Samushia}, L. and {Sanchez}, E. and {Saulder}, C. and {Schlafly}, E.~F. and {Schlegel}, D. and {Schubnell}, M. and {Seo}, H. and {Shafieloo}, A. and {Sharples}, R. and {Silber}, J. and {Slosar}, A. and {Smith}, A. and {Sprayberry}, D. and {Tan}, T. and {Tarl{\'e}}, G. and {Taylor}, P. and {Trusov}, S. and {Vaisakh}, R. and {Valcin}, D. and {Valdes}, F. and {Valogiannis}, G. and {Vargas-Maga{\~n}a}, M. and {Verde}, L. and {Walther}, M. and {Wang}, B. and {Wang}, M.~S. and {Weaver}, B.~A. and {Weaverdyck}, N. and {Wechsler}, R.~H. and {Weinberg}, D.~H. and {White}, M. and {Wilson}, M.~J. and {Yi}, L.},
        title = "{DESI 2024 VII: cosmological constraints from the full-shape modeling of clustering measurements}",
      journal = {\jcap},
         year = 2025,
        month = jul,
       volume = {2025},
       number = {7},
          eid = {028},
        pages = {028},
          doi = {10.1088/1475-7516/2025/07/028},
archivePrefix = {arXiv},
       eprint = {2411.12022},
 primaryClass = {astro-ph.CO},
       adsurl = {https://ui.adsabs.harvard.edu/abs/2025JCAP...07..028A}
}

@ARTICLE{DESI2024_V,
       author = {{Adame}, A.~G. and {Aguilar}, J. and {Ahlen}, S. and {Alam}, S. and {Alexander}, D.~M. and {Alvarez}, M. and {Alves}, O. and {Anand}, A. and {Andrade}, U. and {Armengaud}, E. and {Avila}, S. and {Aviles}, A. and {Awan}, H. and {Bailey}, S. and {Baltay}, C. and {Bault}, A. and {Behera}, J. and {BenZvi}, S. and {Beutler}, F. and {Bianchi}, D. and {Blake}, C. and {Blum}, R. and {Brieden}, S. and {Brodzeller}, A. and {Brooks}, D. and {Buckley-Geer}, E. and {Burtin}, E. and {Calderon}, R. and {Canning}, R. and {Carnero Rosell}, A. and {Cereskaite}, R. and {Cervantes-Cota}, J.~L. and {Chabanier}, S. and {Chaussidon}, E. and {Chaves-Montero}, J. and {Chen}, S. and {Chen}, X. and {Claybaugh}, T. and {Cole}, S. and {Cuceu}, A. and {Davis}, T.~M. and {Dawson}, K. and {de la Macorra}, A. and {de Mattia}, A. and {Deiosso}, N. and {Dey}, A. and {Dey}, B. and {Ding}, Z. and {Doel}, P. and {Edelstein}, J. and {Eftekharzadeh}, S. and {Eisenstein}, D.~J. and {Elliott}, A. and {Fagrelius}, P. and {Fanning}, K. and {Ferraro}, S. and {Ereza}, J. and {Findlay}, N. and {Flaugher}, B. and {Font-Ribera}, A. and {Forero-S{\'a}nchez}, D. and {Forero-Romero}, J.~E. and {Garcia-Quintero}, C. and {Garrison}, L.~H. and {Gazta{\~n}aga}, E. and {Gil-Mar{\'\i}n}, H. and {Gontcho}, S. Gontcho A. and {Gonzalez-Morales}, A.~X. and {Gonzalez-Perez}, V. and {Gordon}, C. and {Green}, D. and {Gruen}, D. and {Gsponer}, R. and {Gutierrez}, G. and {Guy}, J. and {Hadzhiyska}, B. and {Hahn}, C. and {Hanif}, M.~M.~S. and {Herrera-Alcantar}, H.~K. and {Honscheid}, K. and {Howlett}, C. and {Huterer}, D. and {Ir{\v{s}}i{\v{c}}}, V. and {Ishak}, M. and {Juneau}, S. and {Kara{\c{c}}ayl{\i}}, N.~G. and {Kehoe}, R. and {Kent}, S. and {Kirkby}, D. and {Kong}, H. and {Koposov}, S.~E. and {Kremin}, A. and {Krolewski}, A. and {Lai}, Y. and {Lan}, T.-W. and {Landriau}, M. and {Lang}, D. and {Lasker}, J. and {Le Goff}, J.~M. and {Le Guillou}, L. and {Leauthaud}, A. and {Levi}, M.~E. and {Li}, T.~S. and {Lodha}, K. and {Magneville}, C. and {Manera}, M. and {Margala}, D. and {Martini}, P. and {Maus}, M. and {McDonald}, P. and {Medina-Varela}, L. and {Meisner}, A. and {Mena-Fern{\'a}ndez}, J. and {Miquel}, R. and {Moon}, J. and {Moore}, S. and {Moustakas}, J. and {Mueller}, E. and {Mu{\~n}oz-Guti{\'e}rrez}, A. and {Myers}, A.~D. and {Nadathur}, S. and {Napolitano}, L. and {Neveux}, R. and {Newman}, J.~A. and {Nguyen}, N.~M. and {Nie}, J. and {Niz}, G. and {Noriega}, H.~E. and {Padmanabhan}, N. and {Paillas}, E. and {Palanque-Delabrouille}, N. and {Pan}, J. and {Penmetsa}, S. and {Percival}, W.~J. and {Pieri}, M.~M. and {Pinon}, M. and {Poppett}, C. and {Porredon}, A. and {Prada}, F. and {P{\'e}rez-Fern{\'a}ndez}, A. and {P{\'e}rez-R{\`a}fols}, I. and {Rabinowitz}, D. and {Raichoor}, A. and {Ram{\'\i}rez-P{\'e}rez}, C. and {Ramirez-Solano}, S. and {Rashkovetskyi}, M. and {Ravoux}, C. and {Rezaie}, M. and {Rich}, J. and {Rocher}, A. and {Rockosi}, C. and {Rodr{\'\i}guez-Mart{\'\i}nez}, F. and {Roe}, N.~A. and {Rosado-Marin}, A. and {Ross}, A.~J. and {Rossi}, G. and {Ruggeri}, R. and {Ruhlmann-Kleider}, V. and {Samushia}, L. and {Sanchez}, E. and {Saulder}, C. and {Schlafly}, E.~F. and {Schlegel}, D. and {Schubnell}, M. and {Seo}, H. and {Sharples}, R. and {Silber}, J. and {Slosar}, A. and {Smith}, A. and {Sprayberry}, D. and {Tan}, T. and {Tarl{\'e}}, G. and {Trusov}, S. and {Vaisakh}, R. and {Valcin}, D. and {Valdes}, F. and {Vargas-Maga{\~n}a}, M. and {Verde}, L. and {Walther}, M. and {Wang}, B. and {Wang}, M.~S. and {Weaver}, B.~A. and {Weaverdyck}, N. and {Wechsler}, R.~H. and {Weinberg}, D.~H. and {White}, M. and {Wilson}, M.~J. and {Yu}, J. and {Yu}, Y. and {Yuan}, S. and {Y{\`e}che}, C. and {Zaborowski}, E.~A. and {Zarrouk}, P. and {Zhang}, H. and {Zhao}, C. and {Zhao}, R. and {Zhou}, R. and {Zou}, H. and {The DESI collaboration}},
        title = "{DESI 2024 V: Full-Shape galaxy clustering from galaxies and quasars}",
      journal = {\jcap},
         year = 2025,
        month = sep,
       volume = {2025},
       number = {9},
          eid = {008},
        pages = {008},
          doi = {10.1088/1475-7516/2025/09/008},
archivePrefix = {arXiv},
       eprint = {2411.12021},
 primaryClass = {astro-ph.CO},
       adsurl = {https://ui.adsabs.harvard.edu/abs/2025JCAP...09..008A}
}

@ARTICLE{Lodhaetal2025,
       author = {{Lodha}, K. and {Calderon}, R. and {Matthewson}, W.~L. and {Shafieloo}, A. and {Ishak}, M. and {Pan}, J. and {Garcia-Quintero}, C. and {Huterer}, D. and {Valogiannis}, G. and {Ure{\~n}a-L{\'o}pez}, L.~A. and {Kamble}, N.~V. and {Parkinson}, D. and {Kim}, A.~G. and {Zhao}, G.~B. and {Cervantes-Cota}, J.~L. and {Rohlf}, J. and {Lozano-Rodr{\'\i}guez}, F. and {Rom{\'a}n-Herrera}, J.~O. and {Abdul-Karim}, M. and {Aguilar}, J. and {Ahlen}, S. and {Alves}, O. and {Andrade}, U. and {Armengaud}, E. and {Aviles}, A. and {Behera}, J. and {BenZvi}, S. and {Bianchi}, D. and {Brodzeller}, A. and {Brooks}, D. and {Burtin}, E. and {Canning}, R. and {Rosell}, A. Carnero and {Casas}, L. and {Castander}, F.~J. and {Charles}, M. and {Chaussidon}, E. and {Chaves-Montero}, J. and {Chebat}, D. and {Claybaugh}, T. and {Cole}, S. and {Cuceu}, A. and {Dawson}, K.~S. and {de la Macorra}, A. and {de Mattia}, A. and {Deiosso}, N. and {Demina}, R. and {Dey}, Arjun and {Dey}, Biprateep and {Ding}, Z. and {Doel}, P. and {Eisenstein}, D.~J. and {Elbers}, W. and {Ferraro}, S. and {Font-Ribera}, A. and {Forero-Romero}, J.~E. and {Garrison}, Lehman H. and {Gazta{\~n}aga}, E. and {Gil-Mar{\'\i}n}, H. and {Gontcho}, S. Gontcho A. and {Gonzalez-Morales}, A.~X. and {Gutierrez}, G. and {Guy}, J. and {Hahn}, C. and {Herbold}, M. and {Herrera-Alcantar}, H.~K. and {Honscheid}, K. and {Howlett}, C. and {Juneau}, S. and {Kehoe}, R. and {Kirkby}, D. and {Kisner}, T. and {Kremin}, A. and {Lahav}, O. and {Lamman}, C. and {Landriau}, M. and {Le Guillou}, L. and {Leauthaud}, A. and {Levi}, M.~E. and {Li}, Q. and {Magneville}, C. and {Manera}, M. and {Martini}, P. and {Meisner}, A. and {Mena-Fern{\'a}ndez}, J. and {Miquel}, R. and {Moustakas}, J. and {Santos}, D. Mu{\~n}oz and {Mu{\~n}oz-Guti{\'e}rrez}, A. and {Myers}, A.~D. and {Nadathur}, S. and {Niz}, G. and {Noriega}, H.~E. and {Paillas}, E. and {Palanque-Delabrouille}, N. and {Percival}, W.~J. and {Pieri}, Matthew M. and {Poppett}, C. and {Prada}, F. and {P{\'e}rez-Fern{\'a}ndez}, A. and {P{\'e}rez-R{\`a}fols}, I. and {Ram{\'\i}rez-P{\'e}rez}, C. and {Rashkovetskyi}, M. and {Ravoux}, C. and {Ross}, A.~J. and {Rossi}, G. and {Ruhlmann-Kleider}, V. and {Samushia}, L. and {Sanchez}, E. and {Schlegel}, D. and {Schubnell}, M. and {Seo}, H. and {Sinigaglia}, F. and {Sprayberry}, D. and {Tan}, T. and {Tarl{\'e}}, G. and {Taylor}, P. and {Turner}, W. and {Vargas-Maga{\~n}a}, M. and {Walther}, M. and {Weaver}, B.~A. and {Wolfson}, M. and {Y{\`e}che}, C. and {Zarrouk}, P. and {Zhou}, R. and {Zou}, H. and {DESI Collaboration}},
        title = "{Extended dark energy analysis using DESI DR2 BAO measurements}",
      journal = {\prd},
         year = 2025,
        month = oct,
       volume = {112},
       number = {8},
          eid = {083511},
        pages = {083511},
          doi = {10.1103/w4c6-1r5j},
archivePrefix = {arXiv},
       eprint = {2503.14743},
 primaryClass = {astro-ph.CO},
       adsurl = {https://ui.adsabs.harvard.edu/abs/2025PhRvD.112h3511L}
}

@ARTICLE{DESSNe,
       author = {{DES Collaboration} and {Abbott}, T.~M.~C. and {Acevedo}, M. and {Aguena}, M. and {Alarcon}, A. and {Allam}, S. and {Alves}, O. and {Amon}, A. and {Andrade-Oliveira}, F. and {Annis}, J. and {Armstrong}, P. and {Asorey}, J. and {Avila}, S. and {Bacon}, D. and {Bassett}, B.~A. and {Bechtol}, K. and {Bernardinelli}, P.~H. and {Bernstein}, G.~M. and {Bertin}, E. and {Blazek}, J. and {Bocquet}, S. and {Brooks}, D. and {Brout}, D. and {Buckley-Geer}, E. and {Burke}, D.~L. and {Camacho}, H. and {Camilleri}, R. and {Campos}, A. and {Carnero Rosell}, A. and {Carollo}, D. and {Carr}, A. and {Carretero}, J. and {Castander}, F.~J. and {Cawthon}, R. and {Chang}, C. and {Chen}, R. and {Choi}, A. and {Conselice}, C. and {Costanzi}, M. and {da Costa}, L.~N. and {Crocce}, M. and {Davis}, T.~M. and {DePoy}, D.~L. and {Desai}, S. and {Diehl}, H.~T. and {Dixon}, M. and {Dodelson}, S. and {Doel}, P. and {Doux}, C. and {Drlica-Wagner}, A. and {Elvin-Poole}, J. and {Everett}, S. and {Ferrero}, I. and {Fert{\'e}}, A. and {Flaugher}, B. and {Foley}, R.~J. and {Fosalba}, P. and {Friedel}, D. and {Frieman}, J. and {Frohmaier}, C. and {Galbany}, L. and {Garc{\'\i}a-Bellido}, J. and {Gatti}, M. and {Gaztanaga}, E. and {Giannini}, G. and {Glazebrook}, K. and {Graur}, O. and {Gruen}, D. and {Gruendl}, R.~A. and {Gutierrez}, G. and {Hartley}, W.~G. and {Herner}, K. and {Hinton}, S.~R. and {Hollowood}, D.~L. and {Honscheid}, K. and {Huterer}, D. and {Jain}, B. and {James}, D.~J. and {Jeffrey}, N. and {Kasai}, E. and {Kelsey}, L. and {Kent}, S. and {Kessler}, R. and {Kim}, A.~G. and {Kirshner}, R.~P. and {Kovacs}, E. and {Kuehn}, K. and {Lahav}, O. and {Lee}, J. and {Lee}, S. and {Lewis}, G.~F. and {Li}, T.~S. and {Lidman}, C. and {Lin}, H. and {Malik}, U. and {Marshall}, J.~L. and {Martini}, P. and {Mena-Fern{\'a}ndez}, J. and {Menanteau}, F. and {Miquel}, R. and {Mohr}, J.~J. and {Mould}, J. and {Muir}, J. and {M{\"o}ller}, A. and {Neilsen}, E. and {Nichol}, R.~C. and {Nugent}, P. and {Ogando}, R.~L.~C. and {Palmese}, A. and {Pan}, Y.-C. and {Paterno}, M. and {Percival}, W.~J. and {Pereira}, M.~E.~S. and {Pieres}, A. and {Malag{\'o}n}, A.~A. Plazas and {Popovic}, B. and {Porredon}, A. and {Prat}, J. and {Qu}, H. and {Raveri}, M. and {Rodr{\'\i}guez-Monroy}, M. and {Romer}, A.~K. and {Roodman}, A. and {Rose}, B. and {Sako}, M. and {Sanchez}, E. and {Sanchez Cid}, D. and {Schubnell}, M. and {Scolnic}, D. and {Sevilla-Noarbe}, I. and {Shah}, P. and {Smith}, J. Allyn. and {Smith}, M. and {Soares-Santos}, M. and {Suchyta}, E. and {Sullivan}, M. and {Suntzeff}, N. and {Swanson}, M.~E.~C. and {S{\'a}nchez}, B.~O. and {Tarle}, G. and {Taylor}, G. and {Thomas}, D. and {To}, C. and {Toy}, M. and {Troxel}, M.~A. and {Tucker}, B.~E. and {Tucker}, D.~L. and {Uddin}, S.~A. and {Vincenzi}, M. and {Walker}, A.~R. and {Weaverdyck}, N. and {Wechsler}, R.~H. and {Weller}, J. and {Wester}, W. and {Wiseman}, P. and {Yamamoto}, M. and {Yuan}, F. and {Zhang}, B. and {Zhang}, Y.},
        title = "{The Dark Energy Survey: Cosmology Results with {\ensuremath{\sim}}1500 New High-redshift Type Ia Supernovae Using the Full 5 yr Data Set}",
      journal = {\apjl},
         year = 2024,
        month = sep,
       volume = {973},
       number = {1},
          eid = {L14},
        pages = {L14},
          doi = {10.3847/2041-8213/ad6f9f},
archivePrefix = {arXiv},
       eprint = {2401.02929},
 primaryClass = {astro-ph.CO},
       adsurl = {https://ui.adsabs.harvard.edu/abs/2024ApJ...973L..14D}
}

@ARTICLE{Brout22,
       author = {{Brout}, Dillon and {Scolnic}, Dan and {Popovic}, Brodie and {Riess}, Adam G. and {Carr}, Anthony and {Zuntz}, Joe and {Kessler}, Rick and {Davis}, Tamara M. and {Hinton}, Samuel and {Jones}, David and {Kenworthy}, W. D'Arcy and {Peterson}, Erik R. and {Said}, Khaled and {Taylor}, Georgie and {Ali}, Noor and {Armstrong}, Patrick and {Charvu}, Pranav and {Dwomoh}, Arianna and {Meldorf}, Cole and {Palmese}, Antonella and {Qu}, Helen and {Rose}, Benjamin M. and {Sanchez}, Bruno and {Stubbs}, Christopher W. and {Vincenzi}, Maria and {Wood}, Charlotte M. and {Brown}, Peter J. and {Chen}, Rebecca and {Chambers}, Ken and {Coulter}, David A. and {Dai}, Mi and {Dimitriadis}, Georgios and {Filippenko}, Alexei V. and {Foley}, Ryan J. and {Jha}, Saurabh W. and {Kelsey}, Lisa and {Kirshner}, Robert P. and {M{\"o}ller}, Anais and {Muir}, Jessie and {Nadathur}, Seshadri and {Pan}, Yen-Chen and {Rest}, Armin and {Rojas-Bravo}, Cesar and {Sako}, Masao and {Siebert}, Matthew R. and {Smith}, Mat and {Stahl}, Benjamin E. and {Wiseman}, Phil},
        title = "{The Pantheon+ Analysis: Cosmological Constraints}",
      journal = {\apj},
         year = 2022,
        month = oct,
       volume = {938},
       number = {2},
          eid = {110},
        pages = {110},
          doi = {10.3847/1538-4357/ac8e04},
archivePrefix = {arXiv},
       eprint = {2202.04077},
 primaryClass = {astro-ph.CO},
       adsurl = {https://ui.adsabs.harvard.edu/abs/2022ApJ...938..110B}
}

@ARTICLE{Scolnic22,
       author = {{Scolnic}, Dan and {Brout}, Dillon and {Carr}, Anthony and {Riess}, Adam G. and {Davis}, Tamara M. and {Dwomoh}, Arianna and {Jones}, David O. and {Ali}, Noor and {Charvu}, Pranav and {Chen}, Rebecca and {Peterson}, Erik R. and {Popovic}, Brodie and {Rose}, Benjamin M. and {Wood}, Charlotte M. and {Brown}, Peter J. and {Chambers}, Ken and {Coulter}, David A. and {Dettman}, Kyle G. and {Dimitriadis}, Georgios and {Filippenko}, Alexei V. and {Foley}, Ryan J. and {Jha}, Saurabh W. and {Kilpatrick}, Charles D. and {Kirshner}, Robert P. and {Pan}, Yen-Chen and {Rest}, Armin and {Rojas-Bravo}, Cesar and {Siebert}, Matthew R. and {Stahl}, Benjamin E. and {Zheng}, WeiKang},
        title = "{The Pantheon+ Analysis: The Full Data Set and Light-curve Release}",
      journal = {\apj},
         year = 2022,
        month = oct,
       volume = {938},
       number = {2},
          eid = {113},
        pages = {113},
          doi = {10.3847/1538-4357/ac8b7a},
archivePrefix = {arXiv},
       eprint = {2112.03863},
 primaryClass = {astro-ph.CO},
       adsurl = {https://ui.adsabs.harvard.edu/abs/2022ApJ...938..113S}
}

@ARTICLE{DESI_DR2a,
       author = {{Abdul Karim}, M. and {Aguilar}, J. and {Ahlen}, S. and {Allende Prieto}, C. and {Alves}, O. and {Anand}, A. and {Andrade}, U. and {Armengaud}, E. and {Aviles}, A. and {Bailey}, S. and {Bault}, A. and {Behera}, J. and {BenZvi}, S. and {Bianchi}, D. and {Blake}, C. and {Brodzeller}, A. and {Brooks}, D. and {Buckley-Geer}, E. and {Burtin}, E. and {Calderon}, R. and {Canning}, R. and {Carnero Rosell}, A. and {Carrilho}, P. and {Casas}, L. and {Castander}, F.~J. and {Cereskaite}, R. and {Charles}, M. and {Chaussidon}, E. and {Chaves-Montero}, J. and {Chebat}, D. and {Claybaugh}, T. and {Cole}, S. and {Cooper}, A.~P. and {Cuceu}, A. and {Dawson}, K.~S. and {de Belsunce}, R. and {de la Macorra}, A. and {de Mattia}, A. and {Deiosso}, N. and {Della Costa}, J. and {Dey}, A. and {Dey}, B. and {Ding}, Z. and {Doel}, P. and {Edelstein}, J. and {Eisenstein}, D.~J. and {Elbers}, W. and {Fagrelius}, P. and {Fanning}, K. and {Ferraro}, S. and {Font-Ribera}, A. and {Forero-Romero}, J.~E. and {Garcia-Quintero}, C. and {Garrison}, L.~H. and {Gazta{\~n}aga}, E. and {Gil-Mar{\'\i}n}, H. and {Gontcho A Gontcho}, S. and {Gonzalez-Morales}, A.~X. and {Gordon}, C. and {Green}, D. and {Gutierrez}, G. and {Guy}, J. and {Hahn}, C. and {Herbold}, M. and {Herrera-Alcantar}, H.~K. and {Ho}, M. and {Ho}, M.-F. and {Honscheid}, K. and {Howlett}, C. and {Huterer}, D. and {Ishak}, M. and {Juneau}, S. and {Kara{\c{c}}ayl{\i}}, N.~G. and {Kehoe}, R. and {Kent}, S. and {Kirkby}, D. and {Kisner}, T. and {Kitaura}, F.-S. and {Koposov}, S.~E. and {Kremin}, A. and {Lahav}, O. and {Lamman}, C. and {Landriau}, M. and {Lang}, D. and {Lasker}, J. and {Le Goff}, J.~M. and {Le Guillou}, L. and {Leauthaud}, A. and {Levi}, M.~E. and {Li}, Q. and {Li}, T.~S. and {Lodha}, K. and {Lokken}, M. and {Magneville}, C. and {Manera}, M. and {Martini}, P. and {Matthewson}, W.~L. and {McDonald}, P. and {Meisner}, A. and {Mena-Fern{\'a}ndez}, J. and {Miquel}, R. and {Moustakas}, J. and {Mu{\~n}oz-Guti{\'e}rrez}, A. and {Mu{\~n}oz-Santos}, D. and {Myers}, A.~D. and {Newman}, J.~A. and {Niz}, G. and {Noriega}, H.~E. and {Paillas}, E. and {Palanque-Delabrouille}, N. and {Pan}, J. and {Percival}, W.~J. and {P{\'e}rez-R{\`a}fols}, I. and {Pieri}, M.~M. and {Poppett}, C. and {Prada}, F. and {Rabinowitz}, D. and {Raichoor}, A. and {Ram{\'\i}rez-P{\'e}rez}, C. and {Rashkovetskyi}, M. and {Ravoux}, C. and {Rich}, J. and {Rockosi}, C. and {Ross}, A.~J. and {Rossi}, G. and {Ruhlmann-Kleider}, V. and {Sanchez}, E. and {Sanders}, N. and {Satyavolu}, S. and {Schlegel}, D. and {Schubnell}, M. and {Seo}, H. and {Shafieloo}, A. and {Sharples}, R. and {Silber}, J. and {Sinigaglia}, F. and {Sprayberry}, D. and {Tan}, T. and {Tarl{\'e}}, G. and {Taylor}, P. and {Turner}, W. and {Valdes}, F. and {Vargas-Maga{\~n}a}, M. and {Walther}, M. and {Weaver}, B.~A. and {Wolfson}, M. and {Y{\`e}che}, C. and {Zarrouk}, P. and {Zhou}, R. and {Zou}, H. and {DESI Collaboration}},
        title = "{DESI DR2 results. I. Baryon acoustic oscillations from the Lyman alpha forest}",
      journal = {\prd},
         year = 2025,
        month = oct,
       volume = {112},
       number = {8},
          eid = {083514},
        pages = {083514},
          doi = {10.1103/2wwn-xjm5},
archivePrefix = {arXiv},
       eprint = {2503.14739},
 primaryClass = {astro-ph.CO},
       adsurl = {https://ui.adsabs.harvard.edu/abs/2025PhRvD.112h3514A}
}

@ARTICLE{wfirst,
       author = {{Spergel}, D. and {Gehrels}, N. and {Breckinridge}, J. and {Donahue}, M. and {Dressler}, A. and {Gaudi}, B.~S. and {Greene}, T. and {Guyon}, O. and {Hirata}, C. and {Kalirai}, J. and {Kasdin}, N.~J. and {Moos}, W. and {Perlmutter}, S. and {Postman}, M. and {Rauscher}, B. and {Rhodes}, J. and {Wang}, Y. and {Weinberg}, D. and {Centrella}, J. and {Traub}, W. and {Baltay}, C. and {Colbert}, J. and {Bennett}, D. and {Kiessling}, A. and {Macintosh}, B. and {Merten}, J. and {Mortonson}, M. and {Penny}, M. and {Rozo}, E. and {Savransky}, D. and {Stapelfeldt}, K. and {Zu}, Y. and {Baker}, C. and {Cheng}, E. and {Content}, D. and {Dooley}, J. and {Foote}, M. and {Goullioud}, R. and {Grady}, K. and {Jackson}, C. and {Kruk}, J. and {Levine}, M. and {Melton}, M. and {Peddie}, C. and {Ruffa}, J. and {Shaklan}, S.},
        title = "{Wide-Field InfraRed Survey Telescope-Astrophysics Focused Telescope Assets WFIRST-AFTA Final Report}",
      journal = {arXiv e-prints},
         year = 2013,
        month = may,
          eid = {arXiv:1305.5422},
        pages = {arXiv:1305.5422},
          doi = {10.48550/arXiv.1305.5422},
archivePrefix = {arXiv},
       eprint = {1305.5422},
 primaryClass = {astro-ph.IM},
       adsurl = {https://ui.adsabs.harvard.edu/abs/2013arXiv1305.5422S}
}

@ARTICLE{Vera_rubin,
       author = {{Ivezi{\'c}}, {\v{Z}}eljko and {Kahn}, Steven M. and {Tyson}, J. Anthony and {Abel}, Bob and {Acosta}, Emily and {Allsman}, Robyn and {Alonso}, David and {AlSayyad}, Yusra and {Anderson}, Scott F. and {Andrew}, John and {Angel}, James Roger P. and {Angeli}, George Z. and {Ansari}, Reza and {Antilogus}, Pierre and {Araujo}, Constanza and {Armstrong}, Robert and {Arndt}, Kirk T. and {Astier}, Pierre and {Aubourg}, {\'E}ric and {Auza}, Nicole and {Axelrod}, Tim S. and {Bard}, Deborah J. and {Barr}, Jeff D. and {Barrau}, Aurelian and {Bartlett}, James G. and {Bauer}, Amanda E. and {Bauman}, Brian J. and {Baumont}, Sylvain and {Bechtol}, Ellen and {Bechtol}, Keith and {Becker}, Andrew C. and {Becla}, Jacek and {Beldica}, Cristina and {Bellavia}, Steve and {Bianco}, Federica B. and {Biswas}, Rahul and {Blanc}, Guillaume and {Blazek}, Jonathan and {Blandford}, Roger D. and {Bloom}, Josh S. and {Bogart}, Joanne and {Bond}, Tim W. and {Booth}, Michael T. and {Borgland}, Anders W. and {Borne}, Kirk and {Bosch}, James F. and {Boutigny}, Dominique and {Brackett}, Craig A. and {Bradshaw}, Andrew and {Brandt}, William Nielsen and {Brown}, Michael E. and {Bullock}, James S. and {Burchat}, Patricia and {Burke}, David L. and {Cagnoli}, Gianpietro and {Calabrese}, Daniel and {Callahan}, Shawn and {Callen}, Alice L. and {Carlin}, Jeffrey L. and {Carlson}, Erin L. and {Chandrasekharan}, Srinivasan and {Charles-Emerson}, Glenaver and {Chesley}, Steve and {Cheu}, Elliott C. and {Chiang}, Hsin-Fang and {Chiang}, James and {Chirino}, Carol and {Chow}, Derek and {Ciardi}, David R. and {Claver}, Charles F. and {Cohen-Tanugi}, Johann and {Cockrum}, Joseph J. and {Coles}, Rebecca and {Connolly}, Andrew J. and {Cook}, Kem H. and {Cooray}, Asantha and {Covey}, Kevin R. and {Cribbs}, Chris and {Cui}, Wei and {Cutri}, Roc and {Daly}, Philip N. and {Daniel}, Scott F. and {Daruich}, Felipe and {Daubard}, Guillaume and {Daues}, Greg and {Dawson}, William and {Delgado}, Francisco and {Dellapenna}, Alfred and {de Peyster}, Robert and {de Val-Borro}, Miguel and {Digel}, Seth W. and {Doherty}, Peter and {Dubois}, Richard and {Dubois-Felsmann}, Gregory P. and {Durech}, Josef and {Economou}, Frossie and {Eifler}, Tim and {Eracleous}, Michael and {Emmons}, Benjamin L. and {Fausti Neto}, Angelo and {Ferguson}, Henry and {Figueroa}, Enrique and {Fisher-Levine}, Merlin and {Focke}, Warren and {Foss}, Michael D. and {Frank}, James and {Freemon}, Michael D. and {Gangler}, Emmanuel and {Gawiser}, Eric and {Geary}, John C. and {Gee}, Perry and {Geha}, Marla and {Gessner}, Charles J.~B. and {Gibson}, Robert R. and {Gilmore}, D. Kirk and {Glanzman}, Thomas and {Glick}, William and {Goldina}, Tatiana and {Goldstein}, Daniel A. and {Goodenow}, Iain and {Graham}, Melissa L. and {Gressler}, William J. and {Gris}, Philippe and {Guy}, Leanne P. and {Guyonnet}, Augustin and {Haller}, Gunther and {Harris}, Ron and {Hascall}, Patrick A. and {Haupt}, Justine and {Hernandez}, Fabio and {Herrmann}, Sven and {Hileman}, Edward and {Hoblitt}, Joshua and {Hodgson}, John A. and {Hogan}, Craig and {Howard}, James D. and {Huang}, Dajun and {Huffer}, Michael E. and {Ingraham}, Patrick and {Innes}, Walter R. and {Jacoby}, Suzanne H. and {Jain}, Bhuvnesh and {Jammes}, Fabrice and {Jee}, M. James and {Jenness}, Tim and {Jernigan}, Garrett and {Jevremovi{\'c}}, Darko and {Johns}, Kenneth and {Johnson}, Anthony S. and {Johnson}, Margaret W.~G. and {Jones}, R. Lynne and {Juramy-Gilles}, Claire and {Juri{\'c}}, Mario and {Kalirai}, Jason S. and {Kallivayalil}, Nitya J. and {Kalmbach}, Bryce and {Kantor}, Jeffrey P. and {Karst}, Pierre and {Kasliwal}, Mansi M. and {Kelly}, Heather and {Kessler}, Richard and {Kinnison}, Veronica and {Kirkby}, David and {Knox}, Lloyd and {Kotov}, Ivan V. and {Krabbendam}, Victor L. and {Krughoff}, K. Simon and {Kub{\'a}nek}, Petr and {Kuczewski}, John and {Kulkarni}, Shri and {Ku}, John and {Kurita}, Nadine R. and {Lage}, Craig S. and {Lambert}, Ron and {Lange}, Travis and {Langton}, J. Brian and {Le Guillou}, Laurent and {Levine}, Deborah and {Liang}, Ming and {Lim}, Kian-Tat and {Lintott}, Chris J. and {Long}, Kevin E. and {Lopez}, Margaux and {Lotz}, Paul J. and {Lupton}, Robert H. and {Lust}, Nate B. and {MacArthur}, Lauren A. and {Mahabal}, Ashish and {Mandelbaum}, Rachel and {Markiewicz}, Thomas W. and {Marsh}, Darren S. and {Marshall}, Philip J. and {Marshall}, Stuart and {May}, Morgan and {McKercher}, Robert and {McQueen}, Michelle and {Meyers}, Joshua and {Migliore}, Myriam and {Miller}, Michelle and {Mills}, David J.},
        title = "{LSST: From Science Drivers to Reference Design and Anticipated Data Products}",
      journal = {\apj},
         year = 2019,
        month = mar,
       volume = {873},
       number = {2},
          eid = {111},
        pages = {111},
          doi = {10.3847/1538-4357/ab042c},
archivePrefix = {arXiv},
       eprint = {0805.2366},
 primaryClass = {astro-ph},
       adsurl = {https://ui.adsabs.harvard.edu/abs/2019ApJ...873..111I}
}

@ARTICLE{Euclid,
       author = {{Euclid Collaboration} and {Mellier}, Y. and {Abdurro'uf} and {Acevedo Barroso}, J.~A. and {Ach{\'u}carro}, A. and {Adamek}, J. and {Adam}, R. and {Addison}, G.~E. and {Aghanim}, N. and {Aguena}, M. and {Ajani}, V. and {Akrami}, Y. and {Al-Bahlawan}, A. and {Alavi}, A. and {Albuquerque}, I.~S. and {Alestas}, G. and {Alguero}, G. and {Allaoui}, A. and {Allen}, S.~W. and {Allevato}, V. and {Alonso-Tetilla}, A.~V. and {Altieri}, B. and {Alvarez-Candal}, A. and {Alvi}, S. and {Amara}, A. and {Amendola}, L. and {Amiaux}, J. and {Andika}, I.~T. and {Andreon}, S. and {Andrews}, A. and {Angora}, G. and {Angulo}, R.~E. and {Annibali}, F. and {Anselmi}, A. and {Anselmi}, S. and {Arcari}, S. and {Archidiacono}, M. and {Aric{\`o}}, G. and {Arnaud}, M. and {Arnouts}, S. and {Asgari}, M. and {Asorey}, J. and {Atayde}, L. and {Atek}, H. and {Atrio-Barandela}, F. and {Aubert}, M. and {Aubourg}, E. and {Auphan}, T. and {Auricchio}, N. and {Aussel}, B. and {Aussel}, H. and {Avelino}, P.~P. and {Avgoustidis}, A. and {Avila}, S. and {Awan}, S. and {Azzollini}, R. and {Baccigalupi}, C. and {Bachelet}, E. and {Bacon}, D. and {Baes}, M. and {Bagley}, M.~B. and {Bahr-Kalus}, B. and {Balaguera-Antolinez}, A. and {Balbinot}, E. and {Balcells}, M. and {Baldi}, M. and {Baldry}, I. and {Balestra}, A. and {Ballardini}, M. and {Ballester}, O. and {Balogh}, M. and {Ba{\~n}ados}, E. and {Barbier}, R. and {Bardelli}, S. and {Baron}, M. and {Barreiro}, T. and {Barrena}, R. and {Barriere}, J.-C. and {Barros}, B.~J. and {Barthelemy}, A. and {Bartolo}, N. and {Basset}, A. and {Battaglia}, P. and {Battisti}, A.~J. and {Baugh}, C.~M. and {Baumont}, L. and {Bazzanini}, L. and {Beaulieu}, J.-P. and {Beckmann}, V. and {Belikov}, A.~N. and {Bel}, J. and {Bellagamba}, F. and {Bella}, M. and {Bellini}, E. and {Benabed}, K. and {Bender}, R. and {Benevento}, G. and {Bennett}, C.~L. and {Benson}, K. and {Bergamini}, P. and {Bermejo-Climent}, J.~R. and {Bernardeau}, F. and {Bertacca}, D. and {Berthe}, M. and {Berthier}, J. and {Bethermin}, M. and {Beutler}, F. and {Bevillon}, C. and {Bhargava}, S. and {Bhatawdekar}, R. and {Bianchi}, D. and {Bisigello}, L. and {Biviano}, A. and {Blake}, R.~P. and {Blanchard}, A. and {Blazek}, J. and {Blot}, L. and {Bosco}, A. and {Bodendorf}, C. and {Boenke}, T. and {B{\"o}hringer}, H. and {Boldrini}, P. and {Bolzonella}, M. and {Bonchi}, A. and {Bonici}, M. and {Bonino}, D. and {Bonino}, L. and {Bonvin}, C. and {Bon}, W. and {Booth}, J.~T. and {Borgani}, S. and {Borlaff}, A.~S. and {Borsato}, E. and {Bose}, B. and {Botticella}, M.~T. and {Boucaud}, A. and {Bouche}, F. and {Boucher}, J.~S. and {Boutigny}, D. and {Bouvard}, T. and {Bouwens}, R. and {Bouy}, H. and {Bowler}, R.~A.~A. and {Bozza}, V. and {Bozzo}, E. and {Branchini}, E. and {Brando}, G. and {Brau-Nogue}, S. and {Brekke}, P. and {Bremer}, M.~N. and {Brescia}, M. and {Breton}, M.-A. and {Brinchmann}, J. and {Brinckmann}, T. and {Brockley-Blatt}, C. and {Brodwin}, M. and {Brouard}, L. and {Brown}, M.~L. and {Bruton}, S. and {Bucko}, J. and {Buddelmeijer}, H. and {Buenadicha}, G. and {Buitrago}, F. and {Burger}, P. and {Burigana}, C. and {Busillo}, V. and {Busonero}, D. and {Cabanac}, R. and {Cabayol-Garcia}, L. and {Cagliari}, M.~S. and {Caillat}, A. and {Caillat}, L. and {Calabrese}, M. and {Calabro}, A. and {Calderone}, G. and {Calura}, F. and {Camacho Quevedo}, B. and {Camera}, S. and {Campos}, L. and {Ca{\~n}as-Herrera}, G. and {Candini}, G.~P. and {Cantiello}, M. and {Capobianco}, V. and {Cappellaro}, E. and {Cappelluti}, N. and {Cappi}, A. and {Caputi}, K.~I. and {Cara}, C. and {Carbone}, C. and {Cardone}, V.~F. and {Carella}, E. and {Carlberg}, R.~G. and {Carle}, M. and {Carminati}, L. and {Caro}, F. and {Carrasco}, J.~M. and {Carretero}, J. and {Carrilho}, P. and {Carron Duque}, J. and {Carry}, B.},
        title = "{Euclid: I. Overview of the Euclid mission}",
      journal = {\aap},
         year = 2025,
        month = may,
       volume = {697},
          eid = {A1},
        pages = {A1},
          doi = {10.1051/0004-6361/202450810},
archivePrefix = {arXiv},
       eprint = {2405.13491},
 primaryClass = {astro-ph.CO},
       adsurl = {https://ui.adsabs.harvard.edu/abs/2025A&A...697A...1E}
}

@ARTICLE{DESI,
       author = {{DESI Collaboration} and {Abareshi}, B. and {Aguilar}, J. and {Ahlen}, S. and {Alam}, Shadab and {Alexander}, David M. and {Alfarsy}, R. and {Allen}, L. and {Allende Prieto}, C. and {Alves}, O. and {Ameel}, J. and {Armengaud}, E. and {Asorey}, J. and {Aviles}, Alejandro and {Bailey}, S. and {Balaguera-Antol{\'\i}nez}, A. and {Ballester}, O. and {Baltay}, C. and {Bault}, A. and {Beltran}, S.~F. and {Benavides}, B. and {BenZvi}, S. and {Berti}, A. and {Besuner}, R. and {Beutler}, Florian and {Bianchi}, D. and {Blake}, C. and {Blanc}, P. and {Blum}, R. and {Bolton}, A. and {Bose}, S. and {Bramall}, D. and {Brieden}, S. and {Brodzeller}, A. and {Brooks}, D. and {Brownewell}, C. and {Buckley-Geer}, E. and {Cahn}, R.~N. and {Cai}, Z. and {Canning}, R. and {Capasso}, R. and {Carnero Rosell}, A. and {Carton}, P. and {Casas}, R. and {Castander}, F.~J. and {Cervantes-Cota}, J.~L. and {Chabanier}, S. and {Chaussidon}, E. and {Chuang}, C. and {Circosta}, C. and {Cole}, S. and {Cooper}, A.~P. and {da Costa}, L. and {Cousinou}, M.-C. and {Cuceu}, A. and {Davis}, T.~M. and {Dawson}, K. and {de la Cruz-Noriega}, R. and {de la Macorra}, A. and {de Mattia}, A. and {Della Costa}, J. and {Demmer}, P. and {Derwent}, M. and {Dey}, A. and {Dey}, B. and {Dhungana}, G. and {Ding}, Z. and {Dobson}, C. and {Doel}, P. and {Donald-McCann}, J. and {Donaldson}, J. and {Douglass}, K. and {Duan}, Y. and {Dunlop}, P. and {Edelstein}, J. and {Eftekharzadeh}, S. and {Eisenstein}, D.~J. and {Enriquez-Vargas}, M. and {Escoffier}, S. and {Evatt}, M. and {Fagrelius}, P. and {Fan}, X. and {Fanning}, K. and {Fawcett}, V.~A. and {Ferraro}, S. and {Ereza}, J. and {Flaugher}, B. and {Font-Ribera}, A. and {Forero-Romero}, J.~E. and {Frenk}, C.~S. and {Fromenteau}, S. and {G{\"a}nsicke}, B.~T. and {Garcia-Quintero}, C. and {Garrison}, L. and {Gazta{\~n}aga}, E. and {Gerardi}, F. and {Gil-Mar{\'\i}n}, H. and {Gontcho A Gontcho}, S. and {Gonzalez-Morales}, Alma X. and {Gonzalez-de-Rivera}, G. and {Gonzalez-Perez}, V. and {Gordon}, C. and {Graur}, O. and {Green}, D. and {Grove}, C. and {Gruen}, D. and {Gutierrez}, G. and {Guy}, J. and {Hahn}, C. and {Harris}, S. and {Herrera}, D. and {Herrera-Alcantar}, Hiram K. and {Honscheid}, K. and {Howlett}, C. and {Huterer}, D. and {Ir{\v{s}}i{\v{c}}}, V. and {Ishak}, M. and {Jelinsky}, P. and {Jiang}, L. and {Jimenez}, J. and {Jing}, Y.~P. and {Joyce}, R. and {Jullo}, E. and {Juneau}, S. and {Kara{\c{c}}ayl{\i}}, N.~G. and {Karamanis}, M. and {Karcher}, A. and {Karim}, T. and {Kehoe}, R. and {Kent}, S. and {Kirkby}, D. and {Kisner}, T. and {Kitaura}, F. and {Koposov}, S.~E. and {Kov{\'a}cs}, A. and {Kremin}, A. and {Krolewski}, Alex and {L'Huillier}, B. and {Lahav}, O. and {Lambert}, A. and {Lamman}, C. and {Lan}, Ting-Wen and {Landriau}, M. and {Lane}, S. and {Lang}, D. and {Lange}, J.~U. and {Lasker}, J. and {Le Guillou}, L. and {Leauthaud}, A. and {Le Van Suu}, A. and {Levi}, Michael E. and {Li}, T.~S. and {Magneville}, C. and {Manera}, M. and {Manser}, Christopher J. and {Marshall}, B. and {Martini}, Paul and {McCollam}, W. and {McDonald}, P. and {Meisner}, Aaron M. and {Mena-Fern{\'a}ndez}, J. and {Meneses-Rizo}, J. and {Mezcua}, M. and {Miller}, T. and {Miquel}, R. and {Montero-Camacho}, P. and {Moon}, J. and {Moustakas}, J. and {Mueller}, E. and {Mu{\~n}oz-Guti{\'e}rrez}, Andrea and {Myers}, Adam D. and {Nadathur}, S. and {Najita}, J. and {Napolitano}, L. and {Neilsen}, E. and {Newman}, Jeffrey A. and {Nie}, J.~D. and {Ning}, Y. and {Niz}, G. and {Norberg}, P. and {Noriega}, Hern{\'a}n E. and {O'Brien}, T. and {Obuljen}, A. and {Palanque-Delabrouille}, N. and {Palmese}, A. and {Zhiwei}, P. and {Pappalardo}, D. and {PENG}, X. and {Percival}, W.~J. and {Perruchot}, S. and {Pogge}, R. and {Poppett}, C. and {Porredon}, A. and {Prada}, F. and {Prochaska}, J. and {Pucha}, R. and {P{\'e}rez-Fern{\'a}ndez}, A. and {P{\'e}rez-R{\`a}fols}, I. and {Rabinowitz}, D. and {Raichoor}, A.},
        title = "{Overview of the Instrumentation for the Dark Energy Spectroscopic Instrument}",
      journal = {\aj},
         year = 2022,
        month = nov,
       volume = {164},
       number = {5},
          eid = {207},
        pages = {207},
          doi = {10.3847/1538-3881/ac882b},
archivePrefix = {arXiv},
       eprint = {2205.10939},
 primaryClass = {astro-ph.IM},
       adsurl = {https://ui.adsabs.harvard.edu/abs/2022AJ....164..207D}
}

@ARTICLE{Bashinsky_Seljak04,
       author = {{Bashinsky}, Sergei and {Seljak}, Uro{\v{s}}},
        title = "{Signatures of relativistic neutrinos in CMB anisotropy and matter clustering}",
      journal = {\prd},
         year = 2004,
        month = apr,
       volume = {69},
       number = {8},
          eid = {083002},
        pages = {083002},
          doi = {10.1103/PhysRevD.69.083002},
archivePrefix = {arXiv},
       eprint = {astro-ph/0310198},
 primaryClass = {astro-ph},
       adsurl = {https://ui.adsabs.harvard.edu/abs/2004PhRvD..69h3002B}
}

@ARTICLE{SO,
       author = {{Ade}, Peter and {Aguirre}, James and {Ahmed}, Zeeshan and {Aiola}, Simone and {Ali}, Aamir and {Alonso}, David and {Alvarez}, Marcelo A. and {Arnold}, Kam and {Ashton}, Peter and {Austermann}, Jason and {Awan}, Humna and {Baccigalupi}, Carlo and {Baildon}, Taylor and {Barron}, Darcy and {Battaglia}, Nick and {Battye}, Richard and {Baxter}, Eric and {Bazarko}, Andrew and {Beall}, James A. and {Bean}, Rachel and {Beck}, Dominic and {Beckman}, Shawn and {Beringue}, Benjamin and {Bianchini}, Federico and {Boada}, Steven and {Boettger}, David and {Bond}, J. Richard and {Borrill}, Julian and {Brown}, Michael L. and {Bruno}, Sarah Marie and {Bryan}, Sean and {Calabrese}, Erminia and {Calafut}, Victoria and {Calisse}, Paolo and {Carron}, Julien and {Challinor}, Anthony and {Chesmore}, Grace and {Chinone}, Yuji and {Chluba}, Jens and {Cho}, Hsiao-Mei Sherry and {Choi}, Steve and {Coppi}, Gabriele and {Cothard}, Nicholas F. and {Coughlin}, Kevin and {Crichton}, Devin and {Crowley}, Kevin D. and {Crowley}, Kevin T. and {Cukierman}, Ari and {D'Ewart}, John M. and {D{\"u}nner}, Rolando and {de Haan}, Tijmen and {Devlin}, Mark and {Dicker}, Simon and {Didier}, Joy and {Dobbs}, Matt and {Dober}, Bradley and {Duell}, Cody J. and {Duff}, Shannon and {Duivenvoorden}, Adri and {Dunkley}, Jo and {Dusatko}, John and {Errard}, Josquin and {Fabbian}, Giulio and {Feeney}, Stephen and {Ferraro}, Simone and {Flux{\`a}}, Pedro and {Freese}, Katherine and {Frisch}, Josef C. and {Frolov}, Andrei and {Fuller}, George and {Fuzia}, Brittany and {Galitzki}, Nicholas and {Gallardo}, Patricio A. and {Tomas Galvez Ghersi}, Jose and {Gao}, Jiansong and {Gawiser}, Eric and {Gerbino}, Martina and {Gluscevic}, Vera and {Goeckner-Wald}, Neil and {Golec}, Joseph and {Gordon}, Sam and {Gralla}, Megan and {Green}, Daniel and {Grigorian}, Arpi and {Groh}, John and {Groppi}, Chris and {Guan}, Yilun and {Gudmundsson}, Jon E. and {Han}, Dongwon and {Hargrave}, Peter and {Hasegawa}, Masaya and {Hasselfield}, Matthew and {Hattori}, Makoto and {Haynes}, Victor and {Hazumi}, Masashi and {He}, Yizhou and {Healy}, Erin and {Henderson}, Shawn W. and {Hervias-Caimapo}, Carlos and {Hill}, Charles A. and {Hill}, J. Colin and {Hilton}, Gene and {Hilton}, Matt and {Hincks}, Adam D. and {Hinshaw}, Gary and {Hlo{\v{z}}ek}, Ren{\'e}e and {Ho}, Shirley and {Ho}, Shuay-Pwu Patty and {Howe}, Logan and {Huang}, Zhiqi and {Hubmayr}, Johannes and {Huffenberger}, Kevin and {Hughes}, John P. and {Ijjas}, Anna and {Ikape}, Margaret and {Irwin}, Kent and {Jaffe}, Andrew H. and {Jain}, Bhuvnesh and {Jeong}, Oliver and {Kaneko}, Daisuke and {Karpel}, Ethan D. and {Katayama}, Nobuhiko and {Keating}, Brian and {Kernasovskiy}, Sarah S. and {Keskitalo}, Reijo and {Kisner}, Theodore and {Kiuchi}, Kenji and {Klein}, Jeff and {Knowles}, Kenda and {Koopman}, Brian and {Kosowsky}, Arthur and {Krachmalnicoff}, Nicoletta and {Kuenstner}, Stephen E. and {Kuo}, Chao-Lin and {Kusaka}, Akito and {Lashner}, Jacob and {Lee}, Adrian and {Lee}, Eunseong and {Leon}, David and {Leung}, Jason S.-Y. and {Lewis}, Antony and {Li}, Yaqiong and {Li}, Zack and {Limon}, Michele and {Linder}, Eric and {Lopez-Caraballo}, Carlos and {Louis}, Thibaut and {Lowry}, Lindsay and {Lungu}, Marius and {Madhavacheril}, Mathew and {Mak}, Daisy and {Maldonado}, Felipe and {Mani}, Hamdi and {Mates}, Ben and {Matsuda}, Frederick and {Maurin}, Lo{\"\i}c and {Mauskopf}, Phil and {May}, Andrew and {McCallum}, Nialh and {McKenney}, Chris and {McMahon}, Jeff and {Meerburg}, P. Daniel and {Meyers}, Joel and {Miller}, Amber and {Mirmelstein}, Mark and {Moodley}, Kavilan and {Munchmeyer}, Moritz and {Munson}, Charles and {Naess}, Sigurd and {Nati}, Federico and {Navaroli}, Martin and {Newburgh}, Laura and {Nguyen}, Ho Nam and {Niemack}, Michael and {Nishino}, Haruki and {Orlowski-Scherer}, John and {Page}, Lyman and {Partridge}, Bruce and {Peloton}, Julien and {Perrotta}, Francesca and {Piccirillo}, Lucio and {Pisano}, Giampaolo and {Poletti}, Davide and {Puddu}, Roberto and {Puglisi}, Giuseppe and {Raum}, Chris and {Reichardt}, Christian L. and {Remazeilles}, Mathieu and {Rephaeli}, Yoel and {Riechers}, Dominik and {Rojas}, Felipe and {Roy}, Anirban and {Sadeh}, Sharon and {Sakurai}, Yuki and {Salatino}, Maria and {Sathyanarayana Rao}, Mayuri and {Schaan}, Emmanuel and {Schmittfull}, Marcel and {Sehgal}, Neelima and {Seibert}, Joseph},
        title = "{The Simons Observatory: science goals and forecasts}",
      journal = {\jcap},
         year = 2019,
        month = feb,
       volume = {2019},
       number = {2},
          eid = {056},
        pages = {056},
          doi = {10.1088/1475-7516/2019/02/056},
archivePrefix = {arXiv},
       eprint = {1808.07445},
 primaryClass = {astro-ph.CO},
       adsurl = {https://ui.adsabs.harvard.edu/abs/2019JCAP...02..056A}
}

@INPROCEEDINGS{LiteBird,
       author = {{Hazumi}, M. and {Ade}, P.~A.~R. and {Adler}, A. and {Allys}, E. and {Arnold}, K. and {Auguste}, D. and {Aumont}, J. and {Aurlien}, R. and {Austermann}, J. and {Baccigalupi}, C. and {Banday}, A.~J. and {Banjeri}, R. and {Barreiro}, R.~B. and {Basak}, S. and {Beall}, J. and {Beck}, D. and {Beckman}, S. and {Bermejo}, J. and {de Bernardis}, P. and {Bersanelli}, M. and {Bonis}, J. and {Borrill}, J. and {Boulanger}, F. and {Bounissou}, S. and {Brilenkov}, M. and {Brown}, M. and {Bucher}, M. and {Calabrese}, E. and {Campeti}, P. and {Carones}, A. and {Casas}, F.~J. and {Challinor}, A. and {Chan}, V. and {Cheung}, K. and {Chinone}, Y. and {Cliche}, J.~F. and {Colombo}, L. and {Columbro}, F. and {Cubas}, J. and {Cukierman}, A. and {Curtis}, D. and {D'Alessandro}, G. and {Dachlythra}, N. and {De Petris}, M. and {Dickinson}, C. and {Diego-Palazuelos}, P. and {Dobbs}, M. and {Dotani}, T. and {Duband}, L. and {Duff}, S. and {Duval}, J.~M. and {Ebisawa}, K. and {Elleflot}, T. and {Eriksen}, H.~K. and {Errard}, J. and {Essinger-Hileman}, T. and {Finelli}, F. and {Flauger}, R. and {Franceschet}, C. and {Fuskeland}, U. and {Galloway}, M. and {Ganga}, K. and {Gao}, J.~R. and {Genova-Santos}, R. and {Gerbino}, M. and {Gervasi}, M. and {Ghigna}, T. and {Gjerl{\o}w}, E. and {Gradziel}, M.~L. and {Grain}, J. and {Grupp}, F. and {Gruppuso}, A. and {Gudmundsson}, J.~E. and {de Haan}, T. and {Halverson}, N.~W. and {Hargrave}, P. and {Hasebe}, T. and {Hasegawa}, M. and {Hattori}, M. and {Henrot-Versill{\'e}}, S. and {Herman}, D. and {Herranz}, D. and {Hill}, C.~A. and {Hilton}, G. and {Hirota}, Y. and {Hivon}, E. and {Hlozek}, R.~A. and {Hoshino}, Y. and {de la Hoz}, E. and {Hubmayr}, J. and {Ichiki}, K. and {Iida}, T. and {Imada}, H. and {Ishimura}, K. and {Ishino}, H. and {Jaehnig}, G. and {Kaga}, T. and {Kashima}, S. and {Katayama}, N. and {Kato}, A. and {Kawasaki}, T. and {Keskitalo}, R. and {Kisner}, T. and {Kobayashi}, Y. and {Kogiso}, N. and {Kogut}, A. and {Kohri}, K. and {Komatsu}, E. and {Komatsu}, K. and {Konishi}, K. and {Krachmalnicoff}, N. and {Kreykenbohm}, I. and {Kuo}, C.~L. and {Kushino}, A. and {Lamagna}, L. and {Lanen}, J.~V. and {Lattanzi}, M. and {Lee}, A.~T. and {Leloup}, C. and {Levrier}, F. and {Linder}, E. and {Louis}, T. and {Luzzi}, G. and {Maciaszek}, T. and {Maffei}, B. and {Maino}, D. and {Maki}, M. and {Mandelli}, S. and {Martinez-Gonzalez}, E. and {Masi}, S. and {Matsumura}, T. and {Mennella}, A. and {Migliaccio}, M. and {Minami}, Y. and {Mitsuda}, K. and {Montgomery}, J. and {Montier}, L. and {Morgante}, G. and {Mot}, B. and {Murata}, Y. and {Murphy}, J.~A. and {Nagai}, M. and {Nagano}, Y. and {Nagasaki}, T. and {Nagata}, R. and {Nakamura}, S. and {Namikawa}, T. and {Natoli}, P. and {Nerval}, S. and {Nishibori}, T. and {Nishino}, H. and {Noviello}, F. and {O'Sullivan}, C. and {Ogawa}, H. and {Ogawa}, H. and {Oguri}, S. and {Ohsaki}, H. and {Ohta}, I.~S. and {Okada}, N. and {Okada}, N. and {Pagano}, L. and {Paiella}, A. and {Paoletti}, D. and {Patanchon}, G. and {Peloton}, J. and {Piacentini}, F. and {Pisano}, G. and {Polenta}, G. and {Poletti}, D. and {Prouv{\'e}}, T. and {Puglisi}, G. and {Rambaud}, D. and {Raum}, C. and {Realini}, S. and {Reinecke}, M. and {Remazeilles}, M. and {Ritacco}, A. and {Roudil}, G. and {Rubino-Martin}, J.~A. and {Russell}, M. and {Sakurai}, H. and {Sakurai}, Y. and {Sandri}, M. and {Sasaki}, M. and {Savini}, G. and {Scott}, D. and {Seibert}, J. and {Sekimoto}, Y. and {Sherwin}, B. and {Shinozaki}, K. and {Shiraishi}, M. and {Shirron}, P. and {Signorelli}, G. and {Smecher}, G. and {Stever}, S. and {Stompor}, R. and {Sugai}, H. and {Sugiyama}, S. and {Suzuki}, A. and {Suzuki}, J.},
        title = "{LiteBIRD satellite: JAXA's new strategic L-class mission for all-sky surveys of cosmic microwave background polarization}",
    booktitle = {Space Telescopes and Instrumentation 2020: Optical, Infrared, and Millimeter Wave},
         year = 2020,
       editor = {{Lystrup}, Makenzie and {Perrin}, Marshall D.},
       series = {Society of Photo-Optical Instrumentation Engineers (SPIE) Conference Series},
       volume = {11443},
        month = dec,
          eid = {114432F},
        pages = {114432F},
          doi = {10.1117/12.2563050},
archivePrefix = {arXiv},
       eprint = {2101.12449},
 primaryClass = {astro-ph.IM},
       adsurl = {https://ui.adsabs.harvard.edu/abs/2020SPIE11443E..2FH}
}

@ARTICLE{Calabreseetal2011,
       author = {{Calabrese}, Erminia and {Huterer}, Dragan and {Linder}, Eric V. and {Melchiorri}, Alessandro and {Pagano}, Luca},
        title = "{Limits on dark radiation, early dark energy, and relativistic degrees of freedom}",
      journal = {\prd},
         year = 2011,
        month = jun,
       volume = {83},
       number = {12},
          eid = {123504},
        pages = {123504},
          doi = {10.1103/PhysRevD.83.123504},
archivePrefix = {arXiv},
       eprint = {1103.4132},
 primaryClass = {astro-ph.CO},
       adsurl = {https://ui.adsabs.harvard.edu/abs/2011PhRvD..83l3504C}
}

@ARTICLE{Archidiaconoetal2011,
       author = {{Archidiacono}, Maria and {Calabrese}, Erminia and {Melchiorri}, Alessandro},
        title = "{Case for dark radiation}",
      journal = {\prd},
         year = 2011,
        month = dec,
       volume = {84},
       number = {12},
          eid = {123008},
        pages = {123008},
          doi = {10.1103/PhysRevD.84.123008},
archivePrefix = {arXiv},
       eprint = {1109.2767},
 primaryClass = {astro-ph.CO},
       adsurl = {https://ui.adsabs.harvard.edu/abs/2011PhRvD..84l3008A}
}

@ARTICLE{WMAP,
       author = {{Bennett}, C.~L. and {Larson}, D. and {Weiland}, J.~L. and {Jarosik}, N. and {Hinshaw}, G. and {Odegard}, N. and {Smith}, K.~M. and {Hill}, R.~S. and {Gold}, B. and {Halpern}, M. and {Komatsu}, E. and {Nolta}, M.~R. and {Page}, L. and {Spergel}, D.~N. and {Wollack}, E. and {Dunkley}, J. and {Kogut}, A. and {Limon}, M. and {Meyer}, S.~S. and {Tucker}, G.~S. and {Wright}, E.~L.},
        title = "{Nine-year Wilkinson Microwave Anisotropy Probe (WMAP) Observations: Final Maps and Results}",
      journal = {\apjs},
         year = 2013,
        month = oct,
       volume = {208},
       number = {2},
          eid = {20},
        pages = {20},
          doi = {10.1088/0067-0049/208/2/20},
archivePrefix = {arXiv},
       eprint = {1212.5225},
 primaryClass = {astro-ph.CO},
       adsurl = {https://ui.adsabs.harvard.edu/abs/2013ApJS..208...20B}
}

@ARTICLE{ACT,
       author = {{Aiola}, Simone and {Calabrese}, Erminia and {Maurin}, Lo{\"\i}c and {Naess}, Sigurd and {Schmitt}, Benjamin L. and {Abitbol}, Maximilian H. and {Addison}, Graeme E. and {Ade}, Peter A.~R. and {Alonso}, David and {Amiri}, Mandana and {Amodeo}, Stefania and {Angile}, Elio and {Austermann}, Jason E. and {Baildon}, Taylor and {Battaglia}, Nick and {Beall}, James A. and {Bean}, Rachel and {Becker}, Daniel T. and {Bond}, J. Richard and {Bruno}, Sarah Marie and {Calafut}, Victoria and {Campusano}, Luis E. and {Carrero}, Felipe and {Chesmore}, Grace E. and {Cho}, Hsiao-mei and {Choi}, Steve K. and {Clark}, Susan E. and {Cothard}, Nicholas F. and {Crichton}, Devin and {Crowley}, Kevin T. and {Darwish}, Omar and {Datta}, Rahul and {Denison}, Edward V. and {Devlin}, Mark J. and {Duell}, Cody J. and {Duff}, Shannon M. and {Duivenvoorden}, Adriaan J. and {Dunkley}, Jo and {D{\"u}nner}, Rolando and {Essinger-Hileman}, Thomas and {Fankhanel}, Max and {Ferraro}, Simone and {Fox}, Anna E. and {Fuzia}, Brittany and {Gallardo}, Patricio A. and {Gluscevic}, Vera and {Golec}, Joseph E. and {Grace}, Emily and {Gralla}, Megan and {Guan}, Yilun and {Hall}, Kirsten and {Halpern}, Mark and {Han}, Dongwon and {Hargrave}, Peter and {Hasselfield}, Matthew and {Helton}, Jakob M. and {Henderson}, Shawn and {Hensley}, Brandon and {Hill}, J. Colin and {Hilton}, Gene C. and {Hilton}, Matt and {Hincks}, Adam D. and {Hlo{\v{z}}ek}, Ren{\'e}e and {Ho}, Shuay-Pwu Patty and {Hubmayr}, Johannes and {Huffenberger}, Kevin M. and {Hughes}, John P. and {Infante}, Leopoldo and {Irwin}, Kent and {Jackson}, Rebecca and {Klein}, Jeff and {Knowles}, Kenda and {Koopman}, Brian and {Kosowsky}, Arthur and {Lakey}, Vincent and {Li}, Dale and {Li}, Yaqiong and {Li}, Zack and {Lokken}, Martine and {Louis}, Thibaut and {Lungu}, Marius and {MacInnis}, Amanda and {Madhavacheril}, Mathew and {Maldonado}, Felipe and {Mallaby-Kay}, Maya and {Marsden}, Danica and {McMahon}, Jeff and {Menanteau}, Felipe and {Moodley}, Kavilan and {Morton}, Tim and {Namikawa}, Toshiya and {Nati}, Federico and {Newburgh}, Laura and {Nibarger}, John P. and {Nicola}, Andrina and {Niemack}, Michael D. and {Nolta}, Michael R. and {Orlowski-Sherer}, John and {Page}, Lyman A. and {Pappas}, Christine G. and {Partridge}, Bruce and {Phakathi}, Phumlani and {Pisano}, Giampaolo and {Prince}, Heather and {Puddu}, Roberto and {Qu}, Frank J. and {Rivera}, Jesus and {Robertson}, Naomi and {Rojas}, Felipe and {Salatino}, Maria and {Schaan}, Emmanuel and {Schillaci}, Alessandro and {Sehgal}, Neelima and {Sherwin}, Blake D. and {Sierra}, Carlos and {Sievers}, Jon and {Sifon}, Cristobal and {Sikhosana}, Precious and {Simon}, Sara and {Spergel}, David N. and {Staggs}, Suzanne T. and {Stevens}, Jason and {Storer}, Emilie and {Sunder}, Dhaneshwar D. and {Switzer}, Eric R. and {Thorne}, Ben and {Thornton}, Robert and {Trac}, Hy and {Treu}, Jesse and {Tucker}, Carole and {Vale}, Leila R. and {Van Engelen}, Alexander and {Van Lanen}, Jeff and {Vavagiakis}, Eve M. and {Wagoner}, Kasey and {Wang}, Yuhan and {Ward}, Jonathan T. and {Wollack}, Edward J. and {Xu}, Zhilei and {Zago}, Fernando and {Zhu}, Ningfeng},
        title = "{The Atacama Cosmology Telescope: DR4 maps and cosmological parameters}",
      journal = {\jcap},
         year = 2020,
        month = dec,
       volume = {2020},
       number = {12},
          eid = {047},
        pages = {047},
          doi = {10.1088/1475-7516/2020/12/047},
archivePrefix = {arXiv},
       eprint = {2007.07288},
 primaryClass = {astro-ph.CO},
       adsurl = {https://ui.adsabs.harvard.edu/abs/2020JCAP...12..047A}
}

@ARTICLE{Steigman_Schramm_Gunn77,
       author = {{Steigman}, Gary and {Schramm}, David N. and {Gunn}, James E.},
        title = "{Cosmological limits to the number of massive leptons}",
      journal = {Physics Letters B},
         year = 1977,
        month = jan,
       volume = {66},
       number = {2},
        pages = {202-204},
          doi = {10.1016/0370-2693(77)90176-9},
       adsurl = {https://ui.adsabs.harvard.edu/abs/1977PhLB...66..202S}
}

@article{DESI_DR2,
    author = "{DESI Collaboration} and Abdul-Karim, M. and Aguilar, J. and Ahlen, S. and Alam, S. and Allen, L. and Allende Prieto, C. and others",
    title = "{DESI DR2 results II: measurements of baryon acoustic oscillations and cosmological constraints}",
    eprint = "2503.14738",
    archivePrefix = "arXiv",
    year = "2025"
}

@article{DESI_DR1_1,
    author = "{DESI Collaboration} and Adame, A. G. and Aguilar, J. and Ahlen, S. and Alam, S. and Alexander, D. M. and Alvarez, M. and others",
    title = "{DESI 2024 VI: Cosmological Constraints from the Measurements of Baryon Acoustic Oscillations}",
    eprint = "2404.03002",
    archivePrefix = "arXiv",
    year = "2024"
}

@article{BAO,
    author = "Eisenstein, D. J. and Seo, H.-J. and White, M.",
    title = "{On the robustness of the acoustic scale in the low-redshift clustering of matter}",
    journal = "Astrophys. J.",
    volume = "664",
    pages = "660",
    doi = "10.1086/518712",
    eprint = "astro-ph/0604361",
    archivePrefix = "arXiv",
    year = "2007"
}

@article{DR_Notari,
    author = "Allali, I. J. and Notari, A. and Rompineve, F.",
    title = "{Reduced Hubble Tension in Dark Radiation Models after DESI 2024}",
    journal = "JCAP",
    volume = "03",
    pages = "023",
    doi = "10.1088/1475-7516/2025/03/023",
    eprint = "2404.15220",
    archivePrefix = "arXiv",
    year = "2025"
}

@article{DR_CMB,
    author = "Saravanan, M. M. and Brinckmann, T. and Loverde, M. and Weinera, Z. J.",
    title = "{Abundance and properties of dark radiation from the cosmic microwave background}",
    eprint = "2503.04671",
    archivePrefix = "arXiv",
    year = "2025"
}

@article{Baumannetal2016,
    author = "Baumann, D. and Green, D. and Meyers, J. and Wallisch, B.",
    title = "{Phases of New Physics in the CMB}",
    journal = "JCAP",
    volume = "01",
    pages = "007",
    doi = "10.1088/1475-7516/2016/01/007",
    eprint = "1508.06342",
    archivePrefix = "arXiv",
    year = "2016"
}

@article{Planck_2018,
    author = "{PLANCK collaboration}",
    title = "{Planck 2018 results. VI. Cosmological parameters}",
    journal = "Astron. Astrophys.",
    volume = "641",
    pages = "A6",
    doi = "10.1051/0004-6361/201833910",
    eprint = "1807.06209",
    archivePrefix = "arXiv",
    year = "2020"
}

@article{Baumann_2016_axions,
    author = "Baumann, D. and Green, D. and Wallisch, B.",
    title = "{New Target for Cosmic Axion Searches}",
    journal = "Phys. Rev. Lett.",
    volume = "117",
    pages = "171301",
    doi = "10.1103/PhysRevLett.117.171301",
    eprint = "1604.08614",
    archivePrefix = "arXiv",
    year = "2016"
}

@article{Turner_1987_axions,
    author = "Turner, M. S.",
    title = "{Early-Universe thermal production of not-so-invisible axions}",
    journal = "Phys. Rev. Lett.",
    volume = "59",
    pages = "2489",
    doi = "10.1103/PhysRevLett.59.2489",
    year = "1987"
}

@article{CyrRacine_2014_DAO,
    author = "Cyr-Racine, Francis-Yan and de Putter, Roland and Raccanelli, Alvise and Sigurdson, Kris",
    title = "{Constraints on Large-Scale Dark Acoustic Oscillations from Cosmology}",
    journal = "Phys. Rev. D",
    volume = "89",
    pages = "063517",
    doi = "10.1103/PhysRevD.89.063517",
    eprint = "1310.3278",
    archivePrefix = "arXiv",
    year = "2014"
}

@article{Tristram_2024_PlanckPR4,
    author = "Tristram, M. and others",
    title = "{Cosmological parameters derived from the final Planck data release (PR4)}",
    journal = "Astron. Astrophys.",
    volume = "682",
    pages = "A37",
    doi = "10.1051/0004-6361/202348015",
    eprint = "2309.10034",
    archivePrefix = "arXiv",
    year = "2024"
}

@article{Carron_2022_PlanckPR4_lensing,
    author = "Carron, J. and Mirmelstein, M. and Lewis, A.",
    title = "{CMB lensing from Planck PR4 maps}",
    journal = "JCAP",
    volume = "09",
    pages = "039",
    doi = "10.1088/1475-7516/2022/09/039",
    eprint = "2206.07773",
    archivePrefix = "arXiv",
    year = "2022"
}

@article{Blas_2011_CLASS2,
    author = "Blas, D. and Lesgourgues, J. and Tram, T.",
    title = "{The Cosmic Linear Anisotropy Solving System (CLASS) II: Approximation schemes}",
    journal = "JCAP",
    volume = "07",
    pages = "034",
    doi = "10.1088/1475-7516/2011/07/034",
    eprint = "1104.2933",
    archivePrefix = "arXiv",
    year = "2011"
}

@article{Lesgourgues_2011_CLASS1,
    author = "Lesgourgues, J.",
    title = "{The Cosmic Linear Anisotropy Solving System (CLASS) I: Overview}",
    eprint = "1104.2932",
    archivePrefix = "arXiv",
    year = "2011"
}

@article{Lesgourgues_2011_CLASS4,
    author = "Lesgourgues, J. and Tram, T.",
    title = "{The Cosmic Linear Anisotropy Solving System (CLASS) IV: efficient implementation of non-cold relics}",
    journal = "JCAP",
    volume = "09",
    pages = "032",
    doi = "10.1088/1475-7516/2011/09/032",
    eprint = "1104.2935",
    archivePrefix = "arXiv",
    year = "2011"
}

@article{Torrado_2021_Cobaya,
    author = "Torrado, J. and Lewis, A.",
    title = "{Cobaya: Code for Bayesian Analysis of hierarchical physical models}",
    journal = "JCAP",
    volume = "05",
    pages = "057",
    doi = "10.1088/1475-7516/2021/05/057",
    eprint = "2005.05290",
    archivePrefix = "arXiv",
    year = "2021"
}

@misc{Torrado_2019_Cobaya_ASCL,
    author = "Torrado, J. and Lewis, A.",
    title = "{Cobaya: Bayesian analysis in cosmology}",
    howpublished = "Astrophysics Source Code Library, record ascl:1910.019",
    year = "2019",
    month = "Oct."
}

@article{Ghosh_2025_DarkMatter,
    author = "Ghosh, S. and Ho, D. W. R. and Tsai, Y.",
    title = "{Dark Matter-Radiation Scattering Enhances CMB Phase Shift through Dark Matter-loading}",
    journal = "JCAP",
    volume = "01",
    pages = "058",
    doi = "10.1088/1475-7516/2025/01/058",
    eprint = "2405.08064",
    archivePrefix = "arXiv",
    year = "2025"
}

@article{Smith_2024_EDE,
    author = "Smith, Tristan L. and Poulin, Vivian",
    title = "{Current small-scale CMB constraints to axionlike early dark energy}",
    journal = "Phys. Rev. D",
    volume = "109",
    pages = "103506",
    doi = "10.1103/PhysRevD.109.103506",
    eprint = "2309.03265",
    archivePrefix = "arXiv",
    year = "2024"
}

@article{Pisanti_2008_Parthenope,
    author = "Pisanti, O. and Cirillo, A. and Esposito, S. and Iocco, F. and Mangano, G. and Miele, G. and Serpico, P. D.",
    title = "{PArthENoPE: Public Algorithm Evaluating the Nucleosynthesis of Primordial Elements}",
    journal = "Comput. Phys. Commun.",
    volume = "178",
    pages = "956--971",
    doi = "10.1016/j.cpc.2008.02.015",
    eprint = "0705.0290",
    archivePrefix = "arXiv",
    year = "2008"
}

@article{Gelman_1992_Convergence,
    author = "Gelman, A. and Rubin, D. B.",
    title = "{Inference from Iterative Simulation Using Multiple Sequences}",
    journal = "Statist. Sci.",
    volume = "7",
    pages = "457--472",
    doi = "10.1214/ss/1177011136",
    year = "1992"
}

@article{Lewis_2025_GetDist,
    author = "Lewis, A.",
    title = "{GetDist: a Python package for analysing Monte Carlo samples}",
    journal = "JCAP",
    volume = "08",
    pages = "025",
    doi = "10.1088/1475-7516/2025/08/025",
    eprint = "1910.13970",
    archivePrefix = "arXiv",
    year = "2025"
}

@article{Verde_2024_H0,
    author = "Verde, Licia and Schöneberg, Nils and Gil-Marín, Héctor",
    title = "{A Tale of Many $H_0$}",
    journal = "Annu. Rev. Astron. Astrophys.",
    volume = "62",
    pages = "287--331",
    doi = "10.1146/annurev-astro-052622-033813",
    eprint = "2311.13305",
    archivePrefix = "arXiv",
    year = "2024"
}

@article{Riess_2022_SH0ES,
    author = "Riess, Adam G. and Yuan, Wenlong and Macri, Lucas M. and others",
    title = "{A Comprehensive Measurement of the Local Value of the Hubble Constant with 1 km s$^{-1}$ Mpc$^{-1}$ Uncertainty from the Hubble Space Telescope and the SH0ES Team}",
    journal = "Astrophys. J. Lett.",
    volume = "934",
    pages = "L7",
    doi = "10.3847/2041-8213/ac5c5b",
    eprint = "2112.04510",
    archivePrefix = "arXiv",
    year = "2022"
}

@article{Garny_2025_DAO,
    author = "Garny, M. and Niedermann, F. and Sloth, M. S.",
    title = "{Dark Acoustic Oscillations as an Early-Universe Explanation of the DESI Anomaly}",
    eprint = "2512.15870",
    archivePrefix = "arXiv",
    year = "2025"
}

@article{Knox_2020_Hubble,
    author = "Knox, L. and Millea, M.",
    title = "{Hubble constant hunter's guide}",
    journal = "Phys. Rev. D",
    volume = "101",
    pages = "043533",
    doi = "10.1103/PhysRevD.101.043533",
    eprint = "1908.03663",
    archivePrefix = "arXiv",
    year = "2020"
}

@article{Hou_2013_Neutrinos,
    author = "Hou, Z. and Keisler, R. and Knox, L. and Millea, M. and Reichardt, C.",
    title = "{How massless neutrinos affect the cosmic microwave background damping tail}",
    journal = "Phys. Rev. D",
    volume = "87",
    pages = "083008",
    doi = "10.1103/PhysRevD.87.083008",
    eprint = "1104.2333",
    archivePrefix = "arXiv",
    year = "2013"
}

@article{Rosenberg_2022_CamSpecPR4,
    author = "Rosenberg, E. and Gratton, S. and Efstathiou, G.",
    title = "{CMB power spectra and cosmological parameters from Planck PR4 with CamSpec}",
    journal = "Mon. Not. Roy. Astron. Soc.",
    volume = "517",
    pages = "4620",
    doi = "10.1093/mnras/stac2744",
    eprint = "2205.10869",
    archivePrefix = "arXiv",
    year = "2022"
}

@article{Planck_2018_V,
    author = "{Planck Collaboration} and Aghanim, N. and others",
    title = "{Planck 2018 results. V. CMB power spectra and likelihoods}",
    journal = "Astron. Astrophys.",
    volume = "641",
    pages = "A5",
    doi = "10.1051/0004-6361/201936386",
    eprint = "1907.12875",
    archivePrefix = "arXiv",
    year = "2020"
}

@article{Fixsen_2009_CMB,
    author = "Fixsen, D. J.",
    title = "{The Temperature of the Cosmic Microwave Background}",
    journal = "Astrophys. J.",
    volume = "707",
    pages = "916",
    doi = "10.1088/0004-637X/707/2/916",
    eprint = "0911.1955",
    archivePrefix = "arXiv",
    year = "2009"
}

@article{Abazajian_2012_Sterile,
    author = "Abazajian, K. N. and others",
    title = "{Light Sterile Neutrinos: A White Paper}",
    eprint = "1204.5379",
    archivePrefix = "arXiv",
    year = "2012"
}

@article{Baumann_2018_TASI,
    author = "Baumann, D.",
    title = "{TASI Lectures on Primordial Cosmology}",
    eprint = "1807.03098",
    archivePrefix = "arXiv",
    year = "2018"
}

@article{Guy_2007_SALT2,
    author = "Guy, J. and others",
    title = "{SALT2: using distant supernovae to improve the use of type Ia supernovae as distance indicators}",
    journal = "Astron. Astrophys.",
    volume = "466",
    pages = "11--21",
    doi = "10.1051/0004-6361:20066930",
    eprint = "astro-ph/0701828",
    archivePrefix = "arXiv",
    year = "2007"
}

@article{Chevallier_2001_CPL,
    author = "Chevallier, M. and Polarski, D.",
    title = "{Accelerating universes with scaling dark matter}",
    journal = "Int. J. Mod. Phys. D",
    volume = "10",
    pages = "213",
    doi = "10.1142/S0218271801000822",
    eprint = "gr-qc/0009008",
    archivePrefix = "arXiv",
    year = "2001"
}

@article{Linder_2003_CPL,
    author = "Linder, E. V.",
    title = "{Exploring the Expansion History of the Universe}",
    journal = "Phys. Rev. Lett.",
    volume = "90",
    pages = "091301",
    doi = "10.1103/PhysRevLett.90.091301",
    eprint = "astro-ph/0208512",
    archivePrefix = "arXiv",
    year = "2003"
}

@article{Spiegelhalter_2014_DIC,
    author = "Spiegelhalter, D. J. and Best, N. G. and Carlin, B. P. and van der Linde, A.",
    title = "{The deviance information criterion: 12 years on}",
    journal = "J. Roy. Stat. Soc. B",
    volume = "76",
    pages = "485--493",
    doi = "10.1111/rssb.12062",
    year = "2014"
}

@article{Trotta_2008_Bayes,
    author = "Trotta, R.",
    title = "{Bayes in the sky: Bayesian inference and model selection in cosmology}",
    journal = "Contemp. Phys.",
    volume = "49",
    pages = "71--104",
    doi = "10.1080/00107510802066753",
    eprint = "0803.4089",
    archivePrefix = "arXiv",
    year = "2008"
}

@article{Grandis_2016_Tension,
    author = "Grandis, S. and Rapetti, D. and Saro, A. and Mohr, J. J. and Dietrich, J. P.",
    title = "{Quantifying tensions between CMB and distance data sets in models with free curvature or lensing amplitude}",
    journal = "Mon. Not. Roy. Astron. Soc.",
    volume = "463",
    pages = "1416--1430",
    doi = "10.1093/mnras/stw2028",
    eprint = "1604.06463",
    archivePrefix = "arXiv",
    year = "2016"
}

@article{Kass_1995_Bayes,
    author = "Kass, R. E. and Raftery, A. E.",
    title = "{Bayes Factors}",
    journal = "J. Am. Stat. Assoc.",
    volume = "90",
    pages = "773--795",
    doi = "10.1080/01621459.1995.10476572",
    year = "1995"
}

@article{Kenworthy_2021_SALT3,
    author = "Kenworthy, W. D. and others",
    title = "{SALT3: An Improved Type Ia Supernova Model for Measuring Cosmic Distances}",
    journal = "Astrophys. J.",
    volume = "923",
    pages = "265",
    doi = "10.3847/1538-4357/ac30d8",
    eprint = "2104.07795",
    archivePrefix = "arXiv",
    year = "2021"
}

@article{Popovic_2025_Dovekie_Calib,
    author = "Popovic, B. and others",
    title = "{A Reassessment of the Pantheon+ and DES 5YR Calibration Uncertainties: Dovekie}",
    eprint = "2506.05471",
    archivePrefix = "arXiv",
    year = "2025"
}

@article{Popovic_2025_DES_Evolving,
    author = "Popovic, B. and others",
    title = "{The Dark Energy Survey Supernova Program: A Reanalysis Of Cosmology Results And Evidence For Evolving Dark Energy With An Updated Type Ia Supernova Calibration}",
    eprint = "2511.07517",
    archivePrefix = "arXiv",
    year = "2025"
}

@article{Verde_2013_Tension,
    author = "Verde, L. and Protopapas, P. and Jimenez, R.",
    title = "{Planck and the local Universe: quantifying the tension}",
    journal = "Phys. Dark Univ.",
    volume = "2",
    pages = "166--175",
    doi = "10.1016/j.dark.2013.09.001",
    eprint = "1306.6766",
    archivePrefix = "arXiv",
    year = "2013"
}

@article{Chaves_Montero_2026_DESILya,
    author = "Chaves-Montero, J. and others",
    title = "{Cosmological analysis of the DESI DR1 Lyman alpha 1D power spectrum}",
    eprint = "2601.21432",
    archivePrefix = "arXiv",
    year = "2026"
}

@article{Caldwell_2002_Phantom,
    author = "Caldwell, R. R.",
    title = "{A Phantom Menace? Cosmological consequences of a dark energy component with super-negative equation of state}",
    journal = "Phys. Lett. B",
    volume = "545",
    pages = "23--29",
    doi = "10.1016/S0370-2693(02)02589-3",
    eprint = "astro-ph/9908168",
    archivePrefix = "arXiv",
    year = "2002"
}

@article{Carroll_2003_Phantom,
    author = "Carroll, S. M. and Hoffman, M. and Trodden, M.",
    title = "{Can the dark energy equation-of-state parameter w be less than -1?}",
    journal = "Phys. Rev. D",
    volume = "68",
    pages = "023509",
    doi = "10.1103/PhysRevD.68.023509",
    eprint = "astro-ph/0301273",
    archivePrefix = "arXiv",
    year = "2003"
}

@article{Vikman_2005_Phantom,
    author = "Vikman, A.",
    title = "{Can dark energy evolve to the phantom?}",
    journal = "Phys. Rev. D",
    volume = "71",
    pages = "023515",
    doi = "10.1103/PhysRevD.71.023515",
    eprint = "astro-ph/0407107",
    archivePrefix = "arXiv",
    year = "2005"
}

@article{Hu_2005_Phantom,
    author = "Hu, W.",
    title = "{Crossing the phantom divide: Dark energy internal degrees of freedom}",
    journal = "Phys. Rev. D",
    volume = "71",
    pages = "047301",
    doi = "10.1103/PhysRevD.71.047301",
    eprint = "astro-ph/0410680",
    archivePrefix = "arXiv",
    year = "2005"
}

@article{Koussour_2023_Phantom,
    author = "Koussour, M. and Myrzakulov, N. and Alfedeel, A. H. A. and Abebe, A.",
    title = "{Constraining the cosmological model of modified $f(Q)$ gravity: Phantom dark energy and observational insights}",
    journal = "Prog. Theor. Exp. Phys.",
    volume = "2023",
    pages = "113E01",
    doi = "10.1093/ptep/ptad133",
    eprint = "2310.15067",
    archivePrefix = "arXiv",
    year = "2023"
}

@article{Wolf_2025_Phantom,
    author = "Wolf, W. J. and Ferreira, P. G. and Garc\'\i{}a-Garc\'\i{}a, C.",
    title = "{Matching current observational constraints with nonminimally coupled dark energy}",
    journal = "Phys. Rev. D",
    volume = "111",
    pages = "L041303",
    doi = "10.1103/PhysRevD.111.L041303",
    eprint = "2409.17019",
    archivePrefix = "arXiv",
    year = "2025"
}

@article{Clifton_2012_ModifiedGravity,
    author = "Clifton, T. and Ferreira, P. G. and Padilla, A. and Skordis, C.",
    title = "{Modified Gravity and Cosmology}",
    journal = "Phys. Rept.",
    volume = "513",
    pages = "1--189",
    doi = "10.1016/j.physrep.2012.01.001",
    eprint = "1106.2476",
    archivePrefix = "arXiv",
    year = "2012"
}

@article{Ishak_2019_TestingGR,
    author = "Ishak, M.",
    title = "{Testing General Relativity in Cosmology}",
    journal = "Living Rev. Rel.",
    volume = "22",
    pages = "1",
    doi = "10.1007/s41114-018-0017-4",
    eprint = "1806.10122",
    archivePrefix = "arXiv",
    year = "2019"
}

@article{Gelman_2014_WAIC,
    author = "Gelman, A. and Hwang, J. and Vehtari, A.",
    title = "{Understanding predictive information criteria for Bayesian models}",
    journal = "Statist. Comput.",
    volume = "24",
    pages = "997--1016",
    doi = "10.1007/s11222-013-9416-2",
    eprint = "1307.5928",
    archivePrefix = "arXiv",
    year = "2014"
}

@article{Trotta_2007_Applications,
    author = "Trotta, R.",
    title = "{Applications of Bayesian model selection to cosmological parameters}",
    journal = "Mon. Not. Roy. Astron. Soc.",
    volume = "378",
    pages = "72--82",
    eprint = "astro-ph/0504022",
    archivePrefix = "arXiv",
    year = "2007"
}

@article{Garny_2026_ACT_DR6,
    author = "Garny, Mathias and Niedermann, Florian and Sloth, Martin S.",
    title = "{The End of the First Act: Spectral Running, Interacting Dark Radiation, and the Hubble Tension in Light of ACT DR6 Data}",
    journal = "ArXiv e-prints",
    eprint = "2604.26541",
    archivePrefix = "arXiv",
    year = "2026"
}

@article{Garny_2026_DAO,
    author = "Garny, Mathias and Niedermann, Florian and Sloth, Martin S.",
    title = "{Dark Acoustic Oscillations and the Hubble Tension}",
    journal = "ArXiv e-prints",
    eprint = "2602.23895",
    archivePrefix = "arXiv",
    year = "2026"
}

@article{Schoeneberg_2023_SteppedDR,
    author = "Sch{\"o}neberg, N. and Franco Abell\'an, G. and Simon, T. and Bartlett, A. and Patel, Y. and Smith, T. L.",
    title = "{The weak, the strong and the ugly -- A comparative analysis of interacting stepped dark radiation}",
    journal = "Phys. Rev. D",
    volume = "108",
    pages = "123513",
    doi = "10.1103/PhysRevD.108.123513",
    eprint = "2306.12469",
    archivePrefix = "arXiv",
    year = "2023"
}

@article{Aloni_2021_Step,
    author = "Aloni, Daniel and Berlin, Asher and Joseph, Melissa and Schmaltz, Martin and Weiner, Neal",
    title = "{A Step in Understanding the Hubble Tension}",
    journal = "ArXiv e-prints",
    eprint = "2111.00014",
    archivePrefix = "arXiv",
    year = "2021"
}

@article{Chaussidon_2025_DESI_EDE,
    author = "Chaussidon, E. and White, M. and others",
    title = "{Early time solution as an alternative to the late time evolving dark energy with DESI DR2 BAO}",
    journal = "ArXiv e-prints",
    eprint = "2503.24343",
    archivePrefix = "arXiv",
    year = "2025"
}

@article{Poulin_2019_EDE,
    author = "Poulin, V. and Smith, T. L. and Karwal, T. and Kamionkowski, M.",
    title = "{Early Dark Energy Can Resolve The Hubble Tension}",
    journal = "Phys. Rev. Lett.",
    volume = "122",
    pages = "221301",
    doi = "10.1103/PhysRevLett.122.221301",
    eprint = "1811.04083",
    archivePrefix = "arXiv",
    year = "2019"
}

\end{document}